# Fundamental Limits of Communications in Interference Networks

## Part IV: Networks with a Sequence of Less-Noisy Receivers

Reza K. Farsani[1]

Email: reza_khosravi@alum.sharif.ir

***Abstract:*** **In this fourth part of our multi-part papers, some classes of interference networks with a sequence of less-noisy receivers are identified for which a successive decoding scheme achieves the sum-rate capacity. First, the two-receiver networks are analyzed; it is demonstrated that the unified outer bounds derived in Part III of our multi-part papers are sum-rate optimal for network scenarios which satisfy certain less-noisy conditions. Then, the multi-receiver networks are considered. These networks are not well understood. One of the main difficulties in the analysis of such scenarios is how to establish useful capacity outer bounds. In this paper, a novel technique requiring a sequential application of the Csiszar-Korner identity is developed to establish powerful single-letter outer bounds on the sum-rate capacity of multi-receiver interference networks which satisfy certain less-noisy conditions. By using these outer bounds, a full characterization of the sum-rate capacity is derived for general interference networks of arbitrary large sizes with a sequence of less-noisy receivers. Some generalizations of these outer bounds are also presented where we can obtain exact sum-rate capacity for various scenarios.**

[1] Reza K. Farsani was with the department of electrical engineering, Sharif University of Technology. He is by now with the school of cognitive sciences, Institute for Research in Fundamental Sciences (IPM), Tehran, Iran.

**To:**

**THE MAHDI**

# Contents

# I. INTRODUCTION

As defined in Part I [1] of our multi-part papers, the interference network is generally referred to as any single-hop communication scenario which is composed of a number of transmitters and a number of receivers with a given distribution of messages among transmitters and receivers. In recent years, a large body of studies in network information theory has been regarded to characterize fundamental limits of communications in the interference networks. Despite progress for simple scenarios, the interference networks of large sizes are still far less understood [9, p. 6-64]. In fact, the main focus of many researches in this area is to develop approximate capacity results as well as to characterize the degrees of freedom -a parameter which determines the behavior of the capacity in high signal to noise regime- for Gaussian networks mostly relying on the new technique of *interference alignment*. This idea was introduced for the MIMO X-Channel by Maddah-Ali, *et al* in [10] and for the multi-user CIC by Cadambe and Jafar in [11]. In the latter paper, the authors utilized the interference alignment technique to determine the degrees of freedom of the multi-user Gaussian time-varying Classical Interference Channel (CIC). For a detailed discussion regarding the interference alignment approaches refer to [12].

In these series of multi-part papers, we develop many new results and techniques for the interference networks of arbitrary large sizes. Our systematic study was launched in Part I [1] by considering the basic building blocks including the Multiple Access Channel (MAC), the Broadcast Channel (BC), the CIC, and the Cognitive Radio Channel (CRC). We studied in details these basic structures and developed new results. We also provided a detailed review of the existing literature (Sec. I of Part I), which deal with capacity bounds for various interference networks. In Part II [2], we considered the degraded networks where a full characterization of the sum-rate capacity for such networks was derived. In addition, we demonstrated that the transmission of only a certain subset of messages is sufficient to achieve the sum-rate capacity in such networks. We presented an algorithm to exactly determine this desired subset of messages. Using the sum-rate capacity expression for degraded networks, we also established useful outer bounds for the general non-degraded networks. We also applied our algorithm to simplify these outer bounds. Part III [3] of our multi-part papers is related to the study of the information flow in strong interference regime. We developed new approaches to derive strong interference conditions for any interference network of arbitrary large sizes. For this development, we proved some new lemmas which had a central role in our derivations. Indeed, in Part III [3, Th. 11], we derived a capacity result for a general single-hop communication network with strong interference. Now, in this fourth part of our multi-part papers, we identify classes of interference networks with a sequence of less-noisy receivers.

*Less-noisy networks:* The term "less-noisy" is mostly referred to identify a class of two-user BCs [13] for which a superposition coding scheme achieves the capacity. Let $\mathcal{X}$ and $\mathcal{Y}_1$ and $\mathcal{Y}_2$ be arbitrary alphabet sets. A two-user BC with the input variable $X \in \mathcal{X}$ and the output variables $Y_1 \in \mathcal{Y}_1$ and $Y_2 \in \mathcal{Y}_2$ is given by a conditional probability distribution function $\mathbb{P}(y_1, y_2|x)$ on the set $\mathcal{Y}_1 \times \mathcal{Y}_2 \times \mathcal{X}$. Now, let the channel satisfies the following relation:

$$I(U;Y_2) \leq I(U;Y_1), \quad \text{for all distributions} \quad P_{UX}(u,x) \tag{1}$$

In this case, it is said that the receiver $Y_1$ is less-noisy than the receiver $Y_2$. For such a channel, the optimal coding strategy requires that the less-noisy receiver $Y_1$ (stronger receiver) decodes the message corresponding to the receiver $Y_2$ (weaker receiver) as well. Note that a degraded BC, where $U \to Y_1 \to Y_2$ forms a Markov chain, trivially satisfies the condition (1). However, a less-noisy channel may not be necessarily degraded [14]. In other words, the less-noisy BCs strictly include the degraded ones as a subset, although both have the same optimal coding strategy.

In this paper, we adapt the less-noisy concept for other interference networks as follows. Let $\mathcal{X}_1, \dots, \mathcal{X}_{K_1}$ and $\mathcal{Y}_1, \dots, \mathcal{Y}_{K_2}$ be sequences of arbitrary alphabet sets. A general $K_1$-transmitter/$K_2$-receiver interference network with the input variables $X_1, \dots, X_{K_1}$, where $X_i \in \mathcal{X}_i, i = 1, \dots, K_1$, and the output variables $Y_1, \dots, Y_{K_2}$, where $Y_j \in \mathcal{Y}_j, j = 1, \dots, K_2$, is given by a conditional probability distribution $\mathbb{P}(y_1, \dots, y_{K_2}|x_1, \dots, x_{K_1})$ on the set $\mathcal{Y}_1 \times \mathcal{Y}_2 \times \dots \times \mathcal{Y}_{K_2} \times \mathcal{X}_1 \times \dots \times \mathcal{X}_{K_2}$. Let $Y_A$ and $Y_B$ be two arbitrary outputs. Also, let $\mathbb{X}_C$ be an arbitrary subset of the set of the inputs $\mathbb{X} \triangleq \{X_1, \dots, X_{K_1}\}$. Consider the following relation between the outputs $Y_A$ and $Y_B$:

$$I(U;Y_B|\mathbb{X}_C) \leq I(U;Y_A|\mathbb{X}_C), \quad \text{for all joint distributions} \quad P_{U,\mathbb{X}-\mathbb{X}_C} \prod_{X_i \in \mathbb{X}_C} P_{X_i} \tag{2}$$

Note that according to [3, Lemma 4], this condition extends to hold for any arbitrary joint distribution $P_{U\mathbb{X}}$. We generally refer to such a condition as a *less-noisy condition or less-noisy ordering* between the receivers $Y_A$ and $Y_B$. In fact, the condition (2) indicates that, given the inputs $\mathbb{X}_C$, the receiver $Y_A$ is less-noisy than the receiver $Y_B$. In the present part, we identify some classes of interference networks with certain less-noisy conditions for which the receivers can be arranged in a successive order from stronger to weaker (from less-noisy to more-noisy) so that a successive decoding scheme achieves the sum-rate capacity: each receiver decodes its messages as well as all the messages corresponding to receivers weaker than itself in a successive order from weaker receivers to the stronger ones and lastly its own messages. We call such networks as *networks with a sequence of less-noisy receivers*. The less-noisy networks strictly include the degraded networks as a subset. It should also be noted that the fully connected less-noisy networks are associated to a successive decoding scheme; however, for some specific scenarios (networks in which some receivers are unconnected to some transmitters) the corresponding sum-capacity achieving coding scheme is reduced to a simple treating interference as noise strategy. We also identify classes of interference networks with such a characteristic. Such networks are conventionally called *noisy interference networks*. In addition, for some scenarios, combinations of these two strategies, i.e., the successive decoding and treating interference as noise, achieves the sum-rate capacity.

The rest of the paper is organized as follows. We first start with mathematical preliminaries in Section II. Then, in Section III, we will consider the two-receiver networks. We analyze the unified outer bound derived in Part III [3, Sec. IV.A] for these networks and show that, under certain less-noisy conditions, this bound is sum-rate optimal. Thus, classes of less-noisy networks are identified and the explicit sum-rate capacity is established.

In Section IV, we will consider the networks with arbitrary number of receivers. One of the main difficulties in analysis of such scenarios is how to establish useful capacity outer bounds. In Subsection IV.A, we develop a novel technique requiring a sequential application of the Csiszar-Korner identity [15] to establish powerful single-letter outer bounds on the sum-rate capacity of multi-receiver interference networks which satisfy specific less-noisy conditions. Next, using the derived outer bounds, a full characterization of the sum-rate capacity is derived for general interference networks of arbitrary large sizes with a sequence of less-noisy receivers. Finally, in Subsection IV.B, we will present some generalizations of our outer bounds and show that they can be used to establish exact sum-rate capacity for various scenarios.

# II. PRELIMINARIES

In the present part of our multi-part papers, we use the same notations and definitions as Part I [1, Sec. II]. Also, we assume the reader is familiar with the preliminaries provided in Part I [1, Sec. II.B] regarding the interference networks and their structures. Let us briefly review the essentials of these networks. The general interference network model has been shown in Fig. 1.

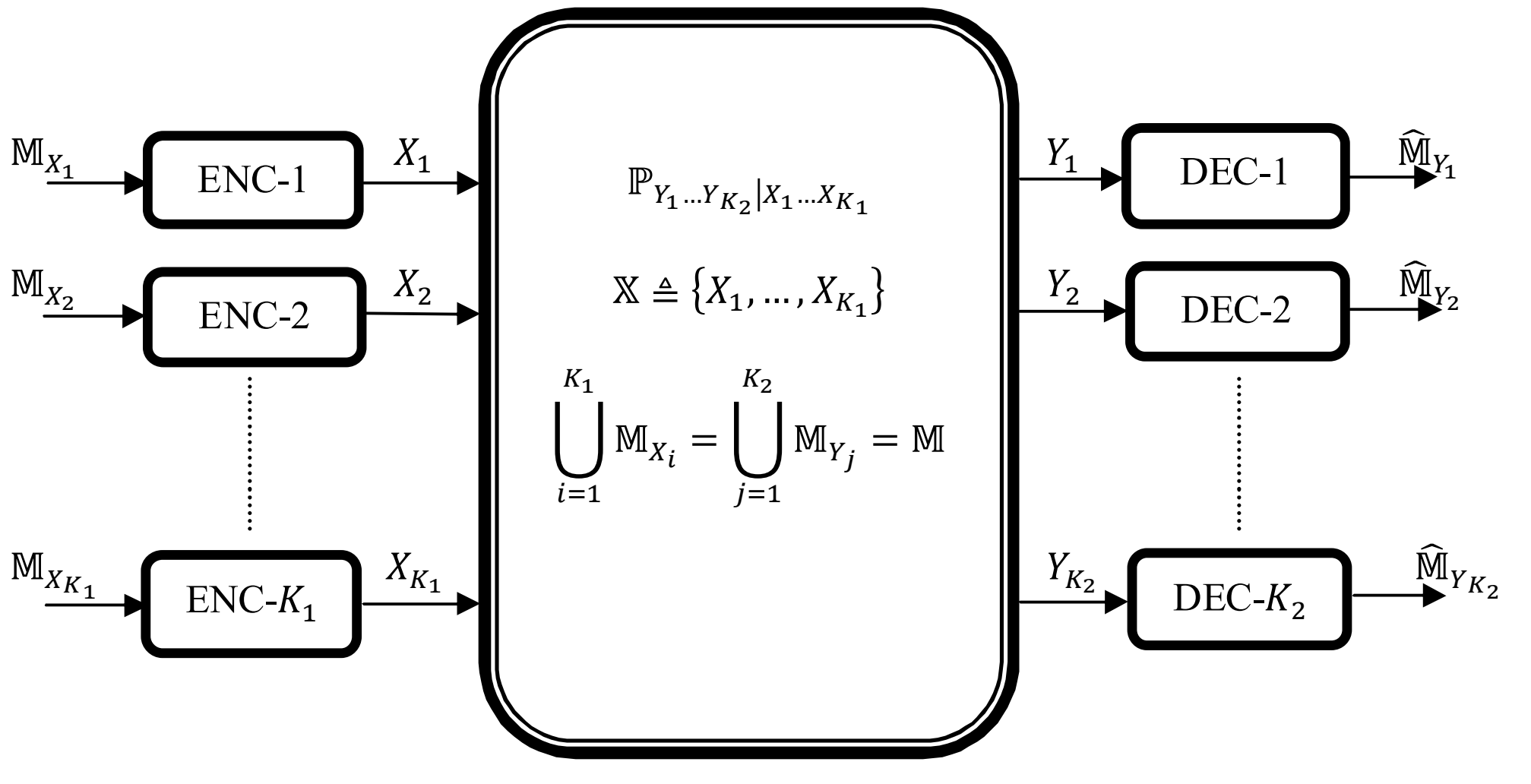

Figure 1. The General Interference Netwrok (GIN).

In this scenario, $K_1$ transmitters send independent messages $\mathbb{M} \triangleq \{M_1, \ldots, M_K\}$ to $K_2$ receivers: the transmitter $X_i$ sends the messages $\mathbb{M}_{X_i}$ over the channel, $i = 1, \ldots, K_1$, and the receiver $Y_j$ decodes the messages $\mathbb{M}_{Y_j}$ for $j = 1, \ldots, K_2$. Therefore, we have:

$$\bigcup_{i=1}^{K_1} \mathbb{M}_{X_i} = \bigcup_{j=1}^{K_2} \mathbb{M}_{Y_j} = \mathbb{M} \tag{3}$$

The network transition probability function $\mathbb{P}_{Y_1\dots Y_{K_2}|X_1\dots X_{K_1}}(y_1,\dots,y_{K_2}|x_1,\dots,x_{K_1})$ describes the relation between the inputs and the outputs. Also, the general Gaussian interference network with real-valued input and output signals is given by:

$$\begin{bmatrix} Y_1 \\ \vdots \\ Y_{K_2} \end{bmatrix} = \begin{bmatrix} a_{11} & \cdots & a_{1K_1} \\ \vdots & \ddots & \vdots \\ a_{K_21} & \cdots & a_{K_2K_1} \end{bmatrix} \begin{bmatrix} X_1 \\ \vdots \\ X_{K_1} \end{bmatrix} + \begin{bmatrix} Z_1 \\ \vdots \\ Z_{K_2} \end{bmatrix} \tag{4}$$

where the parameters $\{a_{ji}\}_{\substack{j=1,\dots,K_2 \\ i=1,\dots,K_1}}$ are (fixed) real-valued numbers, the RVs $\{X_i\}_{i=1}^{K_1}$ are the input symbols and the noise random variables $\{Z_j\}_{j=1}^{K_2}$ are zero-mean unit-variance Gaussian. The $i^{th}$ encoder is subject to an average power constraint as: $\mathbb{E}[X_i^2] \le P_i$, where $P_i \in \mathbb{R}_+$, $i = 1,\dots,K_1$.

We recall the concept of connected and unconnected transmitters with respect to the receivers in an interference network which was explicitly defined in Part I [1, Def. II.9]. Briefly, for a given network, a transmitter is connected to a receiver if its receiving signal statistically depends on the input signal with respect to that transmitter; otherwise, the transmitter is unconnected to the receiver. According to the notations of Part I [1, Def. II.9], for a given receiver $Y_j$, $j = 1,\dots,K_2$ the sets of connected and unconnected transmitters are represented by $\mathbb{X}_{c\to Y_j}$ and $\mathbb{X}_{c\nrightarrow Y_j}$, respectively, where $\mathbb{X}_{c\to Y_j}$ and $\mathbb{X}_{c\nrightarrow Y_j}$ are subsets of the inputs $\mathbb{X} \triangleq \{X_1,\dots,X_{K_1}\}$. As argued in [1, Lemma II.6], for any arbitrary interference, without loss of generality, one can assume that with respect to every receiver there is no message known only at its unconnected transmitters. In other words, we impose that:

$$\mathbb{M}_{Y_j} \subseteq \bigcup_{X_i \in \mathbb{X}_{c\to Y_j}} \mathbb{M}_{X_i}, \qquad j = 1,\dots,K_2 \tag{5}$$

Clearly, for each message belonging to $\mathbb{M}_{Y_j}$, there exists at least one transmitter connected to $Y_j$ which transmits that message.

In Part III [3, Sec. II], some other definitions for the interference networks were introduced which we need to them in the present part as well. These perquisites are given below;

***Definition 1:*** *Consider the general interference network in Fig. 1 with the corresponding message set* $\mathbb{M}$*. Also, for every subset* $\Omega$ *of* $\mathbb{M}$*, we define:*

$$\mathbb{X}_\Omega \triangleq \{X_i \in \mathbb{X} : \mathbb{M}_{X_i} \subseteq \Omega\} \tag{6}$$

*Therefore, the set* $\mathbb{X}_\Omega$ *is composed of all input signals that their messages lie in the set* $\Omega$*.*

***Definition 2:*** **Connected and Unconnected Messages**

*Consider the general interference network in Fig. 1 with the corresponding message set* $\mathbb{M}$*. For a given receiver* $Y_j$*, the set of unconnected messages* $\mathbb{M}_{c\nrightarrow Y_j}$ *is defined as follows:*

$$\mathbb{M}_{c\nrightarrow Y_j} \triangleq \left(\bigcup_{X_i \in \mathbb{X}_{c\nrightarrow Y_j}} \mathbb{M}_{X_i}\right) - \left(\bigcup_{X_i \in \mathbb{X}_{c\to Y_j}} \mathbb{M}_{X_i}\right), \qquad j = 1,\dots,K_2 \tag{7}$$

*where* $\mathbb{X}_{c\to Y_j}$ *and* $\mathbb{X}_{c\nrightarrow Y_j}$ *respectively denote the set of connected and unconnected transmitters corresponding to the receiver* $Y_j$*. Also, the set of connected messages with respect to the receiver* $Y_j$ *is defined as:* $\mathbb{M}_{c\to Y_j} \triangleq \mathbb{M} - \mathbb{M}_{c\nrightarrow Y_j}$.

We remark that for each receiver $Y_j, j = 1, \dots, K_2$, the unconnected messages $\mathbb{M}_{c\nrightarrow Y_j}$ are statistically independent of the received signal $Y_j$, as reported in [3, Observation 1].

In our analysis, we also need to barrow some ideas from Part II [2]. Consider an arbitrary interference network with the message sets $\mathbb{M}, \mathbb{M}_{X_i}, i = 1, \dots, K_1$, and $\mathbb{M}_{Y_j}, j = 1, \dots, K_2$ as shown in Fig. 1. In Part II [2], we established a full characterization of the sum-rate capacity for the general degraded networks where $\mathbb{X} \to Y_1 \to Y_2 \to \cdots \to Y_{K_2}$ form a Markov chain. The sum-rate capacity for these networks is achieved by a successive decoding scheme. We also showed that the transmission of only a certain subset of messages is sufficient to achieve the sum-rate capacity. By using the *MACCM*[2] *plan of messages*, we proposed algorithms to exactly determine this desired subset of messages. Let us briefly review the construction of these plans. Each subset of transmitters sends at most one message to each subset of receivers. There exist $K_1$ transmitters and $K_2$ receivers. Therefore, we can label each message by a nonempty subset of $\{1, \dots, K_1\}$ to denote which transmitters send the message, also a nonempty subset of $\{1, \dots, K_2\}$ to determine to which subset of receivers the message is sent. We represent each message of $\mathbb{M}$ as $M_\Delta^\nabla$, where $\Delta \subseteq \{1, \dots, K_1\}$ and $\nabla \subseteq \{1, \dots, K_2\}$. For example, $M_{\{1,2,3\}}^{\{2,4\}}$ indicates a message which is sent by transmitters 1, 2 and 3 to receivers 2 and 4.

Now, for each $\Delta \subseteq \{1, \dots, K_1\}$ we define:

$$\mathbb{M}_\Delta \triangleq \left\{M_\Delta^\nabla \in \mathbb{M} : \nabla \subseteq \{1, \dots, K_2\}\right\} \tag{8}$$

Using this representation, we arrange the messages into a graph-like illustration as shown in Fig. 2. This illustration is called the MACCM plan of messages. This plan includes $K_1$ columns so that the sets $\mathbb{M}_\Delta, \Delta \subseteq \{1, \dots, K_1\}$ with $\|\Delta\| = i$ are situated in its $i^{th}$ column, $i = 1, \dots, K_1$. Also, the set $\mathbb{M}_{\Delta_1}$ in column $i, i = 2, \dots, K_1$, is connected to the set $\mathbb{M}_{\Delta_2}$ in column $i-1$ provided that $\Delta_2 \subseteq \Delta_1$.

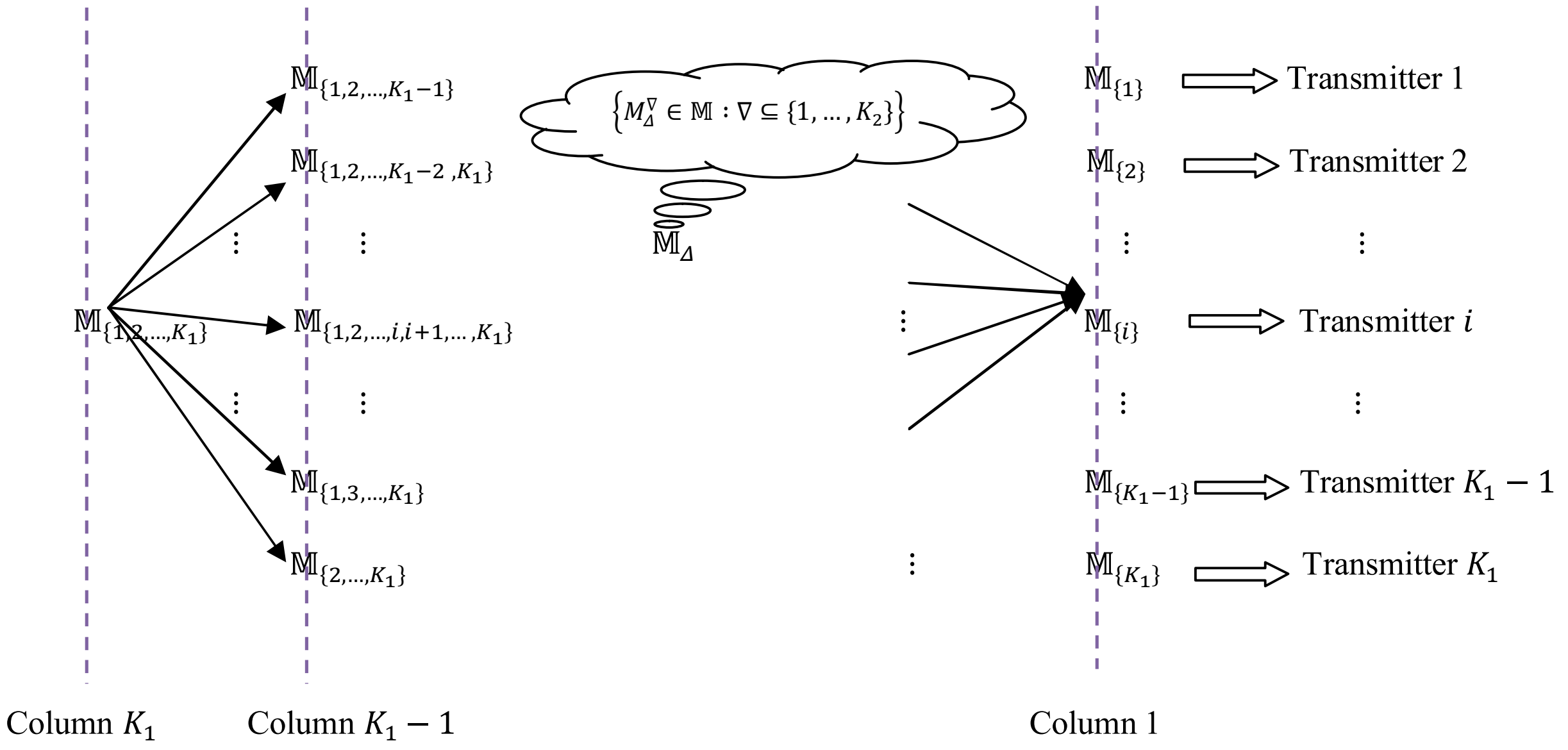

Figure 2. The MACCM plan of messages for an arbitrary interference Network.

Note that according to the MACCM plan, the messages $\mathbb{M}_\Delta$ in (8) are broadcasted by the transmitters $X_i, i \in \Delta$, meanwhile, no transmitter other than those in $\{X_i, i \in \Delta\}$ has access to this messages. For the case where there is only one receiver, i.e., the MAC with common messages, the MACCM plan is reduced to the MACCM message graph. This graph represents the superposition structures among the codewords in the achievability scheme that yields the capacity for the MAC with common messages; every two messages connected to each other by a directed edge are encoded in a superposition fashion so that the codeword conveying the message at the beginning of the directed edge is a cloud center for the codeword conveying the message at the end of the edge (this latter codeword is said to be a satellite for the former). Please refer to Part II [2, Sec. II.A] for details.

[2] Multiple Access Channel with Common Messages

In Part II [2], we proved that the sum-rate capacity of a degraded network with the messages $\mathbb{M}$ is equal to that of the network with the messages $\mathbb{M}^* \subseteq \mathbb{M}$ where $\mathbb{M}^*$ is determined as follows: first, for each $\Delta \subseteq \{1, \dots, K_1\}$, among the messages $\mathbb{M}_\Delta$ (if $\mathbb{M}_\Delta$ is nonempty), one message is selected; this is a message $M_\Delta^{\nabla_{\vartheta_\Delta}}$ with $\max \nabla_{\vartheta_\Delta} = \Theta_\Delta$ where:

$$\Theta_\Delta = \min\{\max \nabla : \quad \nabla \subseteq \{1, \dots, K_2\}, \quad M_\Delta^\nabla \in \mathbb{M}_\Delta\} \tag{9}$$

Note that $\max \nabla$ is the maximum element of the set $\nabla$. Define the sets $\widetilde{\mathbb{M}}, \widetilde{\mathbb{M}}_{X_i}, i = 1, \dots, K_1$, and $\widetilde{\mathbb{M}}_{Y_j}, j = 1, \dots, K_2$, as follows:

$$\begin{cases} \widetilde{\mathbb{M}} \triangleq \mathbb{M} - \bigcup\limits_{\Delta \subseteq \{1,\dots,K_1\}} \left(\mathbb{M}_\Delta - \left\{M_\Delta^{\nabla_{\vartheta_\Delta}}\right\}\right) \\ \widetilde{\mathbb{M}}_{Y_j} \triangleq \mathbb{M}_{Y_j} - \bigcup\limits_{\Delta \subseteq \{1,\dots,K_1\}} \left(\mathbb{M}_\Delta - \left\{M_\Delta^{\nabla_{\vartheta_\Delta}}\right\}\right) \end{cases} \tag{10}$$

The collection of selected messages, i.e., $\widetilde{\mathbb{M}}$, can be viewed as a message set designated for transmission over a MAC. From this collection, considering the messages corresponding to each receiver, the ones which are a satellite for any message that should be decoded at either that receiver or some other stronger receivers are removed. The remaining messages belong to $\mathbb{M}^*$. Precisely, let define $\mathbb{M}^*_{Y_j}$, $j = 1, \dots, K_2$, as follows:

$$\mathbb{M}^*_{Y_j} \triangleq \left\{M_\Delta^\nabla \in \widetilde{\mathbb{M}}_{Y_j} - \left(\widetilde{\mathbb{M}}_{Y_{j+1}} \cup \dots \cup \widetilde{\mathbb{M}}_{Y_{K_2}}\right) : \quad \text{There is no } M_\Gamma^L \in \widetilde{\mathbb{M}} - \left(\widetilde{\mathbb{M}}_{Y_{j+1}} \cup \dots \cup \widetilde{\mathbb{M}}_{Y_{K_2}}\right) \quad \text{with } \Delta \subsetneq \Gamma\right\} \tag{11}$$

Now, we can write:

$$\mathbb{M}^* = \bigcup_{j=1}^{K_2} \mathbb{M}^*_{Y_j} \tag{12}$$

Thus, $\mathbb{M}^*$ is the desired subset of $\mathbb{M}$. To achieve the sum-rate capacity for degraded networks, the transmitters can ignore the messages belonging to $\mathbb{M} - \mathbb{M}^*$ and only transmit those in $\mathbb{M}^*$. Our purpose to bring this issue in attention is that the same conclusion holds for the less-noisy networks.

In Part III [3, Sec. II], we derived some useful lemmas which play a central role to develop our results for the general interference networks with strong interference. To describe the results for the networks with a sequence of less noisy receivers in the present part, these lemmas are still critical. In what follows, for simple accessibility, we present some of these lemmas. The formal proofs for the results are omitted as they can be found in Part III [3, Sec. II].

***Lemma 1)*** **[3]** *Let $\mathcal{Y}_1, \mathcal{Y}_2, \mathcal{X}_1, \mathcal{X}_2, \dots, \mathcal{X}_{\mu_1}, \mathcal{X}_{\mu_1+1}, \dots, \mathcal{X}_{\mu_1+\mu_2}$ be arbitrary sets, where $\mu_1, \mu_2 \in \mathbb{N}$ are arbitrary natural numbers. Let also $\mathbb{P}(y_1, y_2 | x_1, x_2, \dots, x_{\mu_1}, x_{\mu_1+1}, \dots, x_{\mu_1+\mu_2})$ be a given conditional probability distribution defined on the set $\mathcal{Y}_1 \times \mathcal{Y}_2 \times \mathcal{X}_1 \times \mathcal{X}_2 \times \dots \times \mathcal{X}_{\mu_1} \times \mathcal{X}_{\mu_1+1} \times \dots \times \mathcal{X}_{\mu_1+\mu_2}$. Consider the inequality below:*

$$I(X_1, \dots, X_{\mu_1}; Y_1 | X_{\mu_1+1}, \dots, X_{\mu_1+\mu_2}) \leq I(X_1, \dots, X_{\mu_1}; Y_2 | X_{\mu_1+1}, \dots, X_{\mu_1+\mu_2}) \tag{13}$$

*If the inequality (13) holds for all PDFs $P_{X_1 \dots X_{\mu_1} X_{\mu_1+1} \dots X_{\mu_1+\mu_2}}(x_1, \dots, x_{\mu_1}, x_{\mu_1+1}, \dots, x_{\mu_1+\mu_2})$ with the following factorization:*

$$P_{X_1 \dots X_{\mu_1} X_{\mu_1+1} \dots X_{\mu_1+\mu_2}} = P_{X_1 \dots X_{\mu_1}}(x_1, \dots, x_{\mu_1}) P_{X_{\mu_1+1}}(x_{\mu_1+1}) P_{X_{\mu_1+2}}(x_{\mu_1+2}) \dots P_{X_{\mu_1+\mu_2}}(x_{\mu_1+\mu_2}), \tag{14}$$

*then we have:*

$$I(X_1,\dots,X_{\mu_1};Y_1|X_{\mu_1+1},\dots,X_{\mu_1+\mu_2},D)\le I(X_1,\dots,X_{\mu_1};Y_2|X_{\mu_1+1},\dots,X_{\mu_1+\mu_2},D) \tag{15}$$

*for all joint PDFs* $P_{DX_1\dots X_{\mu_1}X_{\mu_1+1}\dots X_{\mu_1+\mu_2}}(d,x_1,\dots,x_{\mu_1},x_{\mu_1+1},\dots,x_{\mu_1+\mu_2})$ *where* $D\to X_1,\dots,X_{\mu_1},X_{\mu_1+1},\dots,X_{\mu_1+\mu_2}\to Y_1,Y_2$ *forms a Markov chain.*

***Corollary 1)*** **[3]** *Let* $\mathcal{L}$ *be an arbitrary subset of* $\{1,\dots,\mu_1\}$. *Denote* $\mathbb{X}_{\mathcal{L}}\triangleq\{X_l : l\in\mathcal{L}\}$. *If the inequality (13) holds for all joint PDFs (14), then we have:*

$$I(\{X_1,\dots,X_{\mu_1}\}-\mathbb{X}_{\mathcal{L}};Y_1|\,\mathbb{X}_{\mathcal{L}},X_{\mu_1+1},\dots,X_{\mu_1+\mu_2},D)\le I(\{X_1,\dots,X_{\mu_1}\}-\mathbb{X}_{\mathcal{L}};Y_2|\,\mathbb{X}_{\mathcal{L}},X_{\mu_1+1},\dots,X_{\mu_1+\mu_2},D) \tag{16}$$

*for all joint PDFs* $P_{DX_1\dots X_{\mu_1}X_{\mu_1+1}\dots X_{\mu_1+\mu_2}}(d,x_1,\dots,x_{\mu_1},x_{\mu_1+1},\dots,x_{\mu_1+\mu_2})$ *where* $D\to X_1,\dots,X_{\mu_1},X_{\mu_1+1},\dots,X_{\mu_1+\mu_2}\to Y_1,Y_2$ *forms a Markov chain.*

***Lemma 2)*** **[3]** *Let* $\mathcal{Y}_1,\mathcal{Y}_2,\mathcal{X}_1,\mathcal{X}_2,\dots,\mathcal{X}_{\mu_1},\mathcal{X}_{\mu_1+1},\dots,\mathcal{X}_{\mu_1+\mu_2}$ *be arbitrary sets, where* $\mu_1,\mu_2\in\mathbb{N}$ *are arbitrary natural numbers. Let also* $\mathbb{P}(y_1,y_2|x_1,x_2,\dots,x_{\mu_1},x_{\mu_1+1},\dots,x_{\mu_1+\mu_2})$ *be a given conditional probability distribution defined on the set* $\mathcal{Y}_1\times\mathcal{Y}_2\times\mathcal{X}_1\times\mathcal{X}_2\times\dots\times\mathcal{X}_{\mu_1}\times\mathcal{X}_{\mu_1+1}\times\dots\times\mathcal{X}_{\mu_1+\mu_2}$. *Consider the inequality below:*

$$I(U;Y_1|X_{\mu_1+1},\dots,X_{\mu_1+\mu_2})\le I(U;Y_2|X_{\mu_1+1},\dots,X_{\mu_1+\mu_2}) \tag{17}$$

*If the inequality (17) holds for all PDFs* $P_{UX_1\dots X_{\mu_1}X_{\mu_1+1}\dots X_{\mu_1+\mu_2}}(x_1,\dots,x_{\mu_1},x_{\mu_1+1},\dots,x_{\mu_1+\mu_2})$ *with the following factorization:*

$$\begin{aligned}&P_{UX_1\dots X_{\mu_1}X_{\mu_1+1}\dots X_{\mu_1+\mu_2}}(u,x_1,\dots,x_{\mu_1},x_{\mu_1+1},\dots,x_{\mu_1+\mu_2})\\&\qquad=P_{UX_1\dots X_{\mu_1}}(u,x_1,\dots,x_{\mu_1})P_{X_{\mu_1+1}}(x_{\mu_1+1})P_{X_{\mu_1+2}}(x_{\mu_1+2})\dots P_{X_{\mu_1+\mu_2}}(x_{\mu_1+\mu_2}),\end{aligned} \tag{18}$$

*then we have:*

$$I(U;Y_1|X_{\mu_1+1},\dots,X_{\mu_1+\mu_2},D)\le I(U;Y_2|X_{\mu_1+1},\dots,X_{\mu_1+\mu_2},D) \tag{19}$$

*for all joint PDFs* $P_{DUX_1\dots X_{\mu_1}X_{\mu_1+1}\dots X_{\mu_1+\mu_2}}(d,u,x_1,\dots,x_{\mu_1},x_{\mu_1+1},\dots,x_{\mu_1+\mu_2})$ *where* $D,U\to X_1,\dots,X_{\mu_1},X_{\mu_1+1},\dots,X_{\mu_1+\mu_2}\to Y_1,Y_2$ *form a Markov chain.*

For the Gaussian networks, we need to the following variations of Lemmas 1 and 2. Let the outputs $Y_1$ and $Y_2$ be given as follows:

$$\begin{cases}Y_1\triangleq a_1X_1+a_2X_2+\dots+a_{\mu_1}X_{\mu_1}+a_{\mu_1+1}X_{\mu_1+1}+\dots+a_{\mu_1+\mu_2}X_{\mu_1+\mu_2}+Z_1\\Y_2\triangleq b_1X_1+b_2X_2+\dots+b_{\mu_1}X_{\mu_1}+b_{\mu_1+1}X_{\mu_1+1}+\dots+b_{\mu_1+\mu_2}X_{\mu_1+\mu_2}+Z_2\end{cases} \tag{20}$$

where $Z_1$ and $Z_2$ are zero-mean unit-variance Gaussian random variables; also, $X_1,X_2,\dots,X_{\mu_1},X_{\mu_1+1},\dots,X_{\mu_1+\mu_2}$ are real-valued power-constrained random variables independent of $(Z_1,Z_2)$ and $a_1,a_2,\dots,a_{\mu_1},a_{\mu_1+1},\dots,a_{\mu_1+\mu_2}$ and $b_1,b_2,\dots,b_{\mu_1},b_{\mu_1+1},\dots,b_{\mu_1+\mu_2}$ are fixed real numbers. We would like to determine sufficient conditions for this setup under which the inequality (15) (or (19)) holds for all joint PDFs $P_{DX_1\dots X_{\mu_1}X_{\mu_1+1}\dots X_{\mu_1+\mu_2}}(d,x_1,\dots,x_{\mu_1},x_{\mu_1+1},\dots,x_{\mu_1+\mu_2})$. The following lemma gives such conditions.

***Lemma 3)*** **[3]** *Consider the Gaussian system in (20). If the following condition satisfies:*

$$\frac{a_1}{b_1} = \frac{a_2}{b_2} = \cdots = \frac{a_{\mu_1}}{b_{\mu_1}} = \alpha, \qquad |\alpha| \leq 1,$$

(21)

*then the inequality (15) holds for all joint PDFs* $P_{DX_1 \ldots X_{\mu_1} X_{\mu_1+1} \ldots X_{\mu_1+\mu_2}}(d, x_1, \ldots, x_{\mu_1}, x_{\mu_1+1}, \ldots, x_{\mu_1+\mu_2})$ *where* $D$ *is independent of* $(Z_1, Z_2)$.

In fact, as discussed in [3, Sec. II] under the condition (21), given $X_{\mu_1+1}, \ldots, X_{\mu_1+\mu_2}$, the signal $Y_1$ is a stochastically degraded version of $Y_2$. Also, we remark that the relation (21) is a sufficient condition under which (15) holds; however, in general the inequality (15) may not be equivalent to (21). It is essential to note that the condition (21) is not derived by evaluating (13) for Gaussian input distributions. Only for the case of $\mu_1 = 1$, the condition (21) can be equivalently derived by evaluating (13) for Gaussian input distributions.

***Lemma 4)*** **[3]** *Consider the Gaussian system in (20). If (21) holds, then the inequality (19) is satisfied for all joint PDFs* $P_{DUX_1 \ldots X_{\mu_1} X_{\mu_1+1} \ldots X_{\mu_1+\mu_2}}(d, u, x_1, \ldots, x_{\mu_1}, x_{\mu_1+1}, \ldots, x_{\mu_1+\mu_2})$ *where* $(D, U)$ *is independent of* $(Z_1, Z_2)$.

By applying these preliminaries, we can present our results for the networks with a sequence of less-noisy receivers in the following subsections.

# III. General Interference Networks with Two Receivers

We launch the study for the interference networks with less-noisy receivers by considering the networks with two receivers in this section. Recall that in Part III [3, Sec. IV.A], we established unified outer bounds for the capacity region of the two-receiver interference networks. As shown in [3, Sec. IV.B], our outer bounds are efficient to derive a general strong interference regime for all two-receiver networks with arbitrary topologies. By using these outer bounds, we now intend to identify classes of less noisy networks for which a successive decoding scheme is sum-rate optimal.

First, let us present our general outer bound derived in [3, Corollary 2].

***Definition 3:*** *Let* $\{R_1, \ldots, R_K\}$ *be a* $K$*-tuple of non-negative real numbers, where* $K$ *is a natural number. Let also* $\mathbb{M} = \{M_1, \ldots, M_K\}$ *be a set of* $K$ *indexed elements. Assume that* $\Omega$ *is an arbitrary subset of* $\mathbb{M}$. *The partial-sum* $\boldsymbol{R}_{\Sigma\Omega}$ *with respect to* $\Omega$ *is defined as follows:*

$$\boldsymbol{R}_{\Sigma\Omega} \triangleq \sum_{l \in \underline{id}_\Omega} R_l$$

(22)

Note that the identification of the set $\Omega$, $\underline{id}_\Omega$, was defined in Part I [1, Def. II.1].

***Theorem 1)*** **[3, Corollary 2]**, *Consider the general two-receiver interference network which is derived from the scenario shown in Fig. 1 by setting* $K_2 = 2$. *Define the rate region* $\mathfrak{R}_o^{GINTR}$ *as follows:*

$$\mathfrak{R}_o^{GINTR} \triangleq \bigcup_{\mathcal{P}_o^{GINTR}} \left\{ \begin{array}{l} (R_1, \dots, R_K) \in \mathbb{R}_+^K: \\ \forall\, \Omega_0, \Omega_1, \Omega_2 : \ \Omega_0 \subseteq \mathbb{M}_{Y_1} \cap \mathbb{M}_{Y_2}, \Omega_1 \subseteq \mathbb{M}_{Y_1}, \Omega_2 \subseteq \mathbb{M}_{Y_2}, \\ \qquad\qquad \text{with} \ \ \langle \Omega_1 \cap \Omega_2 = \Omega_0 \cap \Omega_1 = \Omega_0 \cap \Omega_2 = \emptyset \rangle \\ \qquad\qquad\qquad\qquad\qquad\qquad , \\ \forall\, \Omega_s : \ \Omega_s \subseteq \mathbb{M} - (\Omega_0 \cup \Omega_1 \cup \Omega_2) \\ \qquad\qquad\qquad\qquad\qquad\qquad , \\ \qquad \langle 1 \rangle : \boldsymbol{R}_{\Sigma \Omega_0} + \boldsymbol{R}_{\Sigma \Omega_1} \leq I(Z, \Omega_0, \Omega_1; Y_1 | \Omega_s, Q) \\ \qquad \langle 2 \rangle : \boldsymbol{R}_{\Sigma \Omega_0} + \boldsymbol{R}_{\Sigma \Omega_2} \leq I(Z, \Omega_0, \Omega_2; Y_2 | \Omega_s, Q) \\ \langle 3 \rangle : \boldsymbol{R}_{\Sigma \Omega_0} + \boldsymbol{R}_{\Sigma \Omega_1} + \boldsymbol{R}_{\Sigma \Omega_2} \leq I(\Omega_1; Y_1 | Z, \Omega_0, \Omega_2, \Omega_s, Q) + I(Z, \Omega_0, \Omega_2; Y_2 | \Omega_s, Q) \\ \langle 4 \rangle : \boldsymbol{R}_{\Sigma \Omega_0} + \boldsymbol{R}_{\Sigma \Omega_1} + \boldsymbol{R}_{\Sigma \Omega_2} \leq I(\Omega_2; Y_2 | Z, \Omega_0, \Omega_1, \Omega_s, Q) + I(Z, \Omega_0, \Omega_1; Y_1 | \Omega_s, Q) \\ \qquad\qquad\qquad\qquad\qquad\qquad , \\ \langle 5 \rangle : \boldsymbol{R}_{\Sigma \Omega_0} + \boldsymbol{R}_{\Sigma \Omega_1} + \boldsymbol{R}_{\Sigma \Omega_2} \leq I(\Omega_1; Y_1 | Z, \Omega_0, \Omega_2, \Omega_s, Q) + I(\Omega_2; Y_2 | Z, \Omega_0, \Omega_s, Q) \\ \qquad\qquad\qquad\qquad\qquad\qquad\qquad\qquad + I(Z, \Omega_0; Y_1 | \Omega_s, Q) \\ \langle 6 \rangle : \boldsymbol{R}_{\Sigma \Omega_0} + \boldsymbol{R}_{\Sigma \Omega_1} + \boldsymbol{R}_{\Sigma \Omega_2} \leq I(\Omega_2; Y_2 | Z, \Omega_0, \Omega_1, \Omega_s, Q) + I(\Omega_1; Y_1 | Z, \Omega_0, \Omega_s, Q) \\ \qquad\qquad\qquad\qquad\qquad\qquad\qquad\qquad + I(Z, \Omega_0; Y_2 | \Omega_s, Q) \end{array} \right\} \tag{23}$$

*where $\mathcal{P}_o^{GINTR}$ denotes the set of all joint PDFs $P_{QM_1 \dots M_K Z X_1 \dots X_{K_1}}(q, m_1, \dots, m_K, z, x_1, \dots, x_{K_1})$ satisfying:*

$$P_{QM_1 \dots M_K Z X_1 \dots X_{K_1}} = P_Q \times P_{M_1} \times \dots \times P_{M_K} \times P_{Z|M_1 \dots M_K Q} \times P_{X_1|\mathbb{M}_{X_1}, Q} \times \dots \times P_{X_{K_1}|\mathbb{M}_{X_{K_1}}, Q} \tag{24}$$

*Also, the PDFs $P_{M_l}, l = 1, \dots, K$, are uniformly distributed, and $P_{X_i|\mathbb{M}_{X_i}, Q} \in \{0,1\}$ for $i = 1, \dots, K_1$. The set $\mathfrak{R}_o^{GINTR}$ constitutes an outer bound for the capacity region.*

***Remark 1:*** Note that the unified outer bound $\mathfrak{R}_o^{GINTR}$ in (23) consists of the following parameters:

- The RVs representing the receiver signals, $Y_1, Y_2$.
- The $K$ auxiliary random variables $M_1, \dots, M_K$ which actually represent the messages.
- The time-sharing random variable $Q$.
- The auxiliary random variable $Z$.

In what follows, using the general bound $\mathfrak{R}_o^{GINTR}$ in (23), we first establish specific sum-rate outer bounds for networks which satisfy certain less-noisy conditions. Then, we identify scenarios for which the derived outer bounds coincide with the sum-rate capacity.

Let us concentrate on the constraints $\mathfrak{R}_o^{GINTR}\langle 3 \rangle$ and $\mathfrak{R}_o^{GINTR}\langle 4 \rangle$ in the characterization of (23). By setting $\Omega_0 = \emptyset$, $\Omega_1 = \mathbb{M}_{Y_1} - \mathbb{M}_{Y_2}$, $\Omega_2 = \mathbb{M}_{Y_2}$ and $\Omega_s = \emptyset$ in $\mathfrak{R}_o^{GINTR}\langle 3 \rangle$, we obtain the following constraint on the sum-rate capacity of a general two-receiver interference network:

$$\mathcal{C}_{sum}^{GINTR} = \boldsymbol{R}_{\Sigma \mathbb{M}_{Y_1} - \mathbb{M}_{Y_2}} + \boldsymbol{R}_{\Sigma \mathbb{M}_{Y_2}} \leq I(\mathbb{M}_{Y_1} - \mathbb{M}_{Y_2}; Y_1 | Z, \mathbb{M}_{Y_2}, Q) + I(Z, \mathbb{M}_{Y_2}; Y_2 | Q) = I(\mathbb{M}_{Y_1}; Y_1 | Z, \mathbb{M}_{Y_2}, Q) + I(Z, \mathbb{M}_{Y_2}; Y_2 | Q) \tag{25}$$

Also, by setting $\Omega_0 = \emptyset$, $\Omega_1 = \mathbb{M}_{Y_1}$, $\Omega_2 = \mathbb{M}_{Y_2} - \mathbb{M}_{Y_1}$ and $\Omega_s = \emptyset$ in $\mathfrak{R}_o^{GINTR}\langle 4 \rangle$, we derive:

$$\mathcal{C}_{sum}^{GINTR} = \boldsymbol{R}_{\Sigma \mathbb{M}_{Y_1}} + \boldsymbol{R}_{\Sigma \mathbb{M}_{Y_2} - \mathbb{M}_{Y_1}} \leq I(\mathbb{M}_{Y_2} - \mathbb{M}_{Y_1}; Y_2 | Z, \mathbb{M}_{Y_1}, Q) + I(Z, \mathbb{M}_{Y_1}; Y_1 | Q) = I(\mathbb{M}_{Y_2}; Y_2 | Z, \mathbb{M}_{Y_1}, Q) + I(Z, \mathbb{M}_{Y_1}; Y_1 | Q) \tag{26}$$

where both the bounds (25) and (26) should be evaluated over all joint PDFs of the form (24). Therefore, we have explicit outer bounds (25) and (26) on the sum-rate capacity. In this section, our purpose is to show that under certain conditions these bounds are achievable by a successive decoding scheme; thus, establishing the exact sum-rate capacity. First, we remark that the constraints (25) and (26) both contain the auxiliary random variable "$Z$". Considering the distributions in (24), this auxiliary random variable is undesired from the viewpoint of achievability, i.e., one can rarely propose a coding scheme that achieves either (25) or (26) with the distributions in (24). In the following theorem, we identify scenarios where the auxiliary random variable "$Z$" can be removed from these bounds.

***Theorem 2)*** *Consider the general two-receiver interference network which is given by setting $K_2 = 2$ in the scenario shown in Fig.1.*

*A. If the network transition probability function satisfies the following condition:*

$$I\left(U;Y_2\middle|\mathbb{X}_{\mathbb{M}_{Y_2}}\right) \leq I\left(U;Y_1\middle|\mathbb{X}_{\mathbb{M}_{Y_2}}\right) \quad \textit{for all joint PDFs} \quad P_{U,\mathbb{X}-\mathbb{X}_{\mathbb{M}_{Y_2}}}\prod_{X_i\in\mathbb{X}_{\mathbb{M}_{Y_2}}}P_{X_i}, \tag{27}$$

*then the sum-rate capacity is bounded-above as:*

$$\mathcal{C}_{sum}^{GINTR} \leq \max_{P_QP_{M_1}\dots P_{M_K}\prod_{i=1}^{K_1}P_{X_i|\mathbb{M}_{X_i},Q}} \left(I\left(\mathbb{M}_{Y_1};Y_1\middle|\mathbb{M}_{Y_2},Q\right) + I\left(\mathbb{M}_{Y_2};Y_2\middle|Q\right)\right) \tag{28}$$

*B. If the network transition probability function satisfies the following:*

$$I\left(\mathbb{X}-\mathbb{X}_{\mathbb{M}_{Y_1}};Y_2\middle|\mathbb{X}_{\mathbb{M}_{Y_1}}\right) \leq I\left(\mathbb{X}-\mathbb{X}_{\mathbb{M}_{Y_1}};Y_1\middle|\mathbb{X}_{\mathbb{M}_{Y_1}}\right) \quad \textit{for all joint PDFs} \quad P_{\mathbb{X}-\mathbb{X}_{\mathbb{M}_{Y_1}}}\prod_{X_i\in\mathbb{X}_{\mathbb{M}_{Y_1}}}P_{X_i}, \tag{29}$$

*then the sum-rate capacity is bounded above as:*

$$\mathcal{C}_{sum}^{GINTR} \leq \max_{P_QP_{M_1}\dots P_{M_K}\prod_{i=1}^{K_1}P_{X_i|\mathbb{M}_{X_i},Q}} \left(I\left(\mathbb{M}_{Y_1},\mathbb{M}_{Y_2};Y_1\middle|Q\right)\right) \tag{30}$$

*Proof of Theorem 2)* Consider Part A. First note that, according to Lemma 2, if the condition (27) holds, then we have:

$$I\left(U;Y_2\middle|\mathbb{X}_{\mathbb{M}_{Y_2}},D\right) \leq I\left(U;Y_1\middle|\mathbb{X}_{\mathbb{M}_{Y_2}},D\right) \quad \textit{for all joint PDFs} \quad P_{DU\mathbb{X}} \tag{31}$$

Now, consider the constraint (25) on the sum-rate capacity. One can write:

$$\begin{aligned}
&I\left(\mathbb{M}_{Y_1};Y_1\middle|Z,\mathbb{M}_{Y_2},Q\right) + I\left(Z,\mathbb{M}_{Y_2};Y_2\middle|Q\right)\\
&\quad = I\left(\mathbb{M}_{Y_1};Y_1\middle|Z,\mathbb{M}_{Y_2},Q\right) + I\left(Z;Y_2\middle|\mathbb{M}_{Y_2},Q\right) + I\left(\mathbb{M}_{Y_2};Y_2\middle|Q\right)\\
&\quad = I\left(\mathbb{M}_{Y_1};Y_1\middle|Z,\mathbb{M}_{Y_2},Q\right) + I\left(Z;Y_2\middle|\mathbb{X}_{\mathbb{M}_{Y_2}},\mathbb{M}_{Y_2},Q\right) + I\left(\mathbb{M}_{Y_2};Y_2\middle|Q\right)
\end{aligned} \tag{32}$$

Using the inequality (31), for the second term in (32) we have:

$$I\left(Z;Y_2\middle|\mathbb{X}_{\mathbb{M}_{Y_2}},\mathbb{M}_{Y_2},Q\right) \leq I\left(Z;Y_1\middle|\mathbb{X}_{\mathbb{M}_{Y_2}},\mathbb{M}_{Y_2},Q\right) \tag{33}$$

To derive (33), it is sufficient to replace $U$ by $Z$ and $D$ by $\left(\mathbb{M}_{Y_2},Q\right)$ and consider the joint PDFs in (24) on the corresponding random variables. Note that we have this liberty because (31) holds for any arbitrary distribution on $D$, $U$ and $\mathbb{X}$. By substituting (33) in (32) we obtain the bound in (28). Next, consider Part B. Based on Lemma 1, the condition (29) implies:

$$I\left(\mathbb{X}-\mathbb{X}_{\mathbb{M}_{Y_1}};Y_2\middle|\mathbb{X}_{\mathbb{M}_{Y_1}},D\right) \leq I\left(\mathbb{X}-\mathbb{X}_{\mathbb{M}_{Y_1}};Y_1\middle|\mathbb{X}_{\mathbb{M}_{Y_1}},D\right) \quad \textit{for all joint PDFs} \quad P_{D\mathbb{X}} \tag{34}$$

Now consider the constraint (26) on the sum-rate capacity. We can write:

$$
\begin{aligned}
&I(\mathbb{M}_{Y_2};Y_2|Z,\mathbb{M}_{Y_1},Q)+I(Z,\mathbb{M}_{Y_1};Y_1|Q)\\
&\quad=I\left(\mathbb{X}-\mathbb{X}_{\mathbb{M}_{Y_1}},\mathbb{M}_{Y_2};Y_2\middle|\mathbb{X}_{\mathbb{M}_{Y_1}},Z,\mathbb{M}_{Y_1},Q\right)+I(Z,\mathbb{M}_{Y_1};Y_1|Q)\\
&\quad\overset{(a)}{=}I\left(\mathbb{X}-\mathbb{X}_{\mathbb{M}_{Y_1}};Y_2\middle|\mathbb{X}_{\mathbb{M}_{Y_1}},Z,\mathbb{M}_{Y_1},Q\right)+I(Z,\mathbb{M}_{Y_1};Y_1|Q)\\
&\quad\overset{(b)}{\leq}I\left(\mathbb{X}-\mathbb{X}_{\mathbb{M}_{Y_1}};Y_1\middle|\mathbb{X}_{\mathbb{M}_{Y_1}},Z,\mathbb{M}_{Y_1},Q\right)+I(Z,\mathbb{M}_{Y_1};Y_1|Q)\\
&\quad\overset{(c)}{=}I\left(\mathbb{X}-\mathbb{X}_{\mathbb{M}_{Y_1}},\mathbb{M}_{Y_2};Y_1\middle|\mathbb{X}_{\mathbb{M}_{Y_1}},Z,\mathbb{M}_{Y_1},Q\right)+I(Z,\mathbb{M}_{Y_1};Y_1|Q)\\
&\quad\overset{(d)}{=}I(\mathbb{M}_{Y_2};Y_1|Z,\mathbb{M}_{Y_1},Q)+I(Z,\mathbb{M}_{Y_1};Y_1|Q)\overset{(e)}{=}I(\mathbb{M}_{Y_1},\mathbb{M}_{Y_2};Y_1|Q)
\end{aligned}
\tag{35}
$$

where (a) holds because $Z,\mathbb{M}_{Y_1},\mathbb{M}_{Y_2},Q\to\mathbb{X}\to Y_2$ form a Markov chain, (b) is due to (34), (c) holds because $Z,\mathbb{M}_{Y_1},\mathbb{M}_{Y_2},Q\to\mathbb{X}\to Y_2$ form a Markov chain, (d) holds because the inputs $\mathbb{X}_{\mathbb{M}_{Y_1}}$ are given by deterministic functions of $(\mathbb{M}_{Y_1},Q)$ and $\mathbb{X}$ by deterministic functions of $(\mathbb{M}_{Y_1},\mathbb{M}_{Y_2},Q)$, and (e) holds because $Z\to\mathbb{M}_{Y_1},\mathbb{M}_{Y_2},Q\to Y_1,Y_2$ form a Markov chain. The proof is thus complete. ■

The outer bounds given in (28) and (30) for the sum-rate capacity of the general two-receiver interference network have the desired characteristic that they do not contain any auxiliary random variable other than the messages. This property enables us to easily deal with achievability schemes which potentially lead to rates that coincide with these bounds, thus establishing the explicit sum-rate capacity. We next explore this problem.

First consider the outer bound in (28). We remark that for the degraded networks where $\mathbb{X}\to Y_2\to Y_1$ form a Markov chain, the condition (27) is crucially satisfied. For these networks, in Part II [2] we showed that the rate (28) is achievable as well. A simple achievability scheme is as follows. At the transmitters, the messages are separately encoded using independent codewords. At the weaker receiver $Y_2$, the messages $\mathbb{M}_{Y_2}$ are successively decoded. At the stronger receiver, first the messages $\mathbb{M}_{Y_2}$ and then the messages $\mathbb{M}_{Y_1}-\mathbb{M}_{Y_2}$ are successively decoded. Let $\mathbb{M}_{Y_2}\triangleq\{M_1^2,M_2^2,\dots,M_{\|\mathbb{M}_{Y_2}\|}^2\}$ and $\mathbb{M}_{Y_1}-\mathbb{M}_{Y_2}\triangleq\{M_1^1,M_2^1,\dots,M_{\|\mathbb{M}_{Y_1}-\mathbb{M}_{Y_2}\|}^1\}$. Therefore, the resulting sum-rate corresponding to the achievability scheme is:

$$
\left(\begin{array}{l}
\min\left\{\begin{array}{l}I(M_1^2;Y_1|Q),\\ I(M_1^2;Y_2|Q)\end{array}\right\}+\min\left\{\begin{array}{l}I(M_2^2;Y_1|M_1^2,Q),\\ I(M_2^2;Y_2|M_1^2,Q)\end{array}\right\}+\cdots+\min\left\{\begin{array}{l}I\left(M_{\|\mathbb{M}_{Y_2}\|}^2;Y_1\middle|M_1^2,M_2^2,\dots,M_{\|\mathbb{M}_{Y_2}\|-1}^2,Q\right),\\ I\left(M_{\|\mathbb{M}_{Y_2}\|}^2;Y_2\middle|M_1^2,M_2^2,\dots,M_{\|\mathbb{M}_{Y_2}\|-1}^2,Q\right)\end{array}\right\}+\\
I(M_1^1;Y_1|\mathbb{M}_{Y_2},Q)+I(M_2^1;Y_1|M_1^1,\mathbb{M}_{Y_2},Q)+\cdots+I\left(M_{\|\mathbb{M}_{Y_1}-\mathbb{M}_{Y_2}\|}^1;Y_1\middle|M_1^1,M_2^1,\dots,M_{\|\mathbb{M}_{Y_1}-\mathbb{M}_{Y_2}\|-1}^1,\mathbb{M}_{Y_2},Q\right)
\end{array}\right)
\tag{36}
$$

for a given probability distribution $P_QP_{M_1}\dots P_{M_K}\prod_{i=1}^{K_1}P_{X_i|\mathbb{M}_{X_i},Q}$. If the network is degraded, then we have:

$$
\min\left\{\begin{array}{l}I(M_l^2;Y_1|M_1^2,M_2^2,\dots,M_{l-1}^2,Q),\\ I(M_l^2;Y_2|M_1^2,M_2^2,\dots,M_{l-1}^2,Q)\end{array}\right\}=I(M_l^2;Y_2|M_1^2,M_2^2,\dots,M_{l-1}^2,Q),\qquad l=1,\dots,\|\mathbb{M}_{Y_2}\|
\tag{37}
$$

By substituting (37) in (36), we directly obtain the rate (28). In fact, the degradedness is not mandatory to achieve the outer bound (28). In the next theorem, we derive a weaker condition for this purpose.

***Theorem 3)*** *Consider the general two-receiver interference network which is derived from the network in Fig. 1 by setting $K_2=2$. If the transition probability function of the network satisfies the following less-noisy condition:*

$$
I\left(U;Y_2\middle|\mathbb{X}_{\mathbb{M}_{C\nrightarrow Y_1}}\right)\leq I\left(U;Y_1\middle|\mathbb{X}_{\mathbb{M}_{C\nrightarrow Y_1}}\right)\qquad\text{for all joint PDFs}\qquad P_{U,\mathbb{X}-\mathbb{X}_{\mathbb{M}_{C\nrightarrow Y_1}}}\prod_{X_i\in\mathbb{X}_{\mathbb{M}_{C\nrightarrow Y_1}}}P_{X_i},
\tag{38}
$$

*then the sum-rate capacity of the network is given by:*

$$\max_{P_Q P_{M_1} \dots P_{M_K} \prod_{i=1}^{K_1} P_{X_i|\mathbb{M}_{X_i},Q}} \left( I(\mathbb{M}_{Y_1}; Y_1 | \mathbb{M}_{Y_2}, Q) + I(\mathbb{M}_{Y_2}; Y_2 | Q) \right) \tag{39}$$

*Proof of Theorem 3)* First note that, according to Lemma 2, the condition (38) extends as follows:

$$I\left(U; Y_2 \middle| \mathbb{X}_{\mathbb{M}_{c\nrightarrow Y_1}}, D\right) \leq I\left(U; Y_1 \middle| \mathbb{X}_{\mathbb{M}_{c\nrightarrow Y_1}}, D\right) \qquad \text{for all joint PDFs} \qquad P_{DU\mathbb{X}} \tag{40}$$

Note that $\mathbb{M}_{c\nrightarrow Y_1}$ is a subset of $\mathbb{M}_{Y_2}$. Therefore, $\mathbb{X}_{\mathbb{M}_{c\nrightarrow Y_1}}$ is a subset of $\mathbb{X}_{\mathbb{M}_{Y_2}}$. Now, by substituting $D = \mathbb{X}_{\mathbb{M}_{Y_2}} - \mathbb{X}_{\mathbb{M}_{c\nrightarrow Y_1}}$ in (40), we obtain the condition (27). Thus, the outer bound (28) is also valid for the networks satisfying (38). It remains to show that (39) is achievable as well. The achievability scheme is based on a successive decoding scheme. However, among the messages $\mathbb{M}_{Y_2}$, the receiver $Y_1$ only decodes those which belong to $\mathbb{M}_{Y_2} - \mathbb{M}_{c\nrightarrow Y_1}$. Let $\mathbb{M}_{c\nrightarrow Y_1} \triangleq \left\{M_1^2, M_2^2, \dots, M_{\|\mathbb{M}_{c\nrightarrow Y_1}\|}^2\right\}$, $\mathbb{M}_{Y_2} \triangleq \left\{M_1^2, M_2^2, \dots, M_{\|\mathbb{M}_{Y_2}\|}^2\right\}$, and $\mathbb{M}_{Y_1} - \left(\mathbb{M}_{Y_2} - \mathbb{M}_{c\nrightarrow Y_1}\right) \triangleq \left\{M_1^1, M_2^1, \dots, M_{\|\mathbb{M}_{Y_1} - (\mathbb{M}_{Y_2} - \mathbb{M}_{c\nrightarrow Y_1})\|}^1\right\}$. At the transmitters, the messages are separately encoded using independent codewords. At the receiver $Y_2$, the messages $\left\{M_1^2, M_2^2, \dots, M_{\|\mathbb{M}_{Y_2}\|}^2\right\}$ are successively decoded (first $M_1^2$, then $M_2^2$, and so on). At the receiver $Y_1$, first the messages $\mathbb{M}_{Y_2} - \mathbb{M}_{c\nrightarrow Y_1} = \left\{M_{\|\mathbb{M}_{c\nrightarrow Y_1}\|+1}^2, M_{\|\mathbb{M}_{c\nrightarrow Y_1}\|+2}^2, \dots, M_{\|\mathbb{M}_{Y_2}\|}^2\right\}$ are successively decoded and then the messages $\mathbb{M}_{Y_1} - \left(\mathbb{M}_{Y_2} - \mathbb{M}_{c\nrightarrow Y_1}\right)$. Therefore, the resultant sum-rate corresponding to this achievability scheme is as follows:

$$\left(\begin{array}{c} I(\mathbb{M}_{c\nrightarrow Y_1}; Y_2 | Q) + \min\left\{\begin{array}{c} I\left(M_{\|\mathbb{M}_{c\nrightarrow Y_1}\|+1}^2; Y_1 \middle|, Q\right), \\ I\left(M_{\|\mathbb{M}_{c\nrightarrow Y_1}\|+1}^2; Y_2 \middle| \mathbb{M}_{c\nrightarrow Y_1}, Q\right) \end{array}\right\} \\ + \min\left\{\begin{array}{c} I\left(M_{\|\mathbb{M}_{c\nrightarrow Y_1}\|+2}^2; Y_1 \middle| M_{\|\mathbb{M}_{c\nrightarrow Y_1}\|+1}^2, Q\right), \\ I\left(M_{\|\mathbb{M}_{c\nrightarrow Y_1}\|+2}^2; Y_2 \middle| M_{\|\mathbb{M}_{c\nrightarrow Y_1}\|+1}^2, \mathbb{M}_{c\nrightarrow Y_1}, Q\right) \end{array}\right\} + \dots + \min\left\{\begin{array}{c} I\left(M_{\|\mathbb{M}_{Y_2}\|}^2; Y_1 \middle| M_{\|\mathbb{M}_{c\nrightarrow Y_1}\|+1}^2, \dots, M_{\|\mathbb{M}_{Y_2}\|-1}^2, Q\right), \\ I\left(M_{\|\mathbb{M}_{Y_2}\|}^2; Y_2 \middle| M_{\|\mathbb{M}_{c\nrightarrow Y_1}\|+1}^2, \dots, M_{\|\mathbb{M}_{Y_2}\|-1}^2, \mathbb{M}_{c\nrightarrow Y_1}, Q\right) \end{array}\right\} \\ + I(\mathbb{M}_{Y_1} - (\mathbb{M}_{Y_2} - \mathbb{M}_{c\nrightarrow Y_1}); Y_1 | \mathbb{M}_{Y_2} - \mathbb{M}_{c\nrightarrow Y_1}, Q) \end{array}\right) \tag{41}$$

for a given probability distribution $P_Q P_{M_1} \dots P_{M_K} \prod_{i=1}^{K_1} P_{X_i|\mathbb{M}_{X_i},Q}$. Next, we claim that:

$$\min\left\{\begin{array}{c} I\left(M_l^2; Y_1 \middle| M_{\|\mathbb{M}_{c\nrightarrow Y_1}\|+1}^2, \dots, M_{l-1}^2, Q\right), \\ I\left(M_l^2; Y_2 \middle| M_{\|\mathbb{M}_{c\nrightarrow Y_1}\|+1}^2, \dots, M_{l-1}^2, \mathbb{M}_{c\nrightarrow Y_1}, Q\right) \end{array}\right\} = I\left(M_l^2; Y_2 \middle| M_{\|\mathbb{M}_{c\nrightarrow Y_1}\|+1}^2, \dots, M_{l-1}^2, \mathbb{M}_{c\nrightarrow Y_1}, Q\right), \qquad l = \|\mathbb{M}_{c\nrightarrow Y_1}\| + 1, \dots, \|\mathbb{M}_{Y_2}\| \tag{42}$$

To prove this, first note that for $l = \|\mathbb{M}_{c\nrightarrow Y_1}\| + 1, \dots, \|\mathbb{M}_{Y_2}\|$ we have:

$$\begin{cases} I\left(M_l^2; Y_1 \middle| M_{\|\mathbb{M}_{c\nrightarrow Y_1}\|+1}^2, \dots, M_{l-1}^2, Q\right) \overset{(a)}{=} I\left(M_l^2; Y_1 \middle| M_{\|\mathbb{M}_{c\nrightarrow Y_1}\|+1}^2, \dots, M_{l-1}^2, \mathbb{M}_{c\nrightarrow Y_1}, Q\right) = I\left(M_l^2; Y_1 \middle| \mathbb{X}_{\mathbb{M}_{c\nrightarrow Y_1}}, M_{\|\mathbb{M}_{c\nrightarrow Y_1}\|+1}^2, \dots, M_{l-1}^2, \mathbb{M}_{c\nrightarrow Y_1}, Q\right) \\ I\left(M_l^2; Y_2 \middle| M_{\|\mathbb{M}_{c\nrightarrow Y_1}\|+1}^2, \dots, M_{l-1}^2, \mathbb{M}_{c\nrightarrow Y_1}, Q\right) = I\left(M_l^2; Y_2 \middle| \mathbb{X}_{\mathbb{M}_{c\nrightarrow Y_1}}, M_{\|\mathbb{M}_{c\nrightarrow Y_1}\|+1}^2, \dots, M_{l-1}^2, \mathbb{M}_{c\nrightarrow Y_1}, Q\right) \end{cases} \tag{43}$$

where equality (a) holds because the output $Y_1$ and the messages $\mathbb{M}_{Y_2} - \mathbb{M}_{c\nrightarrow Y_1}$ are independent of $\mathbb{M}_{c\nrightarrow Y_1}$. Now, consider the expressions in the right sides of equalities in (43). The condition (40) implies that:

$$I\left(M_l^2;Y_2\middle|\mathbb{X}_{\mathbb{M}_{c\nrightarrow Y_1}},M^2_{\|\mathbb{M}_{c\nrightarrow Y_1}\|+1},\dots,M^2_{l-1},\mathbb{M}_{c\nrightarrow Y_1},Q\right)\leq I\left(M_l^2;Y_1\middle|\mathbb{X}_{\mathbb{M}_{c\nrightarrow Y_1}},M^2_{\|\mathbb{M}_{c\nrightarrow Y_1}\|+1},\dots,M^2_{l-1},\mathbb{M}_{c\nrightarrow Y_1},Q\right)\tag{44}$$

The inequality (44) is actually derived from (40) by substituting $U\equiv M_l^2$ and $D=\left(M^2_{\|\mathbb{M}_{c\nrightarrow Y_1}\|+1},\dots,M^2_{l-1},\mathbb{M}_{c\nrightarrow Y_1},Q\right)$ for all joint PDFs of the form (24); note that we have this liberty because (40) holds for any distribution $P_{DU\mathbb{X}}$. Therefore, the equalities in (42) hold. By substituting (42) in (41), we obtain that the following achievable rate:

$$\begin{aligned}
&I\left(\mathbb{M}_{c\nrightarrow Y_1};Y_2\middle|Q\right)+I\left(\mathbb{M}_{Y_2};Y_2\middle|\mathbb{M}_{c\nrightarrow Y_1},Q\right)+I\left(\mathbb{M}_{Y_1}-\left(\mathbb{M}_{Y_2}-\mathbb{M}_{c\nrightarrow Y_1}\right);Y_1\middle|\mathbb{M}_{Y_2}-\mathbb{M}_{c\nrightarrow Y_1},Q\right)\\
&\quad=I\left(\mathbb{M}_{c\nrightarrow Y_1},\mathbb{M}_{Y_2};Y_2\middle|Q\right)+I\left(\mathbb{M}_{Y_1};Y_1\middle|\mathbb{M}_{Y_2}-\mathbb{M}_{c\nrightarrow Y_1},Q\right)\\
&\quad\overset{(a)}{=}I\left(\mathbb{M}_{Y_2};Y_2\middle|Q\right)+I\left(\mathbb{M}_{Y_1};Y_1\middle|\mathbb{M}_{Y_2}-\mathbb{M}_{c\nrightarrow Y_1},\mathbb{M}_{c\nrightarrow Y_1},Q\right)\\
&\quad=I\left(\mathbb{M}_{Y_2};Y_2\middle|Q\right)+I\left(\mathbb{M}_{Y_1};Y_1\middle|\mathbb{M}_{Y_2},Q\right)
\end{aligned}$$

where (a) holds because $\mathbb{M}_{c\nrightarrow Y_1}$ is a subset of $\mathbb{M}_{Y_2}$, and also $\mathbb{M}_{Y_1},\mathbb{M}_{Y_2}-\mathbb{M}_{c\nrightarrow Y_1},Y_1$ and $Q$ all are independent of $\mathbb{M}_{c\nrightarrow Y_1}$. The proof is thus complete. ■

***Remarks 2:***

1. Consider the case where the receiver $Y_1$ is connected to all messages, i.e., $\mathbb{M}_{c\nrightarrow Y_1}=\emptyset$. In this case, the less-noisy condition (38) is reduced to:

   $$I(U;Y_2)\leq I(U;Y_1)\qquad\text{for all joint PDFs}\qquad P_{U\mathbb{X}}$$

   Moreover, the sum-rate capacity is achieved by a fully successive decoding scheme: The receiver $Y_2$ successively decodes the messages $\mathbb{M}_{Y_2}$; the receiver $Y_1$ successively decodes all the messages $\mathbb{M}_{Y_2}$ first and then decodes the messages $\mathbb{M}_{Y_1}-\mathbb{M}_{Y_2}$.
2. Consider the case where the receiver $Y_1$ is unconnected to all messages corresponding to the receiver $Y_2$. In other words, let $\mathbb{M}_{c\nrightarrow Y_1}=\mathbb{M}_{Y_2}$. For such networks, under the less noisy condition (38), the coding scheme achieving the sum-rate capacity described in the proof of Theorem 3 (see (41)) is reduced to a simple *treating interference as noise strategy*, i.e., each receiver decodes only its corresponding messages. In this case, the sum-rate capacity is given by:

   $$\max_{P_QP_{M_1}\dots P_{M_K}\prod_{i=1}^{K_1}P_{X_i|\mathbb{M}_{X_i},Q}}\left(I\left(\mathbb{M}_{Y_1};Y_1\middle|Q\right)+I\left(\mathbb{M}_{Y_2};Y_2\middle|Q\right)\right)\tag{45}$$

   Note that since $\mathbb{M}_{c\nrightarrow Y_1}=\mathbb{M}_{Y_2}$, both $Y_1$ and messages $\mathbb{M}_{Y_1}$ are independent of $\mathbb{M}_{Y_2}$; thereby, the first mutual information function in (39) is reduced to $I\left(\mathbb{M}_{Y_1};Y_1\middle|Q\right)$.

Let us provide some examples on our result in Theorem 3. Consider a two-receiver Multiple Access Interference Network (MAIN) as introduced in Part II [2, Sec. IV]. This network is composed of two interfering MACs where two groups of transmitters (each group with an arbitrary size) communicate with two receivers via a common media: each group of transmitters send information to their respective receiver while causing interference to the other receiver. The network is depicted in Fig. 3.

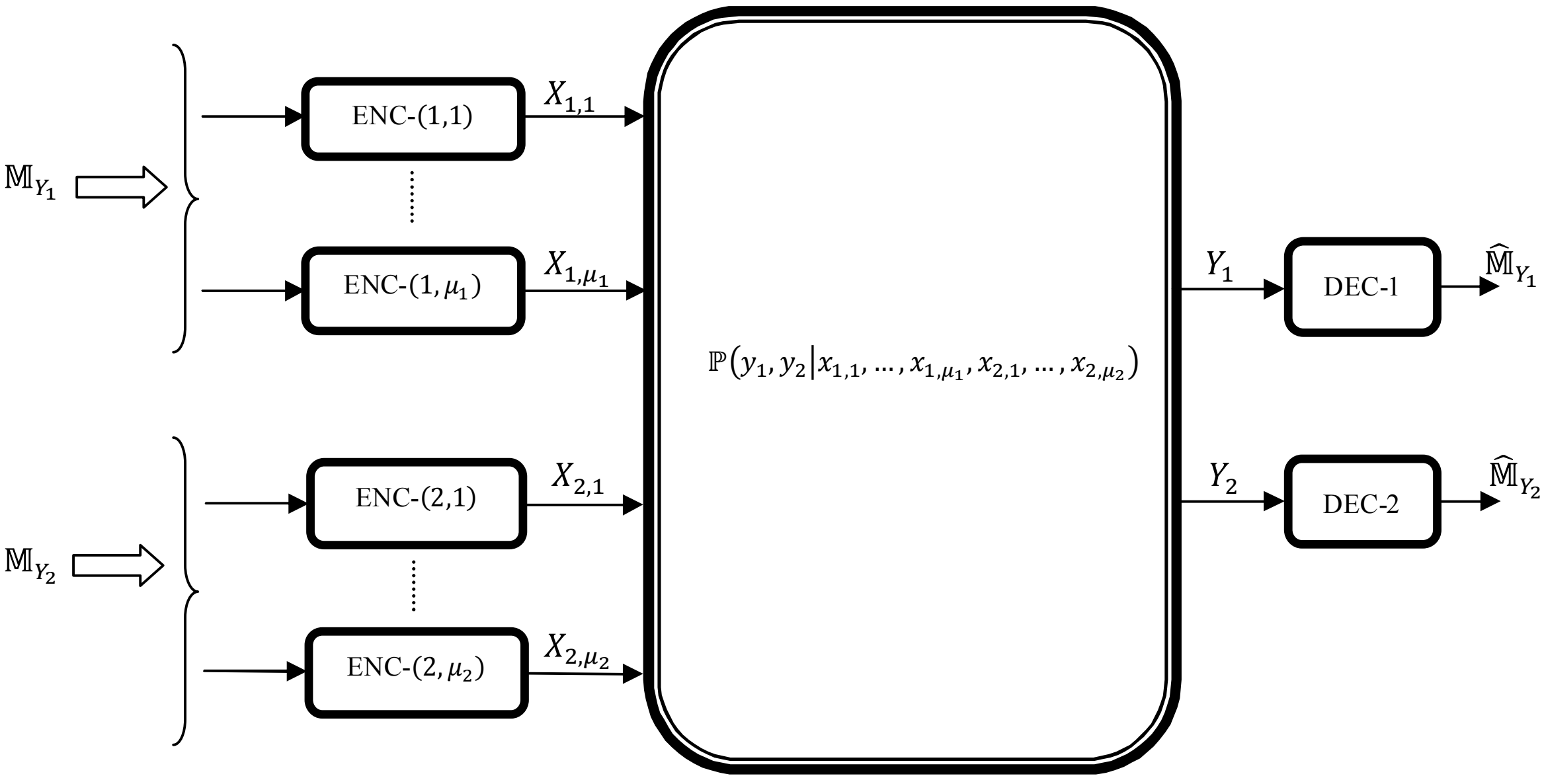

Figure 3. The two-receiver Multiple-Access-Interference Network (MAIN): two groups of transmitters $\mathbb{X}_1 \triangleq \{X_{1,1},\dots,X_{1,\mu_1}\}$ and $\mathbb{X}_2 \triangleq \{X_{2,1},\dots,X_{2,\mu_2}\}$ send the message sets $\mathbb{M}_{Y_1}$ and $\mathbb{M}_{Y_2}$ to the receivers $Y_1$ and $Y_2$, respectively. The arrangement of messages $\mathbb{M}_{Y_1}$ and $\mathbb{M}_{Y_2}$ among the corresponding transmitters is arbitrary.

In this scenario, two groups of transmitters $\mathbb{X}_1 \triangleq \{X_{1,1},\dots,X_{1,\mu_1}\}$ and $\mathbb{X}_2 \triangleq \{X_{2,1},\dots,X_{2,\mu_2}\}$ send the message sets $\mathbb{M}_{Y_1}$ and $\mathbb{M}_{Y_2}$ to the receivers $Y_1$ and $Y_2$, respectively. The parameters $\mu_1$ and $\mu_2$ are arbitrary natural numbers. Also, the arrangement of messages $\mathbb{M}_{Y_1}$ and $\mathbb{M}_{Y_2}$ among the corresponding transmitters is arbitrary. Note that the sets $\mathbb{M}_{Y_1}$ and $\mathbb{M}_{Y_2}$ are disjoint. For the MAIN (with the condition (27)) the outer bound in (28) can be further simplified as:

$$\mathcal{C}_{sum}^{MAIN} \leq \max_{\mathcal{P}_{sum}^{MAIN}} \left( I(\mathbb{X}_1;Y_1|\mathbb{X}_2,Q) + I(\mathbb{X}_2;Y_2|Q) \right) \tag{46}$$

where $\mathcal{P}_{sum}^{MAIN}$ denotes the set of all joint PDFs which are induced on $Q,\mathbb{X}_1,\mathbb{X}_2$ by the following PDFs:

$$P_Q P_{M_1} \dots P_{M_K} \prod_{i=1}^{K_1} P_{X_i|\mathbb{M}_{X_i},Q} \tag{47}$$

The reason is that $\mathbb{X}_1$ and $\mathbb{X}_2$ are respectively given by deterministic functions of $(\mathbb{M}_{Y_1},Q)$ and $(\mathbb{M}_{Y_2},Q)$; moreover, given $(\mathbb{X}_i,Q), i = 1,2$, the outputs $Y_1$ and $Y_2$ are independent of the messages in $\mathbb{M}_{Y_i}$. Similarly, the outer bound (30) is simplified as follows:

$$\mathcal{C}_{sum}^{MAIN} \leq \max_{\mathcal{P}_{sum}^{MAIN}} \left( I(\mathbb{X}_1,\mathbb{X}_2;Y_1|Q) \right) \tag{48}$$

This bound holds if the condition (29) is satisfied. We next examine some special cases of the MAINs.

***Example 1:*** **One-Sided MAIN**

Consider a MAIN where its transition probability function is factorized as follows:

$$\mathbb{P}(y_1,y_2|x_{1,1},\dots,x_{1,\mu_1},x_{2,1},\dots,x_{2,\mu_2}) = \mathbb{P}(y_1|x_{1,1},\dots,x_{1,\mu_1})\mathbb{P}(y_2|x_{1,1},\dots,x_{1,\mu_1},x_{2,1},\dots,x_{2,\mu_2}) \tag{49}$$

In this scenario, only the receiver $Y_1$ experiences interference. Inspired by the two-user one-sided CIC (see Part I [1, Sec. III.A.3]), we call such a network as the *One-Sided MAIN*. According to Theorem 3, for the one-sided MIAN in (49) if the following less-noisy condition holds:

$$I(U;Y_2|\mathbb{X}_2) \leq I(U;Y_1|\mathbb{X}_2) \qquad \text{for all joint PDFs} \qquad P_{U\mathbb{X}_1}\prod_{X_i\in\mathbb{X}_2} P_{X_i},$$

(50)

then the network has noisy interference, i.e., the sum-rate capacity is achieved by treating interference as noise. The sum-rate capacity is given by:

$$\max_{P_Q P_{M_1}\dots P_{M_K}\prod_{i=1}^{K_1} P_{X_i|\mathbb{M}_{X_i},Q}} \left(I(\mathbb{M}_{Y_1};Y_1|Q) + I(\mathbb{M}_{Y_2};Y_2|Q)\right) = \max_{\mathcal{P}_{sum}^{MAIN}} \left(I(\mathbb{X}_1;Y_1|Q) + I(\mathbb{X}_2;Y_2|Q)\right)$$

(51)

where $\mathcal{P}_{sum}^{MAIN}$ denotes the set of all joint PDFs which are induced on $Q,\mathbb{X}_1,\mathbb{X}_2$ by the PDFs (24).

***Example 2:***

Figure 4 depicts a 4-tranmitter/2-reciever MAIN where the transmitters $X_{1,1}$ and $X_{1,2}$, respectively, send independent messages $M_1$ and $M_2$ to the receiver $Y_1$, and the transmitters $X_{2,1}$ and $X_{2,2}$, respectively, send $M_3$ and $M_4$ to the receiver $Y_2$.

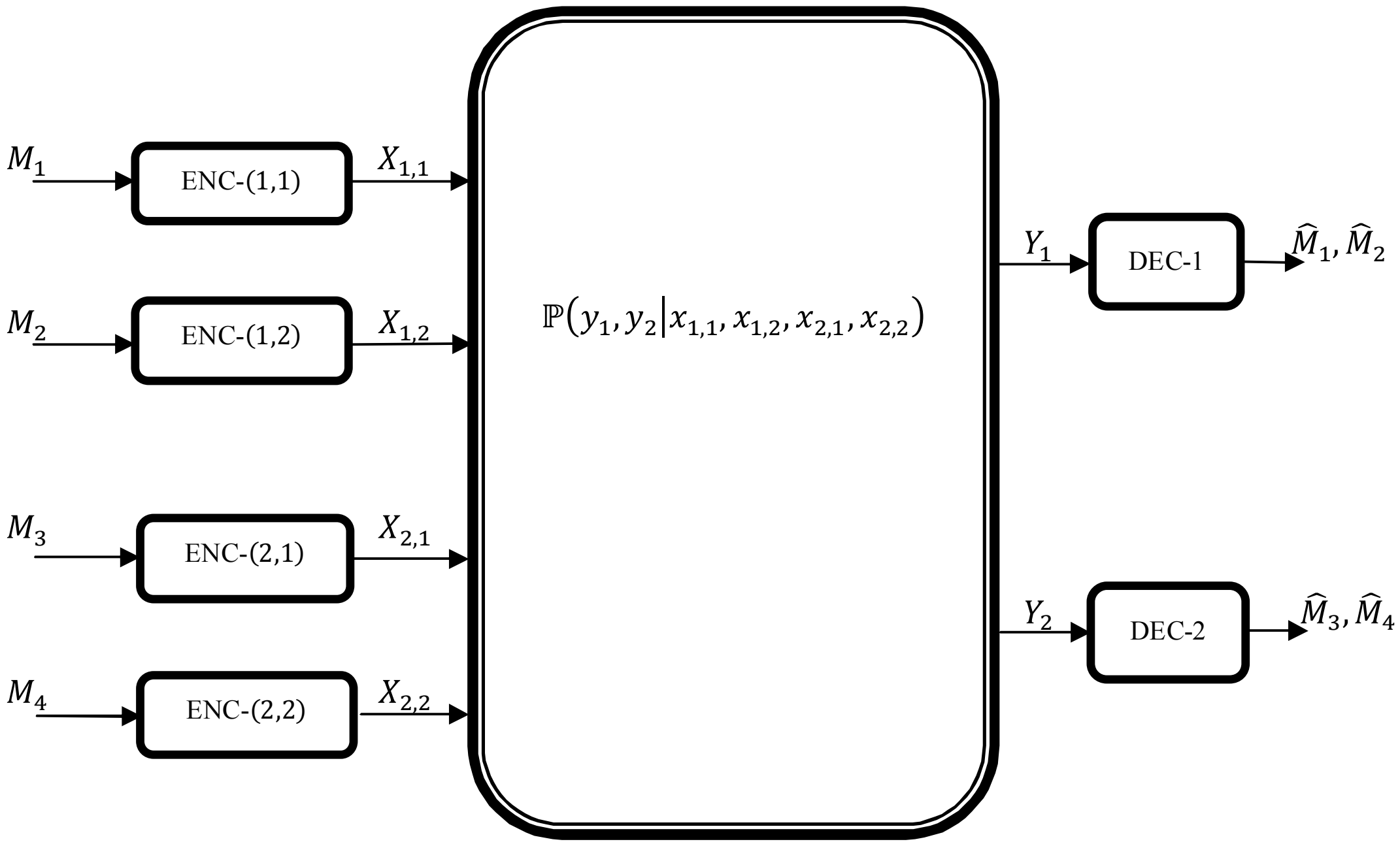

Figure 4. A two-receiver MAIN.

Consider the case where the transition probability function of the network is given by:

$$\mathbb{P}(y_1,y_2|x_{1,1},x_{1,2},x_{2,1},x_{2,2}) = \mathbb{P}(y_1|x_{1,1},x_{1,2},x_{2,1})\mathbb{P}(y_2|x_{1,1},x_{1,2},x_{2,1},x_{2,2})$$

(52)

Therefore, the message $M_3$ is unconnected to the receiver $Y_1$. According to Theorem 3, if the network satisfies the following less-noisy condition:

$$I(U;Y_2|X_{2,1}) \leq I(U;Y_1|X_{2,1}) \quad \text{for all joint PDFs} \quad P_{UX_{1,1}X_{1,2}X_{2,2}}P_{X_{2,1}},$$

(53)

then the sum-rate capacity is given by:

$$\max_{P_Q P_{X_{1,1}|Q} P_{X_{1,2}|Q} P_{X_{2,1}|Q} P_{X_{2,2}|Q}} \left(I(X_{1,1},X_{1,2};Y_1|X_{2,1},X_{2,2},Q) + I(X_{2,1},X_{2,2};Y_2|Q)\right)$$

(54)

Note that the first mutual information in (54) is actually equal to $I\left(X_{1,1},X_{1,2};Y_1\middle|X_{2,2},Q\right)$ because the receiver $Y_1$ is unconnected to the transmitter $X_{2,1}$.

We next derive scenarios where the incorporation of the outer bounds (28) and (30) is optimal. First, we remark that the messages $\mathbb{M}=\mathbb{M}_{Y_1}\cup\mathbb{M}_{Y_2}$ in characterization of these outer bounds are actually auxiliary random variables. The expressions in (28) and (30) should be optimized over all these auxiliaries. Let us consider a degraded network where $\mathbb{X}\to Y_1\to Y_2$ form a Markov chain. For such a network, it is clear that the condition (38) is satisfied; thereby, the outer bound (28) is optimal and the sum-rate capacity is given by (39). As discussed in introduction, in Part II [2], we proved that for the degraded network the sum-rate capacity is the same as when the messages $\mathbb{M}^*=\mathbb{M}_{Y_1}^*\cup\mathbb{M}_{Y_2}^*$ are transmitted where $\mathbb{M}_{Y_1}^*,\mathbb{M}_{Y_2}^*$ and $\mathbb{M}^*$ are given by (11)-(12). In other words, if the network is degraded, then a solution to the maximization in (28) (or (39)) is to nullify all the axillaries belonging to $\mathbb{M}-\mathbb{M}^*$ and replace $\mathbb{M}_{Y_j}$ by $\mathbb{M}_{Y_j}^*$, $j=1,2$. Note that the outer bound (28) holds if the less-noisy condition (27) is satisfied. This less-noisy condition is weaker than the degradedness; nonetheless, by exploiting it, one can still show that a solution to the maximization (28) is to nullify all the axillaries belonging to $\mathbb{M}-\mathbb{M}^*$. The proof of this fact is in essence similar to the one we presented in Part II [2] for degraded networks; however, it is somewhat more difficult in its description; the details are omitted for brevity. It is remarkable that the same conclusion still holds when we consider the incorporation of the outer bounds (28) and (30) for those networks which satisfy both the conditions (27) and (29) simultaneously. This is given in the following theorem.

***Theorem 4)*** *Consider the general two-receiver interference network which is obtained by setting $K_2=2$ in the scenario shown in Fig.1. If the transition probability function of the network satisfies the following conditions:*

$$\begin{cases} I\left(U;Y_2\middle|\mathbb{X}_{\mathbb{M}_{Y_2}}\right)\le I\left(U;Y_1\middle|\mathbb{X}_{\mathbb{M}_{Y_2}}\right), & \text{for all joint PDFs } P_{U,\mathbb{X}-\mathbb{X}_{\mathbb{M}_{Y_2}}}\prod_{X_i\in\mathbb{X}_{\mathbb{M}_{Y_2}}}P_{X_i} \\ I\left(\mathbb{X}-\mathbb{X}_{\mathbb{M}_{Y_1}};Y_2\middle|\mathbb{X}_{\mathbb{M}_{Y_1}}\right)\le I\left(\mathbb{X}-\mathbb{X}_{\mathbb{M}_{Y_1}};Y_1\middle|\mathbb{X}_{\mathbb{M}_{Y_1}}\right), & \text{for all joint PDFs } P_{\mathbb{X}-\mathbb{X}_{\mathbb{M}_{Y_1}}}\prod_{X_i\in\mathbb{X}_{\mathbb{M}_{Y_1}}}P_{X_i} \end{cases},\tag{55}$$

*then the sum-rate capacity is bounded-above as:*

$$\mathcal{C}_{sum}^{GINTR}\le\max_{P_Q\prod_{M_\Delta^\nabla\in\mathbb{M}^*}P_{M_\Delta^\nabla}\prod_{i=1}^{K_1}P_{X_i|\mathbb{M}_{X_i}^*,Q}}\min\begin{Bmatrix} I\left(\mathbb{M}_{Y_1}^*;Y_1\middle|\mathbb{M}_{Y_2}^*,Q\right)+I\left(\mathbb{M}_{Y_2}^*;Y_2\middle|Q\right), \\ I\left(\mathbb{M}_{Y_1}^*,\mathbb{M}_{Y_2}^*;Y_1\middle|Q\right)\end{Bmatrix}\tag{56}$$

*where the set $\mathbb{M}^*$ and its subsets $\mathbb{M}_{Y_1}^*,\mathbb{M}_{Y_2}^*$ are derived from* (11)-(12) *for $K_2=2$, and the sets $\mathbb{M}_{X_i}^*$ are given by $\mathbb{M}_{X_i}^*\triangleq\mathbb{M}_{X_i}\cap\mathbb{M}^*$, $i=1,\ldots,K_1$. Moreover, if $\left\|\mathbb{M}_{Y_2}^*\right\|=1$, i.e., there exists only one message in $\mathbb{M}^*$ for the receiver $Y_2$, the bound (56) is achievable; therefore, it coincides with the sum-rate capacity.*

*Proof of Theorem 4)* By combining the results of Parts A and B of Theorem 2, we deduce that if the conditions (55) hold simultaneously, then the following constitutes an outer bound on the sum-rate capacity:

$$\mathcal{C}_{sum}^{GINTR}\le\max_{P_Q\prod_{M_\Delta^\nabla\in\mathbb{M}}P_{M_\Delta^\nabla}\prod_{i=1}^{K_1}P_{X_i|\mathbb{M}_{X_i},Q}}\min\begin{Bmatrix} I\left(\mathbb{M}_{Y_1};Y_1\middle|\mathbb{M}_{Y_2},Q\right)+I\left(\mathbb{M}_{Y_2};Y_2\middle|Q\right), \\ I\left(\mathbb{M}_{Y_1},\mathbb{M}_{Y_2};Y_1\middle|Q\right)\end{Bmatrix}\tag{57}$$

As discussed before, a solution to the maximization (57) is to nullify all the axillaries belonging to $\mathbb{M}-\mathbb{M}^*$. This fact is due to the first condition of (55). The outer bound (56) is obtained by replacing $\mathbb{M}_{Y_1}$ by $\mathbb{M}_{Y_1}^*$ and $\mathbb{M}_{Y_2}$ by $\mathbb{M}_{Y_2}^*$ in (57).

It remains to show that if $\left\|\mathbb{M}_{Y_2}^*\right\|=1$, then (56) is achievable as well. The achievability scheme is derived by a simple successive decoding scheme. At the transmitters, the messages $\mathbb{M}-\mathbb{M}^*$ are ignored and the remaining messages, i.e., those belonging to $\mathbb{M}^*$, are separately encoded using independent codewords. At the receiver $Y_2$, the single message belonging to $\mathbb{M}_{Y_2}^*$ is decoded. This message is also decoded at the receiver $Y_1$ first. The rate cost due to these two steps is given by:

$$\min\left\{I\left(\mathbb{M}_{Y_2}^*;Y_1\middle|Q\right),I\left(\mathbb{M}_{Y_2}^*;Y_2\middle|Q\right)\right\}\tag{58}$$

After decoding the message in $\mathbb{M}_{Y_2}^*$, the receiver $Y_1$ successively decodes all the messages belonging to $\mathbb{M}_{Y_1}^*$ in an arbitrary order. The rate cost due to this step is,

$$I\left(\mathbb{M}_{Y_1}^*; Y_1 \middle| \mathbb{M}_{Y_2}^*, Q\right) \tag{59}$$

We remark that the auxiliary random variable $Q$ in (58) and (59) is the time-sharing parameter. By combining (58) and (59), we derive the achievability of (56). The proof is thus complete. ■

***Remarks 3:***

1. Note that if $\left\|\mathbb{M}_{Y_2}^*\right\| \geq 1$, then the sum-rate achievable by the successive decoding scheme does not necessarily coincide with the bound (56).
2. One can easily check that Theorems 4 and 6 of Part III [3, Sec. IV.A] are special cases of Theorem 4.

Let us provide an example to illustrate our result in Theorem 4. Consider again the MAIN in Fig. 3. Suppose that the arrangement of messages among the transmitters is arbitrary except that all the transmitters $X_{2,1}, \dots, X_{2,\mu_2}$ cooperatively send a common message to their respective receiver, i.e., $Y_2$, in addition to other messages which possibly transmit. This condition implies that $\left\|\mathbb{M}_{Y_2}^*\right\| = 1$ because, considering the MACCM plan in Fig. 2, the message that is cooperatively transmitted by $X_{2,1}, \dots, X_{2,\mu_2}$ is a cloud center for all other possible messages among these transmitters. Also, assume that the network transition probability function satisfies the following conditions:

$$\begin{cases} I(U; Y_2|\mathbb{X}_2) \leq I(U; Y_1|\mathbb{X}_2), & \text{for all joint PDFs} \quad P_{U\mathbb{X}_1}\prod_{X_i\in\mathbb{X}_2} P_{X_i} \\ I(\mathbb{X}_2; Y_2|\mathbb{X}_1) \leq I(\mathbb{X}_2; Y_1|\mathbb{X}_1), & \text{for all joint PDFs} \quad P_{\mathbb{X}_2}\prod_{X_i\in\mathbb{X}_1} P_{X_i} \end{cases} \tag{60}$$

Therefore, for such a network the conditions of Theorem 4 are satisfied and the sum-rate capacity is given by (56) which is simplified below:

$$\max_{\mathcal{P}_{Q\mathbb{X}_1\mathbb{X}_2}^*} \min \begin{Bmatrix} I(\mathbb{X}_1; Y_1|\mathbb{X}_2, Q) + I(\mathbb{X}_2; Y_2|Q), \\ I(\mathbb{X}_1, \mathbb{X}_2; Y_1|Q) \end{Bmatrix} \tag{61}$$

where $\mathcal{P}_{Q\mathbb{X}_1\mathbb{X}_2}^*$ denotes the set of all joint PDFs which are induced on $Q, \mathbb{X}_1$ and $\mathbb{X}_2$ by the following:

$$P_Q \prod_{M_\Delta^\nabla \in \mathbb{M}_{Y_1}^*} P_{M_\Delta^\nabla} \prod_{i=1}^{\mu_1} P_{X_{1,i}|\mathbb{M}_{X_{1,i}}^*, Q} P_{\mathbb{X}_2|Q} \tag{62}$$

We remark that the outer bounds given in this section can be used to establish the exact sum-rate capacity for many cases other than those derived in Theorems 3 and 4. Let us concentrate on the outer bound (56) which holds for those networks satisfying the conditions (55). In Theorem 4, we proved that if there is only one message in $\mathbb{M}^*$ for the receiver $Y_2$, this bound is also achievable which yields the sum-rate capacity. The latter condition is in fact a sufficient condition for this purpose. Clearly, for a network satisfying (55) if $\left\|\mathbb{M}_{Y_2}^*\right\| = 1$, the sum-rate (56) is achieved by the successive decoding scheme without requiring to introduce any new condition on the network transition probability function. Nonetheless, the bound (56) is achievable (and thereby, the sum-rate capacity) for a broad range of other network scenarios. We describe the procedure to this development below.

First, let us discuss a more efficient successive decoding scheme. Clearly, in the previous scheme given by (36), the receiver $Y_2$ decodes its corresponding messages successively. The receiver $Y_1$ also successively decodes the messages corresponding to the receiver $Y_2$ first and then successively decodes its own messages. This achievability scheme can be improved by applying a jointly decoding technique at each step. Specifically, consider the following achievability scheme for the general two-receiver interference network with the associated message set $\mathbb{M}$. At the transmitters, all the messages in $\mathbb{M}$ are separately encoded using independent codewords. At the receiver $Y_2$, the messages $\mathbb{M}_{Y_2}$ are jointly decoded. The receiver $Y_1$ also jointly decodes the messages $\mathbb{M}_{Y_2}$ first.

After decoding the messages $\mathbb{M}_{Y_2}$, this receiver jointly decodes those in $\mathbb{M}_{Y_1}-\mathbb{M}_{Y_2}$. One can show that the sum-rate achievable by this scheme is given by:

$$\max_{P_Q\prod_{M_\Delta^\nabla\in\mathbb{M}}P_{M_\Delta^\nabla}\prod_{i=1}^{K_1}P_{X_i|\mathbb{M}_{X_i},Q}}\left(I\left(\mathbb{M}_{Y_1};Y_1\middle|\mathbb{M}_{Y_2},Q\right)+\min_{\Omega\subseteq\mathbb{M}_{Y_2}}\left\{I\left(\Omega;Y_1\middle|\mathbb{M}_{Y_2}-\Omega\right)+I\left(\mathbb{M}_{Y_2}-\Omega;Y_2\middle|\Omega\right)\right\}\right)$$

(63)

This scheme, which we call "*successive-joint decoding scheme*", in general yields a larger achievable sum-rate than the case where the messages at each step are decoded successively, i.e., the scheme given by (36).

Now consider the achievable sum-rate (63). If $\left\|\mathbb{M}_{Y_2}^*\right\|=1$, then it coincides directly with (56) which yields the exact sum-rate capacity as given in Theorem 4. Otherwise, for this purpose, it is required to impose additional appropriate conditions on the network probability function. We conclude this section by providing an example.

Again consider the 4-tranmitter/2-reciever MAIN in Fig. 4. Assume that the network is fully connected. According to Theorem 4, if the following conditions hold:

$$\begin{cases} I\left(U;Y_2\middle|X_{2,1},X_{2,2}\right)\leq I\left(U;Y_1\middle|X_{2,1},X_{2,2}\right), & \text{for all joint PDFs } P_{UX_{1,1}X_{1,2}}P_{X_{2,1}}P_{X_{2,2}} \\ I\left(X_{2,1},X_{2,2};Y_2\middle|X_{1,1},X_{1,2}\right)\leq I\left(X_{2,1},X_{2,2};Y_1\middle|X_{1,1},X_{1,2}\right), & \text{for all joint PDFs } P_{X_{2,1}X_{2,2}}P_{X_{1,1}}P_{X_{1,2}}\end{cases},$$

(64)

then the sum-rate capacity is outer bounded by:

$$\max_{P_QP_{X_{1,1}|Q}P_{X_{1,2}|Q}P_{X_{2,1}|Q}P_{X_{2,2}|Q}}\min\left\{\begin{matrix} I\left(X_{1,1},X_{1,2};Y_1\middle|X_{2,1},X_{2,2},Q\right)+I\left(X_{2,1},X_{2,2};Y_2\middle|Q\right), \\ I\left(X_{1,1},X_{1,2},X_{2,1},X_{2,2};Y_1\middle|Q\right)\end{matrix}\right\}$$

(65)

On the one hand, by the successive-joint decoding scheme given by (63), we obtain the following achievable sum-rate:

$$\max_{P_QP_{X_{1,1}|Q}P_{X_{1,2}|Q}P_{X_{2,1}|Q}P_{X_{2,2}|Q}}\min\left\{\begin{matrix} I\left(X_{1,1},X_{1,2},X_{2,1};Y_1\middle|X_{2,2},Q\right)+I\left(X_{2,2};Y_2\middle|X_{2,1},Q\right), \\ I\left(X_{1,1},X_{1,2};Y_1\middle|X_{2,1},X_{2,2},Q\right)+I\left(X_{2,1},X_{2,2};Y_2\middle|Q\right), \\ I\left(X_{1,1},X_{1,2},X_{2,2};Y_1\middle|X_{2,1},Q\right)+I\left(X_{2,1};Y_2\middle|X_{2,2},Q\right), \\ I\left(X_{1,1},X_{1,2},X_{2,1},X_{2,2};Y_1\middle|Q\right),\end{matrix}\right\}$$

(66)

Since this network does not satisfy the condition $\left\|\mathbb{M}_{Y_2}^*\right\|=1$, the expressions (65) and (66) do not agree in general. Nonetheless, let us impose that the network transition probability function also satisfies the following conditions:

$$\begin{cases} I\left(X_{2,1};Y_2\middle|Q\right)\leq I\left(X_{2,1};Y_1\middle|X_{2,2},Q\right) \\ I\left(X_{2,2};Y_2\middle|Q\right)\leq I\left(X_{2,2};Y_1\middle|X_{2,1},Q\right)\end{cases}$$

(67)

for all joint PDFs $P_QP_{X_{1,1}|Q}P_{X_{1,2}|Q}P_{X_{2,1}|Q}P_{X_{2,2}|Q}$ which are a solution to the maximization (65). Under these conditions, one can readily show that (66) coincides with (65). Thus, the MAIN in Fig. 4 if satisfies both the conditions (64) and (67), its sum-rate capacity is given by (65). In fact, one can show that if both inequalities in (67) simultaneously hold in the reverse direction, the outer bound (65) and the achievable sum-rate (66) still coincide. In this case, the sum-rate capacity is reduced to the following:

$$\max_{P_QP_{X_{1,1}|Q}P_{X_{1,2}|Q}P_{X_{2,1}|Q}P_{X_{2,2}|Q}}\left(I\left(X_{1,1},X_{1,2},X_{2,1},X_{2,2};Y_1\middle|Q\right)\right)$$

(68)

Now, let us consider the Gaussian version of the network in Fig. 4 which is given by:

$$\begin{cases} Y_1 = a_{1,1}X_{1,1} + a_{1,2}X_{1,2} + a_{2,1}X_{2,1} + a_{2,2}X_{2,2} + Z_1 \\ Y_2 = b_{1,1}X_{1,1} + b_{1,2}X_{1,2} + b_{2,1}X_{2,1} + b_{2,2}X_{2,2} + Z_2 \end{cases} \tag{69}$$

where $Z_1$ and $Z_2$ are zero-mean unit-variance Gaussian RVs, and the inputs are subject to power constraints $\mathbb{E}[X_{i,j}^2] \leq P_{i,j},\ i,j = 1,2$. According to Lemmas 3 and 4, if the network gains satisfy the following:

$$\begin{cases} \frac{b_{1,1}}{a_{1,1}} = \frac{b_{1,2}}{a_{1,2}} = \alpha & |\alpha| \leq 1 \\ \frac{b_{2,1}}{a_{2,1}} = \frac{b_{2,2}}{a_{2,2}} = \beta & |\beta| \leq 1 \end{cases}, \tag{70}$$

then the conditions (64) hold. Therefore, (65) constitutes an outer bound on the sum-rate capacity. For the Gaussian network (69) satisfying (70), this outer bound actually coincides with the sum-rate capacity (without any additional condition on the network gains). Let us prove this fact. First note that, by using the entropy power inequality, one can show that under the conditions (70), the Gaussian input distributions without time-sharing ($Q \equiv \emptyset$) is the solution to the maximization (65). Therefore, if both the inequalities (67) are satisfied for Gaussian input distributions, then (65) and (66) coincide which yield the sum-rate capacity. Consider the case where one of the inequalities (67) holds in the reverse direction (for Gaussian distributions). Without loss of generality, assume that:

$$I(X_{2,2}; Y_2) > I(X_{2,2}; Y_1|X_{2,1}),\quad \text{for Gaussian distributions} \tag{71}$$

Considering (70), one can readily show that (71) implies the following:

$$I(X_{2,1}; Y_2|X_{2,2}) > I(X_{2,1}; Y_1),\quad \text{for Gaussian distributions} \tag{72}$$

This is derived by a simple evaluation of (71) and (72) for Gaussian input distributions. Now, using (71) and (72), we can obtain that (65) and (66) coincide, which yield the sum-rate capacity. Thus, if the network parameters satisfy the conditions (70), its sum-rate capacity is given by:

$$\min\begin{Bmatrix} \psi(a_{1,1}^2P_{1,1} + a_{1,2}^2P_{1,2}) + \psi\left(\frac{b_{2,1}^2P_{2,1} + b_{2,2}^2P_{2,2}}{b_{1,1}^2P_{1,1} + b_{1,2}^2P_{1,2} + 1}\right) \\ \psi(a_{1,1}^2P_{1,1} + a_{1,2}^2P_{1,2} + a_{2,1}^2P_{2,1} + a_{2,2}^2P_{2,2}) \end{Bmatrix}, \tag{73}$$

which is achieved by the successive-joint decoding scheme.

Indeed, the above approach can be followed to obtain the exact sum-rate capacity for many other scenarios.

# IV. General Multi-Receiver Networks

Now, we intend to extend our results to the interference networks with arbitrary number of receivers. In what follows, we present new techniques to derive powerful capacity outer bounds for the interference networks of arbitrary large sizes. As we will see, these outer bounds can be used to prove explicit sum capacity results for a broad range of network topologies. To clarify the importance of our results, it is sufficient to note that the multi-receiver interference networks are far less understood [9, p. 6-64] so that even there are a very few cases where the sum-rate capacity is known. In fact, most of the existing researches for these networks are regarded to derive approximate capacity results and the degrees of freedom region for Gaussian networks while for discrete networks the research results are scarce.

## IV.A) Outer bounds and sum-rate capacities

In this subsection, we establish a unified outer bound on the sum-rate capacity of the general multi-receiver interference networks which satisfy certain sequentially less-noisy conditions. The proof of this outer bound includes a novel interesting technique requiring a sequential application of the Csiszar-Korner identity [15]. We then obtain scenarios for which the derived outer bound is also achievable which yields the exact sum-rate capacity. Our general outer bound is given in the next theorem.

***Theorem 5)*** *Consider the general interference network with $K_1$ transmitter and $K_2$ receivers with the associated message set $\mathbb{M}$ as shown in Fig. 1. Assume that the network transition probability function satisfies the following less-noisy conditions:*

$$I\left(U;Y_j\middle|\mathbb{X}_{\cup_{l=j}^{K_2}\mathbb{M}_{Y_l}}\right) \leq I\left(U;Y_{j-1}\middle|\mathbb{X}_{\cup_{l=j}^{K_2}\mathbb{M}_{Y_l}}\right),$$

$$\textit{for all joint PDFs} \quad P_{U,\mathbb{X}-\mathbb{X}_{\cup_{l=j}^{K_2}\mathbb{M}_{Y_l}}} \prod_{X_i\in\mathbb{X}_{\cup_{l=j}^{K_2}\mathbb{M}_{Y_l}}} P_{X_i}, \qquad j=2,\ldots,K_2 \tag{74}$$

*The sum-rate capacity of the network is bounded-above by:*

$$\mathcal{C}_{sum}^{GIN} \leq \max_{\mathcal{P}^{GIN}}\left(I\left(\mathbb{M}_{Y_1};Y_1\middle|\mathbb{M}_{Y_2},\ldots,\mathbb{M}_{Y_{K_2-1}},\mathbb{M}_{Y_{K_2}},Q\right)+\cdots+I\left(\mathbb{M}_{Y_{K_2-1}};Y_{K_2-1}\middle|\mathbb{M}_{Y_{K_2}},Q\right)+I\left(\mathbb{M}_{Y_{K_2}};Y_{K_2}\middle|Q\right)\right) \tag{75}$$

*where $\mathcal{P}^{GIN}$ denotes the set of all joint PDFs given below:*

$$P_Q P_{M_1}\ldots P_{M_K}\prod_{i=1}^{K_1} P_{X_i|\mathbb{M}_{X_i},Q}, \quad P_{X_i|\mathbb{M}_{X_i},Q}\in\{0,1\}, \; i=1,\ldots,K_1 \tag{76}$$

*Proof of Theorem 5)* Consider a code of length $n$ for the network with the rates $R_1,R_2,\ldots,R_K$ corresponding to the messages $M_1,M_2,\ldots,M_K$. Assume that the conditions (74) hold. According to Lemma 2, these conditions imply that:

$$I\left(U;Y_j\middle|\mathbb{X}_{\cup_{l=j}^{K_2}\mathbb{M}_{Y_l}},D\right) \leq I\left(U;Y_{j-1}\middle|\mathbb{X}_{\cup_{l=j}^{K_2}\mathbb{M}_{Y_l}},D\right), \quad \textit{for all joint PDFs} \quad P_{DU\mathbb{X}}, \quad j=2,\ldots,K_2 \tag{77}$$

Define the sets $\mathbb{M}_{\overset{\mapsto}{Y_j}}, j=1,\ldots,K_2$ as follows:

$$\mathbb{M}_{\overset{\mapsto}{Y_j}} \triangleq \mathbb{M}_{Y_j}-\left(\mathbb{M}_{Y_{j+1}}\cup\ldots\cup\mathbb{M}_{Y_{K_2}}\right), \qquad j=1,\ldots,K_2, \tag{78}$$

where $\mathbb{M}_{Y_{K_2+1}}\triangleq\emptyset$. Let us recall that in the following analysis, for a given subset of messages $\Omega$, the notation $\underline{id}_\Omega$ denotes the identification of the set $\Omega$, as defined in Part I [1, Def. II.1]. Using Fano's inequality, we have:

$$\begin{aligned} n\left(\sum_{l\in\underline{id}_{\mathbb{M}_{\overset{\mapsto}{Y_j}}}} R_l\right) &\leq I\left(\mathbb{M}_{\overset{\mapsto}{Y_j}};Y_j^n\right)+n\epsilon_{j,n} \\ &\overset{(a)}{\leq} I\left(\mathbb{M}_{\overset{\mapsto}{Y_j}};Y_j^n\middle|\mathbb{M}_{Y_{j+1}}\cup\ldots\cup\mathbb{M}_{Y_{K_2}}\right)+n\epsilon_{j,n} \\ &= I\left(\mathbb{M}_{Y_j};Y_j^n\middle|\mathbb{M}_{Y_{j+1}}\cup\ldots\cup\mathbb{M}_{Y_{K_2}}\right)+n\epsilon_{j,n} \end{aligned} \tag{79}$$

where $\epsilon_{j,n} \to 0$ as $n \to \infty$. Note that the inequality (a) in (79) holds because the messages that belong to $\mathbb{M}_{\overset{\mapsto}{Y_j}}$ are independent of those in $\mathbb{M}_{Y_{j+1}} \cup \ldots \cup \mathbb{M}_{Y_{K_2}}$. By adding the two sides of (79) for $j = 1, \ldots, K_2$, we obtain:

$$nC_{sum}^{GIN} = n\left(\sum_{j=1}^{K_2} \sum_{l \in \underline{id}_{\mathbb{M}_{\overset{\mapsto}{Y_j}}}} R_l\right) \leq I\left(\mathbb{M}_{Y_1}; Y_1^n \middle| \mathbb{M}_{Y_2}, \ldots, \mathbb{M}_{Y_{K_2-1}}, \mathbb{M}_{Y_{K_2}}\right) + \cdots + I\left(\mathbb{M}_{Y_{K_2-1}}; Y_{K_2-1}^n \middle| \mathbb{M}_{Y_{K_2}}\right) + I\left(\mathbb{M}_{Y_{K_2}}; Y_{K_2}^n\right) + n\epsilon_n \tag{80}$$

where $\epsilon_n \to 0$ as $n \to \infty$. Now using the conditions in (77), we derive a single-letter form of the right side of (80). Consider the last two mutual information functions in (80). We have:

$$\begin{cases} I\left(\mathbb{M}_{Y_{K_2-1}}; Y_{K_2-1}^n \middle| \mathbb{M}_{Y_{K_2}}\right) = \sum_{t=1}^{n} I\left(\mathbb{M}_{Y_{K_2-1}}; Y_{K_2-1,t} \middle| \mathbb{M}_{Y_{K_2}}, Y_{K_2-1}^{t-1}\right) \\ I\left(\mathbb{M}_{Y_{K_2}}; Y_{K_2}^n\right) = \sum_{t=1}^{n} I\left(\mathbb{M}_{Y_{K_2}}; Y_{K_2,t} \middle| Y_{K_2,t+1}^n\right) \leq \sum_{t=1}^{n} I\left(\mathbb{M}_{Y_{K_2}}, Y_{K_2,t+1}^n; Y_{K_2,t}\right) \end{cases} \tag{81}$$

Please precisely observe the style of applying the chain rule in each of the equations (81). Now, from (81) we derive:

$$\begin{aligned}
& I\left(\mathbb{M}_{Y_{K_2-1}}; Y_{K_2-1}^n \middle| \mathbb{M}_{Y_{K_2}}\right) + I\left(\mathbb{M}_{Y_{K_2}}; Y_{K_2}^n\right) \\
&\leq \textstyle\sum_{t=1}^{n} I\left(\mathbb{M}_{Y_{K_2-1}}; Y_{K_2-1,t} \middle| \mathbb{M}_{Y_{K_2}}, Y_{K_2-1}^{t-1}\right) + \sum_{t=1}^{n} I\left(\mathbb{M}_{Y_{K_2}}, Y_{K_2,t+1}^n; Y_{K_2,t}\right) \\
&= \textstyle\sum_{t=1}^{n} I\left(\mathbb{M}_{Y_{K_2-1}}, Y_{K_2,t+1}^n; Y_{K_2-1,t} \middle| \mathbb{M}_{Y_{K_2}}, Y_{K_2-1}^{t-1}\right) + \sum_{t=1}^{n} I\left(\mathbb{M}_{Y_{K_2}}, Y_{K_2-1}^{t-1}, Y_{K_2,t+1}^n; Y_{K_2,t}\right) \\
&\quad - \textstyle\sum_{t=1}^{n} I\left(Y_{K_2,t+1}^n; Y_{K_2-1,t} \middle| \mathbb{M}_{Y_{K_2-1}}, \mathbb{M}_{Y_{K_2}}, Y_{K_2-1}^{t-1}\right) - \sum_{t=1}^{n} I\left(Y_{K_2-1}^{t-1}; Y_{K_2,t} \middle| \mathbb{M}_{Y_{K_2}}, Y_{K_2,t+1}^n\right) \\
&= \textstyle\sum_{t=1}^{n} I\left(\mathbb{M}_{Y_{K_2-1}}; Y_{K_2-1,t} \middle| \mathbb{M}_{Y_{K_2}}, Y_{K_2-1}^{t-1}, Y_{K_2,t+1}^n\right) + \sum_{t=1}^{n} I\left(Y_{K_2,t+1}^n; Y_{K_2-1,t} \middle| \mathbb{M}_{Y_{K_2}}, Y_{K_2-1}^{t-1}\right) \\
&\quad + \textstyle\sum_{t=1}^{n} I\left(Y_{K_2-1}^{t-1}, Y_{K_2,t+1}^n; Y_{K_2,t} \middle| \mathbb{M}_{Y_{K_2}}\right) + \sum_{t=1}^{n} I\left(\mathbb{M}_{Y_{K_2}}; Y_{K_2,t}\right) \\
&\quad - \textstyle\sum_{t=1}^{n} I\left(Y_{K_2,t+1}^n; Y_{K_2-1,t} \middle| \mathbb{M}_{Y_{K_2-1}}, \mathbb{M}_{Y_{K_2}}, Y_{K_2-1}^{t-1}\right) - \sum_{t=1}^{n} I\left(Y_{K_2-1}^{t-1}; Y_{K_2,t} \middle| \mathbb{M}_{Y_{K_2}}, Y_{K_2,t+1}^n\right) \\
&\overset{(a)}{=} \textstyle\sum_{t=1}^{n} I\left(\mathbb{M}_{Y_{K_2-1}}; Y_{K_2-1,t} \middle| \mathbb{M}_{Y_{K_2}}, Y_{K_2-1}^{t-1}, Y_{K_2,t+1}^n\right) + \sum_{t=1}^{n} I\left(Y_{K_2-1}^{t-1}, Y_{K_2,t+1}^n; Y_{K_2,t} \middle| \mathbb{M}_{Y_{K_2}}\right) \\
&\quad + \textstyle\sum_{t=1}^{n} I\left(\mathbb{M}_{Y_{K_2}}; Y_{K_2,t}\right) - \sum_{t=1}^{n} I\left(Y_{K_2,t+1}^n; Y_{K_2-1,t} \middle| \mathbb{M}_{Y_{K_2-1}}, \mathbb{M}_{Y_{K_2}}, Y_{K_2-1}^{t-1}\right)
\end{aligned} \tag{82}$$

where the equality (a) holds because, according to the Csiszar-Korner identity, the $2^{nd}$ and the $6^{th}$ ensembles in the left hand side of (a) are equal. Now consider the second ensemble in right side of (a) in (82). We claim that:

$$\sum_{t=1}^{n} I\left(Y_{K_2-1}^{t-1}, Y_{K_2,t+1}^n; Y_{K_2,t} \middle| \mathbb{M}_{Y_{K_2}}\right) \leq \sum_{t=1}^{n} I\left(Y_{K_2-1}^{t-1}, Y_{K_2,t+1}^n; Y_{K_2-1,t} \middle| \mathbb{M}_{Y_{K_2}}\right) \tag{83}$$

To verify the inequality (83), first note that we have:

$$\begin{cases} \sum_{t=1}^{n} I\left(Y_{K_2-1}^{t-1}, Y_{K_2,t+1}^{n}; Y_{K_2,t} \middle| \mathbb{M}_{Y_{K_2}}\right) = \sum_{t=1}^{n} I\left(Y_{K_2-1}^{t-1}, Y_{K_2,t+1}^{n}; Y_{K_2,t} \middle| \mathbb{X}_{\mathbb{M}_{Y_{K_2}},t}, \mathbb{M}_{Y_{K_2}}\right) \\ \sum_{t=1}^{n} I\left(Y_{K_2-1}^{t-1}, Y_{K_2,t+1}^{n}; Y_{K_2-1,t} \middle| \mathbb{M}_{Y_{K_2}}\right) = \sum_{t=1}^{n} I\left(Y_{K_2-1}^{t-1}, Y_{K_2,t+1}^{n}; Y_{K_2-1,t} \middle| \mathbb{X}_{\mathbb{M}_{Y_{K_2}},t}, \mathbb{M}_{Y_{K_2}}\right) \end{cases} \tag{84}$$

Considering (84), the inequality (83) is derived from the condition (77) for $j = K_2, U \equiv \left(Y_{K_2-1}^{t-1}, Y_{K_2,t+1}^{n}\right)$ and $D \equiv \mathbb{M}_{Y_{K_2}}$. Now, by substituting (83) into (82), we obtain:

$$\begin{aligned} & I\left(\mathbb{M}_{Y_{K_2-1}}; Y_{K_2-1}^{n} \middle| \mathbb{M}_{Y_{K_2}}\right) + I\left(\mathbb{M}_{Y_{K_2}}; Y_{K_2}^{n}\right) \\ & \le \sum_{t=1}^{n} I\left(\mathbb{M}_{Y_{K_2-1}}; Y_{K_2-1,t} \middle| \mathbb{M}_{Y_{K_2}}, Y_{K_2-1}^{t-1}, Y_{K_2,t+1}^{n}\right) + \sum_{t=1}^{n} I\left(Y_{K_2-1}^{t-1}, Y_{K_2,t+1}^{n}; Y_{K_2-1,t} \middle| \mathbb{M}_{Y_{K_2}}\right) \\ & \quad + \sum_{t=1}^{n} I\left(\mathbb{M}_{Y_{K_2}}; Y_{K_2,t}\right) - \sum_{t=1}^{n} I\left(Y_{K_2,t+1}^{n}; Y_{K_2-1,t} \middle| \mathbb{M}_{Y_{K_2-1}}, \mathbb{M}_{Y_{K_2}}, Y_{K_2-1}^{t-1}\right) \\ & = \sum_{t=1}^{n} I\left(\mathbb{M}_{Y_{K_2-1}}, Y_{K_2-1}^{t-1}, Y_{K_2,t+1}^{n}; Y_{K_2-1,t} \middle| \mathbb{M}_{Y_{K_2}}\right) + \sum_{t=1}^{n} I\left(\mathbb{M}_{Y_{K_2}}; Y_{K_2,t}\right) - \sum_{t=1}^{n} I\left(Y_{K_2,t+1}^{n}; Y_{K_2-1,t} \middle| \mathbb{M}_{Y_{K_2-1}}, \mathbb{M}_{Y_{K_2}}, Y_{K_2-1}^{t-1}\right) \\ & = \sum_{t=1}^{n} I\left(\mathbb{M}_{Y_{K_2-1}}, Y_{K_2-1}^{t-1}; Y_{K_2-1,t} \middle| \mathbb{M}_{Y_{K_2}}\right) + \sum_{t=1}^{n} I\left(\mathbb{M}_{Y_{K_2}}; Y_{K_2,t}\right) \end{aligned} \tag{85}$$

Next, consider the third mutual information from the right side in (80). We have:

$$I\left(\mathbb{M}_{Y_{K_2-2}}; Y_{K_2-2}^{n} \middle| \mathbb{M}_{Y_{K_2-1}}, \mathbb{M}_{Y_{K_2}}\right) = \sum_{t=1}^{n} I\left(\mathbb{M}_{Y_{K_2-2}}; Y_{K_2-2,t} \middle| \mathbb{M}_{Y_{K_2-1}}, \mathbb{M}_{Y_{K_2}}, Y_{K_2-2,t+1}^{n}\right) \tag{86}$$

Note that the style of applying the chain rule in (86) is similar to the second relation in (81); it changes alternately among the mutual information functions in (80). Now consider the ensemble in (86) and the first ensemble in the last equality of (85); we can write:

$$\begin{aligned} & \sum_{t=1}^{n} I\left(\mathbb{M}_{Y_{K_2-2}}; Y_{K_2-2,t} \middle| \mathbb{M}_{Y_{K_2-1}}, \mathbb{M}_{Y_{K_2}}, Y_{K_2-2,t+1}^{n}\right) + \sum_{t=1}^{n} I\left(\mathbb{M}_{Y_{K_2-1}}, Y_{K_2-1}^{t-1}; Y_{K_2-1,t} \middle| \mathbb{M}_{Y_{K_2}}\right) \\ & = \sum_{t=1}^{n} I\left(\mathbb{M}_{Y_{K_2-2}}, Y_{K_2-1}^{t-1}; Y_{K_2-2,t} \middle| \mathbb{M}_{Y_{K_2-1}}, \mathbb{M}_{Y_{K_2}}, Y_{K_2-2,t+1}^{n}\right) + \sum_{t=1}^{n} I\left(\mathbb{M}_{Y_{K_2-1}}, Y_{K_2-2,t+1}^{n}, Y_{K_2-1}^{t-1}; Y_{K_2-1,t} \middle| \mathbb{M}_{Y_{K_2}}\right) \\ & \quad - \sum_{t=1}^{n} I\left(Y_{K_2-1}^{t-1}; Y_{K_2-2,t} \middle| \mathbb{M}_{Y_{K_2-2}}, \mathbb{M}_{Y_{K_2-1}}, \mathbb{M}_{Y_{K_2}}, Y_{K_2-2,t+1}^{n}\right) - \sum_{t=1}^{n} I\left(Y_{K_2-2,t+1}^{n}; Y_{K_2-1,t} \middle| \mathbb{M}_{Y_{K_2-1}}, \mathbb{M}_{Y_{K_2}}, Y_{K_2-1}^{t-1}\right) \\ & = \sum_{t=1}^{n} I\left(\mathbb{M}_{Y_{K_2-2}}; Y_{K_2-2,t} \middle| \mathbb{M}_{Y_{K_2-1}}, \mathbb{M}_{Y_{K_2}}, Y_{K_2-1}^{t-1}, Y_{K_2-2,t+1}^{n}\right) + \sum_{t=1}^{n} I\left(Y_{K_2-1}^{t-1}; Y_{K_2-2,t} \middle| \mathbb{M}_{Y_{K_2-1}}, \mathbb{M}_{Y_{K_2}}, Y_{K_2-2,t+1}^{n}\right) \\ & \quad + \sum_{t=1}^{n} I\left(Y_{K_2-2,t+1}^{n}, Y_{K_2-1}^{t-1}; Y_{K_2-1,t} \middle| \mathbb{M}_{Y_{K_2-1}}, \mathbb{M}_{Y_{K_2}}\right) + \sum_{t=1}^{n} I\left(\mathbb{M}_{Y_{K_2-1}}; Y_{K_2-1,t} \middle| \mathbb{M}_{Y_{K_2}}\right) \\ & \quad - \sum_{t=1}^{n} I\left(Y_{K_2-1}^{t-1}; Y_{K_2-2,t} \middle| \mathbb{M}_{Y_{K_2-2}}, \mathbb{M}_{Y_{K_2-1}}, \mathbb{M}_{Y_{K_2}}, Y_{K_2-2,t+1}^{n}\right) - \sum_{t=1}^{n} I\left(Y_{K_2-2,t+1}^{n}; Y_{K_2-1,t} \middle| \mathbb{M}_{Y_{K_2-1}}, \mathbb{M}_{Y_{K_2}}, Y_{K_2-1}^{t-1}\right) \\ & \stackrel{(a)}{=} \sum_{t=1}^{n} I\left(\mathbb{M}_{Y_{K_2-2}}; Y_{K_2-2,t} \middle| \mathbb{M}_{Y_{K_2-1}}, \mathbb{M}_{Y_{K_2}}, Y_{K_2-1}^{t-1}, Y_{K_2-2,t+1}^{n}\right) + \sum_{t=1}^{n} I\left(Y_{K_2-2,t+1}^{n}, Y_{K_2-1}^{t-1}; Y_{K_2-1,t} \middle| \mathbb{M}_{Y_{K_2-1}}, \mathbb{M}_{Y_{K_2}}\right) \\ & \quad + \sum_{t=1}^{n} I\left(\mathbb{M}_{Y_{K_2-1}}; Y_{K_2-1,t} \middle| \mathbb{M}_{Y_{K_2}}\right) - \sum_{t=1}^{n} I\left(Y_{K_2-1}^{t-1}; Y_{K_2-2,t} \middle| \mathbb{M}_{Y_{K_2-2}}, \mathbb{M}_{Y_{K_2-1}}, \mathbb{M}_{Y_{K_2}}, Y_{K_2-2,t+1}^{n}\right) \end{aligned} \tag{87}$$

where the equality (a) holds because, according to the Csiszar-Korner identity, the $2^{nd}$ and the $6^{th}$ ensembles in the left hand side of (a) are equal. Now consider the second ensemble in the right side of (a) in (87). We have:

$$\sum_{t=1}^{n} I\left(Y_{K_2-2,t+1}^{n}, Y_{K_2-1}^{t-1}; Y_{K_2-1,t} \middle| \mathbb{M}_{Y_{K_2-1}}, \mathbb{M}_{Y_{K_2}}\right) \le \sum_{t=1}^{n} I\left(Y_{K_2-2,t+1}^{n}, Y_{K_2-1}^{t-1}; Y_{K_2-2,t} \middle| \mathbb{M}_{Y_{K_2-1}}, \mathbb{M}_{Y_{K_2}}\right) \tag{88}$$

To prove this inequality, first note that we have:

$$\begin{cases}\sum_{t=1}^{n} I\left(Y_{K_2-2,t+1}^{n}, Y_{K_2-1}^{t-1}; Y_{K_2-1,t}\middle|\mathbb{M}_{Y_{K_2-1}}, \mathbb{M}_{Y_{K_2}}\right) = \sum_{t=1}^{n} I\left(Y_{K_2-2,t+1}^{n}, Y_{K_2-1}^{t-1}; Y_{K_2-1,t}\middle|\mathbb{X}_{\bigcup_{l=K_2-1}^{K_2}\mathbb{M}_{Y_l},t}, \mathbb{M}_{Y_{K_2-1}}, \mathbb{M}_{Y_{K_2}}\right) \\ \sum_{t=1}^{n} I\left(Y_{K_2-2,t+1}^{n}, Y_{K_2-1}^{t-1}; Y_{K_2-2,t}\middle|\mathbb{M}_{Y_{K_2-1}}, \mathbb{M}_{Y_{K_2}}\right) = \sum_{t=1}^{n} I\left(Y_{K_2-2,t+1}^{n}, Y_{K_2-1}^{t-1}; Y_{K_2-2,t}\middle|\mathbb{X}_{\bigcup_{l=K_2-1}^{K_2}\mathbb{M}_{Y_l},t}, \mathbb{M}_{Y_{K_2-1}}, \mathbb{M}_{Y_{K_2}}\right)\end{cases} \tag{89}$$

By considering these equalities, (88) is derived from the condition (77) for $j = K_2 - 1, U \equiv \left(Y_{K_2-2,t+1}^{n}, Y_{K_2-1}^{t-1}\right)$ and $D \equiv \left(\mathbb{M}_{Y_{K_2-1}}, \mathbb{M}_{Y_{K_2}}\right)$. By substituting (88) in (87), we get:

$$\begin{aligned}
&\sum_{t=1}^{n} I\left(\mathbb{M}_{Y_{K_2-2}}; Y_{K_2-2,t}\middle|\mathbb{M}_{Y_{K_2-1}}, \mathbb{M}_{Y_{K_2}}, Y_{K_2-2,t+1}^{n}\right) + \sum_{t=1}^{n} I\left(\mathbb{M}_{Y_{K_2-1}}, Y_{K_2-1}^{t-1}; Y_{K_2-1,t}\middle|\mathbb{M}_{Y_{K_2}}\right) \\
&\quad\le \sum_{t=1}^{n} I\left(\mathbb{M}_{Y_{K_2-2}}; Y_{K_2-2,t}\middle|\mathbb{M}_{Y_{K_2-1}}, \mathbb{M}_{Y_{K_2}}, Y_{K_2-1}^{t-1}, Y_{K_2-2,t+1}^{n}\right) + \sum_{t=1}^{n} I\left(Y_{K_2-2,t+1}^{n}, Y_{K_2-1}^{t-1}; Y_{K_2-2,t}\middle|\mathbb{M}_{Y_{K_2-1}}, \mathbb{M}_{Y_{K_2}}\right) \\
&\qquad + \sum_{t=1}^{n} I\left(\mathbb{M}_{Y_{K_2-1}}; Y_{K_2-1,t}\middle|\mathbb{M}_{Y_{K_2}}\right) - \sum_{t=1}^{n} I\left(Y_{K_2-1}^{t-1}; Y_{K_2-2,t}\middle|\mathbb{M}_{Y_{K_2-2}}, \mathbb{M}_{Y_{K_2-1}}, \mathbb{M}_{Y_{K_2}}, Y_{K_2-2,t+1}^{n}\right) \\
&\quad= \sum_{t=1}^{n} I\left(\mathbb{M}_{Y_{K_2-2}}, Y_{K_2-1}^{t-1}, Y_{K_2-2,t+1}^{n}; Y_{K_2-2,t}\middle|\mathbb{M}_{Y_{K_2-1}}, \mathbb{M}_{Y_{K_2}}\right) + \sum_{t=1}^{n} I\left(\mathbb{M}_{Y_{K_2-1}}; Y_{K_2-1,t}\middle|\mathbb{M}_{Y_{K_2}}\right) \\
&\qquad - \sum_{t=1}^{n} I\left(Y_{K_2-1}^{t-1}; Y_{K_2-2,t}\middle|\mathbb{M}_{Y_{K_2-2}}, \mathbb{M}_{Y_{K_2-1}}, \mathbb{M}_{Y_{K_2}}, Y_{K_2-2,t+1}^{n}\right) \\
&\quad= \sum_{t=1}^{n} I\left(\mathbb{M}_{Y_{K_2-2}}, Y_{K_2-2,t+1}^{n}; Y_{K_2-2,t}\middle|\mathbb{M}_{Y_{K_2-1}}, \mathbb{M}_{Y_{K_2}}\right) + \sum_{t=1}^{n} I\left(\mathbb{M}_{Y_{K_2-1}}; Y_{K_2-1,t}\middle|\mathbb{M}_{Y_{K_2}}\right)
\end{aligned} \tag{90}$$

Therefore, by combining (85), (86) and (90), we have:

$$\begin{aligned}
&I\left(\mathbb{M}_{Y_{K_2-2}}; Y_{K_2-2}^{n}\middle|\mathbb{M}_{Y_{K_2-1}}, \mathbb{M}_{Y_{K_2}}\right) + I\left(\mathbb{M}_{Y_{K_2-1}}; Y_{K_2-1}^{n}\middle|\mathbb{M}_{Y_{K_2}}\right) + I\left(\mathbb{M}_{Y_{K_2}}; Y_{K_2}^{n}\right) \\
&\quad\le \sum_{t=1}^{n} I\left(\mathbb{M}_{Y_{K_2-2}}, Y_{K_2-2,t+1}^{n}; Y_{K_2-2,t}\middle|\mathbb{M}_{Y_{K_2-1}}, \mathbb{M}_{Y_{K_2}}\right) + \sum_{t=1}^{n} I\left(\mathbb{M}_{Y_{K_2-1}}; Y_{K_2-1,t}\middle|\mathbb{M}_{Y_{K_2}}\right) + \sum_{t=1}^{n} I\left(\mathbb{M}_{Y_{K_2}}; Y_{K_2,t}\right)
\end{aligned} \tag{91}$$

This procedure can be followed sequentially to manipulate other mutual information functions in (80). At last, we derive:

$$I\left(\mathbb{M}_{Y_1}; Y_1^{n}\middle|\mathbb{M}_{Y_2}, \dots, \mathbb{M}_{Y_{K_2-1}}, \mathbb{M}_{Y_{K_2}}\right) + \cdots + I\left(\mathbb{M}_{Y_{K_2-1}}; Y_{K_2-1}^{n}\middle|\mathbb{M}_{Y_{K_2}}\right) + I\left(\mathbb{M}_{Y_{K_2}}; Y_{K_2}^{n}\right) \le \Xi \tag{92}$$

where:

- ✓ If $K_2$ is even, $\Xi$ is given by:

$$\Xi = \sum_{t=1}^{n} I\left(\mathbb{M}_{Y_1}, Y_1^{t-1}; Y_{1,t}\middle|\mathbb{M}_{Y_2}, \dots, \mathbb{M}_{Y_{K_2-1}}, \mathbb{M}_{Y_{K_2}}\right) + \cdots + \sum_{t=1}^{n} I\left(\mathbb{M}_{Y_{K_2-1}}; Y_{K_2-1,t}\middle|\mathbb{M}_{Y_{K_2}}\right) + \sum_{t=1}^{n} I\left(\mathbb{M}_{Y_{K_2}}; Y_{K_2,t}\right) \tag{93}$$

- ✓ If $K_2$ is odd, $\Xi$ is given by:

$$\Xi = \sum_{t=1}^{n} I\left(\mathbb{M}_{Y_1}, Y_{1,t+1}^{n}; Y_{1,t}\middle|\mathbb{M}_{Y_2}, \dots, \mathbb{M}_{Y_{K_2-1}}, \mathbb{M}_{Y_{K_2}}\right) + \cdots + \sum_{t=1}^{n} I\left(\mathbb{M}_{Y_{K_2-1}}; Y_{K_2-1,t}\middle|\mathbb{M}_{Y_{K_2}}\right) + \sum_{t=1}^{n} I\left(\mathbb{M}_{Y_{K_2}}; Y_{K_2,t}\right) \tag{94}$$

The expressions (93) and (94) both are actually identical and equal to:

$$\Xi = \sum_{t=1}^{n} I\left(\mathbb{M}_{Y_1}; Y_{1,t}\middle|\mathbb{M}_{Y_2}, \dots, \mathbb{M}_{Y_{K_2-1}}, \mathbb{M}_{Y_{K_2}}\right) + \cdots + \sum_{t=1}^{n} I\left(\mathbb{M}_{Y_{K_2-1}}; Y_{K_2-1,t}\middle|\mathbb{M}_{Y_{K_2}}\right) + \sum_{t=1}^{n} I\left(\mathbb{M}_{Y_{K_2}}; Y_{K_2,t}\right) \tag{95}$$

The reason is that:

$$\sum_{t=1}^{n} I\left(Y_1^{t-1};Y_{1,t}\middle|\underbrace{\mathbb{M}_{Y_1},\mathbb{M}_{Y_2},\dots,\mathbb{M}_{Y_{K_2-1}},\mathbb{M}_{Y_{K_2}}}_{\mathbb{M}}\right)=\sum_{t=1}^{n} I\left(Y_{1,t+1}^{n};Y_{1,t}\middle|\underbrace{\mathbb{M}_{Y_1},\mathbb{M}_{Y_2},\dots,\mathbb{M}_{Y_{K_2-1}},\mathbb{M}_{Y_{K_2}}}_{\mathbb{M}}\right)=0 \quad (96)$$

Thus, by applying a standard time-sharing argument, we derive the desired outer bound as given in (75). ■

Consider the degraded network for which $\mathbb{X}\to Y_1\to Y_2\to\cdots\to Y_{K_2}$ form a Markov chain. For this network all the less-noisy conditions in (74) are satisfied. Thus, Theorem 5 provides an alternative proof for our result in Part II [2, Lemma 2]. However, for degraded networks, the Markov relation $\mathbb{X}\to Y_1\to Y_2\to\cdots\to Y_{K_2}$ enables us to derive a simpler proof as given in Part II [2, Appendix] and there is no need to the complex arguments presented above. As discussed in Section II, for the degraded networks, the bound (75) can be made substantially simpler using the MACCM plan of messages. The key is that the sum-rate capacity for a degraded network with the associated message set $\mathbb{M}$ is identical to that of the network with the message set $\mathbb{M}^*$ where $\mathbb{M}^*$ is given by (12). The remarkable point is that the same conclusion still holds for the less-noisy networks in Theorem 5. Precisely, one can show that a solution to the maximization in (75) is to nullify all the auxiliaries belonging to $\mathbb{M}-\mathbb{M}^*$ and replace everywhere $\mathbb{M}_{Y_j}$ by $\mathbb{M}^*_{Y_j}$, $j=1,2,\dots,K_2$. Specifically, for the MAIN networks introduced in Part II [2] (see also Fig. 3 for the two-receiver case) the messages with respect to each receiver could be replaced by the corresponding input signals. Here, we do not discuss in details this problem.

Let us concentrate on the less-noisy conditions (74) under which the outer bound (75) holds for the sum-rate capacity of the general interference network. The structure of these conditions is as follows:

$$I(U;Y_A|\mathbb{X}_C)\leq I(U;Y_B|\mathbb{X}_C)$$

for all joint PDFs $P_{U,\mathbb{X}-\mathbb{X}_C}\prod_{X_i\in\mathbb{X}_C}P_{X_i}$, where $\mathbb{X}_C$ is a subset of the input signals, i.e., $\mathbb{X}_C\subseteq\mathbb{X}=\{X_1,X_2,\dots,X_{K_1}\}$. According to Lemma 2, such a condition extends to hold for any arbitrary joint PDF $P_{U\mathbb{X}}(u,x_1,\dots,x_{K_1})$. Such characterization of the required conditions for establishing the outer bound (75) is useful when treating large network topologies due to its compactness. However, for specific scenarios, one can establish the same outer bound under conditions even weaker than those in (74). Let provide an example. Consider the K-user CIC where K transmitters send separately independent messages to their respective receivers. Fig. 5 depicts the channel model.

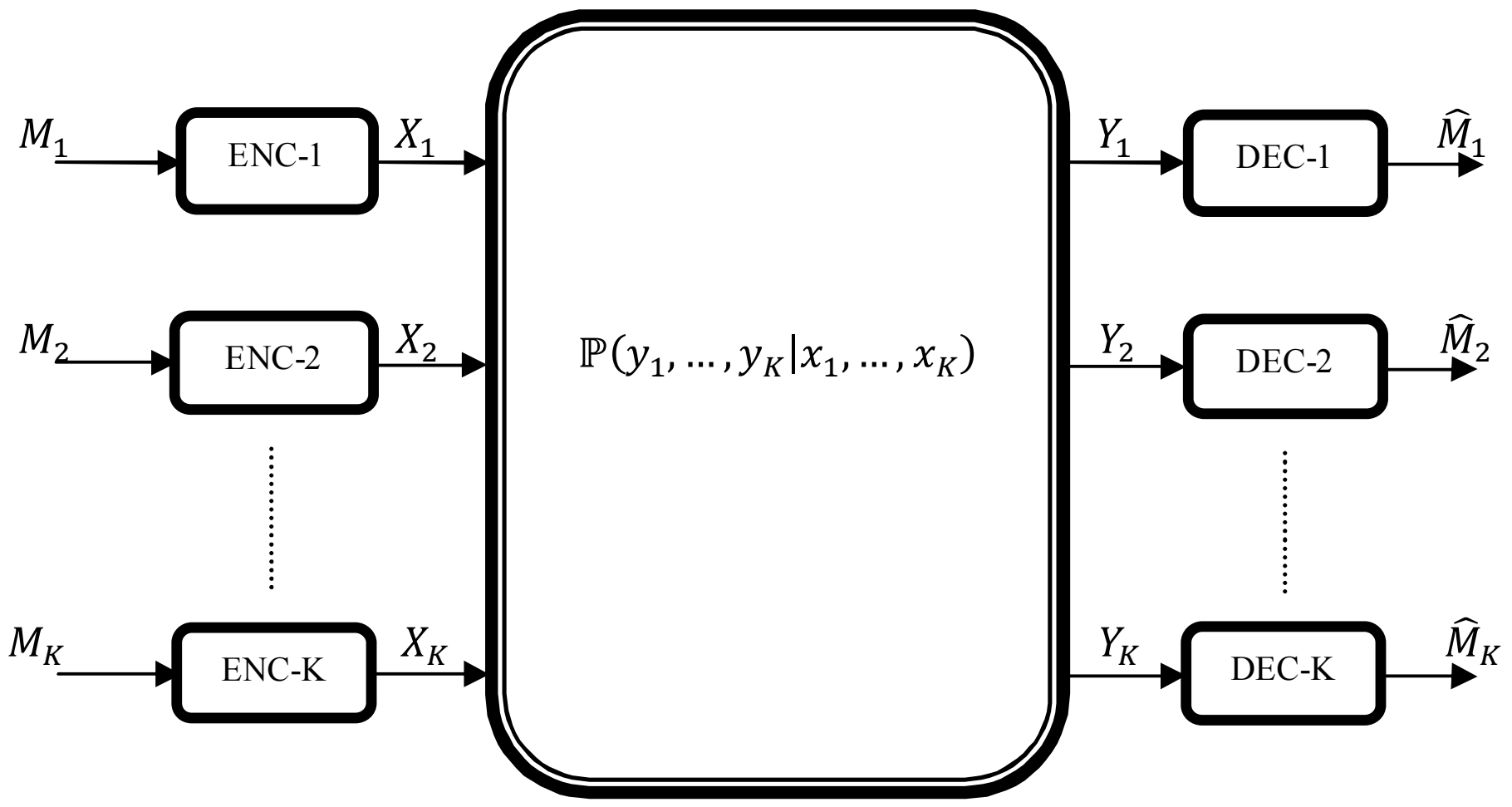

Figure 5. The K-user Classical Interference Channel (CIC).

This channel is derived from the general interference network given in Fig. 1 by setting $\mathbb{M}=\{M_1,M_2,\dots,M_K\}$, $K_1=K_2=K$, $\mathbb{M}_{X_i}=\{M_i\}$, $i=1,\dots,K$, and $\mathbb{M}_{Y_j}=\{M_j\}$, $j=1,\dots,K$.

Theorem 5 implies that if the following less-noisy conditions hold:

$$I(U;Y_j|X_j,X_{j+1},\dots,X_K) \leq I(U;Y_{j-1}|X_j,X_{j+1},\dots,X_K), \qquad \text{for all joint PDFs} \quad P_{UX_1\dots X_{j-1}}P_{X_j}P_{X_{j+1}}\dots P_{X_K}, \quad j=2,\dots,K, \tag{97}$$

then the sum-rate capacity is bounded above by:

$$\mathcal{C}_{sum}^{CIC:K-user} \leq \max_{P_QP_{X_1|Q}P_{X_2|Q}\dots P_{X_K|Q}} \left(I(X_1;Y_1|X_2,X_3,\dots,X_K,Q)+I(X_2;Y_2|X_3,X_4,\dots,X_K,Q)+\dots+I(X_K;Y_K|Q)\right) \tag{98}$$

In the next theorem, we prove that this outer bound still holds under conditions weaker than those in (97).

***Theorem 6)*** *Consider the K-user CIC shown in Fig. 5. Assume that the network transition probability function satisfies the following conditions:*

$$I(U;Y_j|X_j,X_{j+1},\dots,X_K) \leq I(U;Y_{j-1}|X_j,X_{j+1},\dots,X_K), \qquad \textit{for all joint PDFs} \quad P_{X_1}P_{X_2}\dots P_{X_K}P_{U|X_1X_2\dots X_K}, \quad j=2,\dots,K \tag{99}$$

*The sum-rate capacity is bounded as (98).*

*Proof of Theorem 6)* Let us again consider the proof of Theorem 5 specialized for the K-user CIC. One can easily verify that, to derive the desired outer bound in (98), the following inequalities should hold:

$$\sum_{t=1}^{n} I(Y_{K_2-2,t+1}^{n},Y_j^{t-1};Y_{j,t}|M_j,\dots,M_K,X_{j,t},\dots,X_{K,t}) \leq \sum_{t=1}^{n} I(U_t;Y_{j-1,t}|M_j,\dots,M_K,X_{j,t},\dots,X_{K,t}), \qquad j=2,\dots,K \tag{100}$$

where $U_t$ is either $(Y_{j-1,t+1}^{n},Y_j^{t-1})$ or $(Y_{j-1}^{t},Y_{j,t+1}^{n})$. By incorporating the time-sharing parameter $Q$, we deduce that if the following inequalities are satisfied:

$$I(U;Y_j|M_j,\dots,M_K,X_j,\dots,X_K,Q) \leq I(U;Y_{j-1}|M_j,\dots,M_K,X_j,\dots,X_K,Q), \qquad j=2,\dots,K \tag{101}$$

for all joint PDFs given by:

$$P_QP_{M_j}P_{M_{j+1}}\dots P_{M_K}P_{X_1|Q}P_{X_2|Q}\dots P_{X_{j-1}|Q}P_{X_j|M_j,Q}\dots P_{X_K|M_K,Q}P_{U|M_jM_{j+1}\dots M_KX_1X_2\dots X_KQ}, \tag{102}$$

then the ones given in (100) also hold. Now consider the conditions (99). One can show that these conditions imply that:

$$I(U;Y_j|X_j,\dots,X_K,D) \leq I(U;Y_{j-1}|X_j,\dots,X_K,D), \qquad \text{for all joint PDFs} \quad P_DP_{X_1|D}P_{X_2|D}\dots P_{X_K|D}P_{U|X_1X_2\dots X_KD}, \quad j=2,\dots,K \tag{103}$$

In order to derive (101), it is sufficient to set $D \equiv (Q,M_j,\dots,M_K)$ in (103); note that by this substitution the joint PDFs in (103) include those in (102) as a subset. The proof is thus complete. ■

***Remark 4:*** Note that, according to Lemma 2, the conditions (97) extends to hold for all arbitrary joint PDFs $P_{UX_1X_2\dots X_K}$, specifically, those in (99). In other words, (97) imply (99). However, the inverse is not true in general. Therefore, the conditions (99) are weaker than those in (97), i.e., they represent a larger class of CICs.

The outer bound (75) for the sum-rate capacity of the general interference network has the remarkable characteristic that it does not contain any auxiliary random variable other than messages. Similar to the derivations in Section III for two-receiver networks, this characteristic enables us to prove important capacity results. In fact, the outer bound (75) is optimal (it coincides with the sum-rate capacity) for a broad range of network scenarios. Some major classes are identified in the next theorem.

***Theorem 7)*** *Consider the general interference network with $K_1$ transmitters and $K_2$ receivers with the associated message set $\mathbb{M}$ as shown in Fig. 1. Define the sets $\mathbb{M}_{\overset{\mapsto}{Y_j}}, j = 1, \dots, K_2$, as follows:*

$$\mathbb{M}_{\overset{\mapsto}{Y_j}} \triangleq \mathbb{M}_{Y_j} - \left(\mathbb{M}_{Y_{j+1}} \cup \dots \cup \mathbb{M}_{Y_{K_2}}\right), \qquad j = 1, \dots, K_2 \tag{104}$$

*For $j = 2, \dots, K_2$, let $\lambda_j(.)$ be an arbitrary permutation function defined over the set $\{1, \dots, j-1\}$. Also, define:*

$$\mathbb{M}_{\lambda_j}^{\theta} \triangleq \mathbb{M}_{\overset{\mapsto}{Y_j}} \cap \mathbb{M}_{c \nrightarrow Y_{\lambda_j(\theta)}} \cap \mathbb{M}_{c \nrightarrow Y_{\lambda_j(\theta-1)}} \cap \dots \cap \mathbb{M}_{c \nrightarrow Y_{\lambda_j(1)}}, \qquad \theta = 1, \dots, j-1 \tag{105}$$

*Assume that the network transition probability function satisfies the following less-noisy conditions; for $j = 2, \dots, K_2$:*

- ✓ *For $\theta = 1, \dots, j-1$, if $\mathbb{M}_{\lambda_j}^{\theta} \neq \mathbb{M}_{\overset{\mapsto}{Y_j}}$* $\Rightarrow$

$$I\left(U; Y_j \middle| \mathbb{X}_{\cup_{l=j+1}^{K_2} \mathbb{M}_{Y_l} \cup \mathbb{M}_{\lambda_j}^{\theta}}\right) \leq I\left(U; Y_{\lambda_j(\theta)} \middle| \mathbb{X}_{\cup_{l=j+1}^{K_2} \mathbb{M}_{Y_l} \cup \mathbb{M}_{\lambda_j}^{\theta}}\right),$$
$$\textit{for all joint PDFs} \quad P_{U,\mathbb{X}-\mathbb{X}_{\cup_{l=j+1}^{K_2} \mathbb{M}_{Y_l} \cup \mathbb{M}_{\lambda_j}^{\theta}}} \prod_{X_i \in \mathbb{X}_{\cup_{l=j+1}^{K_2} \mathbb{M}_{Y_l} \cup \mathbb{M}_{\lambda_j}^{\theta}}} P_{X_i}. \tag{106}$$

- ✓ *If $\mathbb{M}_{\lambda_j}^{\lambda_j^{-1}(j-1)} = \mathbb{M}_{\overset{\mapsto}{Y_j}}$* $\Rightarrow$

$$I\left(U; Y_j \middle| \mathbb{X}_{\cup_{l=j}^{K_2} \mathbb{M}_{Y_l}}\right) \leq I\left(U; Y_{j-1} \middle| \mathbb{X}_{\cup_{l=j}^{K_2} \mathbb{M}_{Y_l}}\right),$$
$$\textit{for all joint PDFs} \quad P_{U,\mathbb{X}-\mathbb{X}_{\cup_{l=j}^{K_2} \mathbb{M}_{Y_l}}} \prod_{X_i \in \mathbb{X}_{\cup_{l=j}^{K_2} \mathbb{M}_{Y_l}}} P_{X_i}. \tag{107}$$

*The sum-rate capacity is given by:*

$$\max_{\substack{P_Q P_{M_1} \dots P_{M_K}, \\ X_i = f_i(\mathbb{M}_{X_i}, Q), i=1,\dots,K_1}} \left( I\left(\mathbb{M}_{\overset{\mapsto}{Y_1}}; Y_1 \middle| \mathbb{M}_{\overset{\mapsto}{Y_2}}, \dots, \mathbb{M}_{\overset{\mapsto}{Y_{K_2-1}}}, \mathbb{M}_{\overset{\mapsto}{Y_{K_2}}}, Q\right) + \dots + I\left(\mathbb{M}_{\overset{\mapsto}{Y_{K_2-1}}}; Y_{K_2-1} \middle| \mathbb{M}_{\overset{\mapsto}{Y_{K_2}}}, Q\right) + I\left(\mathbb{M}_{\overset{\mapsto}{Y_{K_2}}}; Y_{K_2} \middle| Q\right) \right) \tag{108}$$

*Proof of Theorem 7)* This theorem can be viewed as a generalization of Theorem 3 to the multi-receiver case. For the two-receiver, a detailed proof was given in Theorem 3. Here, we present a more compact proof for the general case. For $j = 2, \dots, K_2$, if $\mathbb{M}_{\lambda_j}^{\lambda_j^{-1}(j-1)} = \mathbb{M}_{\overset{\mapsto}{Y_j}}$, then the corresponding condition of (74) is directly given in (107). Otherwise, consider the conditions (106); according to Lemma 2, these conditions imply that (for $\theta = 1, \dots, j-1$):

$$I\left(U; Y_j \middle| \mathbb{X}_{\cup_{l=j+1}^{K_2} \mathbb{M}_{Y_l} \cup \mathbb{M}_{\lambda_j}^{\theta}}, D\right) \leq I\left(U; Y_{\lambda_j(\theta)} \middle| \mathbb{X}_{\cup_{l=j+1}^{K_2} \mathbb{M}_{Y_l} \cup \mathbb{M}_{\lambda_j}^{\theta}}, D\right), \quad \textit{for all joint PDFs} \quad P_{DU\mathbb{X}} \tag{109}$$

By setting $\theta = \lambda_j^{-1}(j-1)$ in (109), we derive:

$$I\left(U; Y_j \middle| \mathbb{X}_{\cup_{l=j+1}^{K_2} \mathbb{M}_{Y_l} \cup \mathbb{M}_{\lambda_j}^{\lambda_j^{-1}(j-1)}}, D\right) \leq I\left(U; Y_{j-1} \middle| \mathbb{X}_{\cup_{l=j+1}^{K_2} \mathbb{M}_{Y_l} \cup \mathbb{M}_{\lambda_j}^{\lambda_j^{-1}(j-1)}}, D\right), \quad \textit{for all joint PDFs} \quad P_{DU\mathbb{X}}, \quad j = 2, \dots, K_2 \tag{110}$$

According to the Definitions (105), the set $\mathbb{M}_{\lambda_j}^{\lambda_j^{-1}(j-1)}$ is a subset of $\mathbb{M}_{Y_j}^{\mapsto}$, and thereby, $\bigcup_{l=j+1}^{K_2}\mathbb{M}_{Y_l}\cup\mathbb{M}_{\lambda_j}^{\lambda_j^{-1}(j-1)}$ is also a subset of $\bigcup_{l=j}^{K_2}\mathbb{M}_{Y_l}$. Thus, $\mathbb{X}_{\bigcup_{l=j+1}^{K_2}\mathbb{M}_{Y_l}\cup\mathbb{M}_{\lambda_j}^{\lambda_j^{-1}(j-1)}}$ is a subset of $\mathbb{X}_{\bigcup_{l=j}^{K_2}\mathbb{M}_{Y_l}}$. Now by substituting $D\equiv\mathbb{X}_{\bigcup_{l=j}^{K_2}\mathbb{M}_{Y_l}}-\mathbb{X}_{\bigcup_{l=j+1}^{K_2}\mathbb{M}_{Y_l}\cup\mathbb{M}_{\lambda_j}^{\lambda_j^{-1}(j-1)}}$ in (110), we derive the condition (74). Therefore, the conditions of Theorem 5 are satisfied and (108) is an outer bound on the sum-rate capacity (note that (108) is identical to (75)). It remains to prove the achievability of this bound. Let us describe the achievability scheme. At the transmitters, all the messages are encoded using independent codewords. The decoding procedure is as follows. First, note that the definitions (105) imply that:

$$\mathbb{M}_{\lambda_j}^{j-1}\subseteq\mathbb{M}_{\lambda_j}^{j-2}\subseteq\cdots\subseteq\mathbb{M}_{\lambda_j}^{1}\subseteq\mathbb{M}_{Y_j}^{\mapsto},\qquad j=2,\ldots,K_2$$

(111)

Moreover, the messages $\mathbb{M}_{\lambda_j}^{\theta},\theta=1,\ldots,j-1$, are unconnected to the receiver $Y_{\lambda_j(\theta)}$, and therefore they are statistically independent of this receiver.

*Decoding the messages $\mathbb{M}_{Y_{K_2}}^{\mapsto}$ at the receivers $Y_1,Y_2,\ldots,Y_{K_2}$:*

- ✓ At the receiver $Y_{K_2}$, the messages $\mathbb{M}_{\lambda_{K_2}}^{K_2-1}$ are successively decoded first, then the messages $\mathbb{M}_{\lambda_{K_2}}^{K_2-2}-\mathbb{M}_{\lambda_{K_2}}^{K_2-1}$, then those in $\mathbb{M}_{\lambda_{K_2}}^{K_2-3}-\mathbb{M}_{\lambda_{K_2}}^{K_2-2}$ and so forth; at last, this receiver successively decodes the messages $\mathbb{M}_{Y_{K_2}}^{\mapsto}-\mathbb{M}_{\lambda_{K_2}}^{1}$. The partial sum-rate due to these steps is given by:

$$I\left(\mathbb{M}_{\lambda_{K_2}}^{K_2-1};Y_{K_2}\right)+I\left(\mathbb{M}_{\lambda_{K_2}}^{K_2-2};Y_{K_2}\middle|\mathbb{M}_{\lambda_{K_2}}^{K_2-1}\right)+I\left(\mathbb{M}_{\lambda_{K_2}}^{K_2-3};Y_{K_2}\middle|\mathbb{M}_{\lambda_{K_2}}^{K_2-2}\right)+\cdots+I\left(\mathbb{M}_{Y_{K_2}}^{\mapsto};Y_{K_2}\middle|\mathbb{M}_{\lambda_{K_2}}^{1}\right)=I\left(\mathbb{M}_{Y_{K_2}}^{\mapsto};Y_{K_2}\right)$$

(112)

- ✓ For $\theta=1,\ldots,K_2-1$, if $\mathbb{M}_{\lambda_{K_2}}^{\theta}\neq\mathbb{M}_{Y_{K_2}}^{\mapsto}$, the receiver $Y_{\lambda_{K_2}(\theta)}$ successively decodes the messages $\mathbb{M}_{\lambda_{K_2}}^{\theta-1}-\mathbb{M}_{\lambda_{K_2}}^{\theta}$ first, then the messages $\mathbb{M}_{\lambda_{K_2}}^{\theta-2}-\mathbb{M}_{\lambda_{K_2}}^{\theta-1}$, and so forth; at last, this receiver decodes the messages $\mathbb{M}_{Y_{K_2}}^{\mapsto}-\mathbb{M}_{\lambda_{K_2}}^{1}$. These steps do not introduce any new rate cost due to the following conditions:

$$I\left(U;Y_{K_2}\middle|\mathbb{X}_{\mathbb{M}_{\lambda_{K_2}}^{\theta}},D\right)\leq I\left(U;Y_{\lambda_{K_2}(\theta)}\middle|\mathbb{X}_{\mathbb{M}_{\lambda_{K_2}}^{\theta}},D\right),\quad \textit{for all joint PDFs}\quad P_{DU\mathbb{X}},\quad \theta=1,\ldots,K_2-1$$

(113)

  Note that there is no need to decode the messages that belonging to $\mathbb{M}_{\lambda_{K_2}}^{\theta}$ at the receiver $Y_{\lambda_{K_2}(\theta)}$ because they are unconnected to this receiver. Also, if $\mathbb{M}_{\lambda_{K_2}}^{\theta}=\mathbb{M}_{Y_{K_2}}^{\mapsto}$, all the messages that belonging to $\mathbb{M}_{Y_{K_2}}^{\mapsto}$ are unconnected to the receiver $Y_{\lambda_{K_2}(\theta)}$; thus, none of them is decoded at this receiver.

*Decoding the messages $\mathbb{M}_{Y_{K_2-1}}^{\mapsto}$ at the receivers $Y_1,Y_2,\ldots,Y_{K_2-1}$:*

First note that for any receiver $Y_l,l=1,\ldots,K_2-1$, each of the messages belonging to $\mathbb{M}_{Y_{K_2}}^{\mapsto}=\mathbb{M}_{Y_{K_2}}$ either has been previously decoded at the receiver or is unconnected to it. Thus, given the messages $\mathbb{M}_{Y_{K_2}}$:

- ✓ At the receiver $Y_{K_2-1}$, the messages $\mathbb{M}_{\lambda_{K_2-1}}^{K_2-2}$ are successively decoded first, then the messages $\mathbb{M}_{\lambda_{K_2-1}}^{K_2-3}-\mathbb{M}_{\lambda_{K_2-1}}^{K_2-2}$, then those in $\mathbb{M}_{\lambda_{K_2-1}}^{K_2-4}-\mathbb{M}_{\lambda_{K_2-1}}^{K_2-3}$, and so forth; at last, the receiver successively decodes the messages $\mathbb{M}_{Y_{K_2-1}}^{\mapsto}-\mathbb{M}_{\lambda_{K_2-1}}^{1}$. The rate cost due to these steps is given by:

$$I\left(\mathbb{M}_{\lambda_{K_2-1}}^{K_2-2};Y_{K_2-1}\middle|\mathbb{M}_{Y_{K_2}}\right)+I\left(\mathbb{M}_{\lambda_{K_2-1}}^{K_2-3};Y_{K_2-1}\middle|\mathbb{M}_{\lambda_{K_2-1}}^{K_2-2},\mathbb{M}_{Y_{K_2}}\right)+\cdots+I\left(\mathbb{M}_{Y_{K_2-1}}^{\mapsto};Y_{K_2-1}\middle|\mathbb{M}_{\lambda_{K_2-1}}^{1},\mathbb{M}_{Y_{K_2}}\right)=I\left(\mathbb{M}_{Y_{K_2-1}}^{\mapsto};Y_{K_2-1}\middle|\mathbb{M}_{Y_{K_2}}\right)$$

(114)

- ✓ For $\theta=1,\dots,K_2-2$, if $\mathbb{M}^{\theta}_{\lambda_{K_2-1}}\neq\mathbb{M}_{\overset{\mapsto}{Y_{K_2-1}}}$, the receiver $Y_{\lambda_{K_2-1}}(\theta)$ successively decodes the messages $\mathbb{M}^{\theta-1}_{\lambda_{K_2-1}}-\mathbb{M}^{\theta}_{\lambda_{K_2-1}}$ first, then the messages $\mathbb{M}^{\theta-2}_{\lambda_{K_2-1}}-\mathbb{M}^{\theta-1}_{\lambda_{K_2-1}}$, and so forth; at last, this receiver decodes the messages $\mathbb{M}_{\overset{\mapsto}{Y_{K_2-1}}}-\mathbb{M}^{1}_{\lambda_{K_2-1}}$. These steps do not introduce any new rate cost due to the following conditions:

$$I\left(U;Y_{K_2-1}\middle|\mathbb{X}_{\mathbb{M}_{Y_{K_2}}\cup\mathbb{M}^{\theta}_{\lambda_{K_2-1}}},D\right)\leq I\left(U;Y_{\lambda_{K_2-1}(\theta)}\middle|\mathbb{X}_{\mathbb{M}_{Y_{K_2}}\cup\mathbb{M}^{\theta}_{\lambda_{K_2-1}}},D\right),\quad \textit{for all joint PDFs}\quad P_{DU\mathbb{X}},\quad \theta=1,\dots,K_2-2 \tag{115}$$

  Also, if $\mathbb{M}^{\theta}_{\lambda_{K_2-1}}=\mathbb{M}_{\overset{\mapsto}{Y_{K_2-1}}}$, all the messages belonging to $\mathbb{M}_{\overset{\mapsto}{Y_{K_2-1}}}$ are unconnected to the receiver $Y_{\lambda_{K_2-1}}(\theta)$; thus, none of them is decoded at this receiver.

*Decoding the messages* $\mathbb{M}_{\overset{\mapsto}{Y_j}}$ $(1<j<K_2-1)$ *at the receivers* $Y_1,Y_2,\dots,Y_j$*:*

For any receiver $Y_l, l=1,\dots,j$, each of the messages belonging to $\mathbb{M}_{\overset{\mapsto}{Y_{j+1}}}\cup\mathbb{M}_{\overset{\mapsto}{Y_{j+2}}}\cup\dots\cup\mathbb{M}_{\overset{\mapsto}{Y_{K_2}}}=\mathbb{M}_{Y_{j+1}}\cup\mathbb{M}_{Y_{j+2}}\cup\dots\cup\mathbb{M}_{Y_{K_2}}$ either has been previously decoded at the receiver or is unconnected to it. Thus, given the messages $\mathbb{M}_{Y_{j+1}}\cup\mathbb{M}_{Y_{j+2}}\cup\dots\cup\mathbb{M}_{Y_{K_2}}$:

- ✓ At the receiver $Y_j$, the messages $\mathbb{M}^{j-1}_{\lambda_j}$ are successively decoded first, then the messages $\mathbb{M}^{j-2}_{\lambda_j}-\mathbb{M}^{j-1}_{\lambda_j}$, then those in $\mathbb{M}^{j-3}_{\lambda_j}-\mathbb{M}^{j-2}_{\lambda_j}$, and so forth; at last, the receiver decodes the messages in $\mathbb{M}_{\overset{\mapsto}{Y_j}}-\mathbb{M}^{1}_{\lambda_j}$. The rate cost due to this step is given by:

$$\begin{pmatrix} I\left(\mathbb{M}^{j-1}_{\lambda_j};Y_j\middle|\mathbb{M}_{Y_{j+1}}\cup\dots\cup\mathbb{M}_{Y_{K_2}}\right)+I\left(\mathbb{M}^{j-2}_{\lambda_j};Y_j\middle|\mathbb{M}^{j-1}_{\lambda_j},\mathbb{M}_{Y_{j+1}}\cup\dots\cup\mathbb{M}_{Y_{K_2}}\right)\\ +\dots+I\left(\mathbb{M}_{\overset{\mapsto}{Y_j}};Y_j\middle|\mathbb{M}^{1}_{\lambda_j},\mathbb{M}_{Y_{j+1}}\cup\dots\cup\mathbb{M}_{Y_{K_2}}\right)\end{pmatrix}=I\left(\mathbb{M}_{\overset{\mapsto}{Y_j}};Y_j\middle|\mathbb{M}_{Y_{j+1}}\cup\dots\cup\mathbb{M}_{Y_{K_2}}\right) \tag{116}$$

- ✓ For $\theta=1,\dots,j-1$, if $\mathbb{M}^{\theta}_{\lambda_j}\neq\mathbb{M}_{\overset{\mapsto}{Y_j}}$, the receiver $Y_{\lambda_j}(\theta)$ successively decodes the messages $\mathbb{M}^{\theta-1}_{\lambda_j}-\mathbb{M}^{\theta}_{\lambda_j}$ first, then the messages $\mathbb{M}^{\theta-2}_{\lambda_j}-\mathbb{M}^{\theta-1}_{\lambda_j}$, and so forth; at last, this receiver decodes the messages $\mathbb{M}_{\overset{\mapsto}{Y_j}}-\mathbb{M}^{1}_{\lambda_j}$. These steps do not introduce any new rate cost due to the following conditions:

$$I\left(U;Y_j\middle|\mathbb{X}_{\bigcup_{l=j+1}^{K_2}\mathbb{M}_{Y_l}\cup\mathbb{M}^{\theta}_{\lambda_j}},D\right)\leq I\left(U;Y_{\lambda_j(\theta)}\middle|\mathbb{X}_{\bigcup_{l=j+1}^{K_2}\mathbb{M}_{Y_l}\cup\mathbb{M}^{\theta}_{\lambda_j}},D\right),\quad \textit{for all joint PDFs}\quad P_{DU\mathbb{X}},\quad \theta=1,\dots,j-1 \tag{117}$$

  Also, if $\mathbb{M}^{\theta}_{\lambda_j}=\mathbb{M}_{\overset{\mapsto}{Y_j}}$, then all the messages belonging to $\mathbb{M}_{\overset{\mapsto}{Y_j}}$ are unconnected to the receiver $Y_{\lambda_j}(\theta)$; thus, none of them is decoded at this receiver.

*Decoding the messages* $\mathbb{M}_{\overset{\mapsto}{Y_1}}$ *at the receiver* $Y_1$*:*

Each of the messages belonging to $\mathbb{M}_{\overset{\mapsto}{Y_2}}\cup\mathbb{M}_{\overset{\mapsto}{Y_3}}\cup\dots\cup\mathbb{M}_{\overset{\mapsto}{Y_{K_2}}}=\mathbb{M}_{Y_2}\cup\mathbb{M}_{Y_3}\cup\dots\cup\mathbb{M}_{Y_{K_2}}$ either has been previously decoded at the receiver $Y_1$ or is unconnected to it. Thus, given the messages $\mathbb{M}_{Y_2}\cup\mathbb{M}_{Y_3}\cup\dots\cup\mathbb{M}_{Y_{K_2}}$, the receiver $Y_1$ successively decodes all the messages in $\mathbb{M}_{\overset{\mapsto}{Y_1}}$. The partial sum-rate due to this step is given by:

$$I\left(\mathbb{M}_{\overset{\mapsto}{Y_1}};Y_1\middle|\mathbb{M}_{Y_2}\cup\mathbb{M}_{Y_3}\cup\dots\cup\mathbb{M}_{Y_{K_2}}\right) \tag{118}$$

Now by combing (112), (114), (116) and (118), we derive the achievability of (108). The proof is thus complete. ■

***Corollary 2:*** Let us consider a fully connected network where each transmitter is connected to all the receivers. For such a network, we have: $\mathbb{M}_{c\nrightarrow Y_j} \equiv \emptyset$ for $j = 1, \ldots, K_2$. In this case, the conditions given in Theorem 7 are reduced as follows:

$$I\left(U; Y_j \middle| \mathbb{X}_{\bigcup_{l=j+1}^{K_2} \mathbb{M}_{Y_l}}\right) \leq \min\left\{I\left(U; Y_1 \middle| \mathbb{X}_{\bigcup_{l=j+1}^{K_2} \mathbb{M}_{Y_l}}\right), I\left(U; Y_2 \middle| \mathbb{X}_{\bigcup_{l=j+1}^{K_2} \mathbb{M}_{Y_l}}\right), \ldots, I\left(U; Y_{j-1} \middle| \mathbb{X}_{\bigcup_{l=j+1}^{K_2} \mathbb{M}_{Y_l}}\right)\right\}, \qquad j = 2, \ldots, K_2$$

$$\text{for all joint PDFs} \quad P_{U,\mathbb{X}-\mathbb{X}_{\bigcup_{l=j+1}^{K_2} \mathbb{M}_{Y_l}}} \prod_{X_i \in \mathbb{X}_{\bigcup_{l=j+1}^{K_2} \mathbb{M}_{Y_l}}} P_{X_i} \tag{119}$$

If these less-noisy conditions hold, then the simple successive decoding scheme, where the receiver $Y_j$ decodes all the messages corresponding to the receivers $Y_l$ with $j \leq l$, achieves the sum-rate capacity. The sum-rate capacity is given by (108).

Let us provide an example on our result in Theorem 7. Consider the interference network shown in Fig. 6.

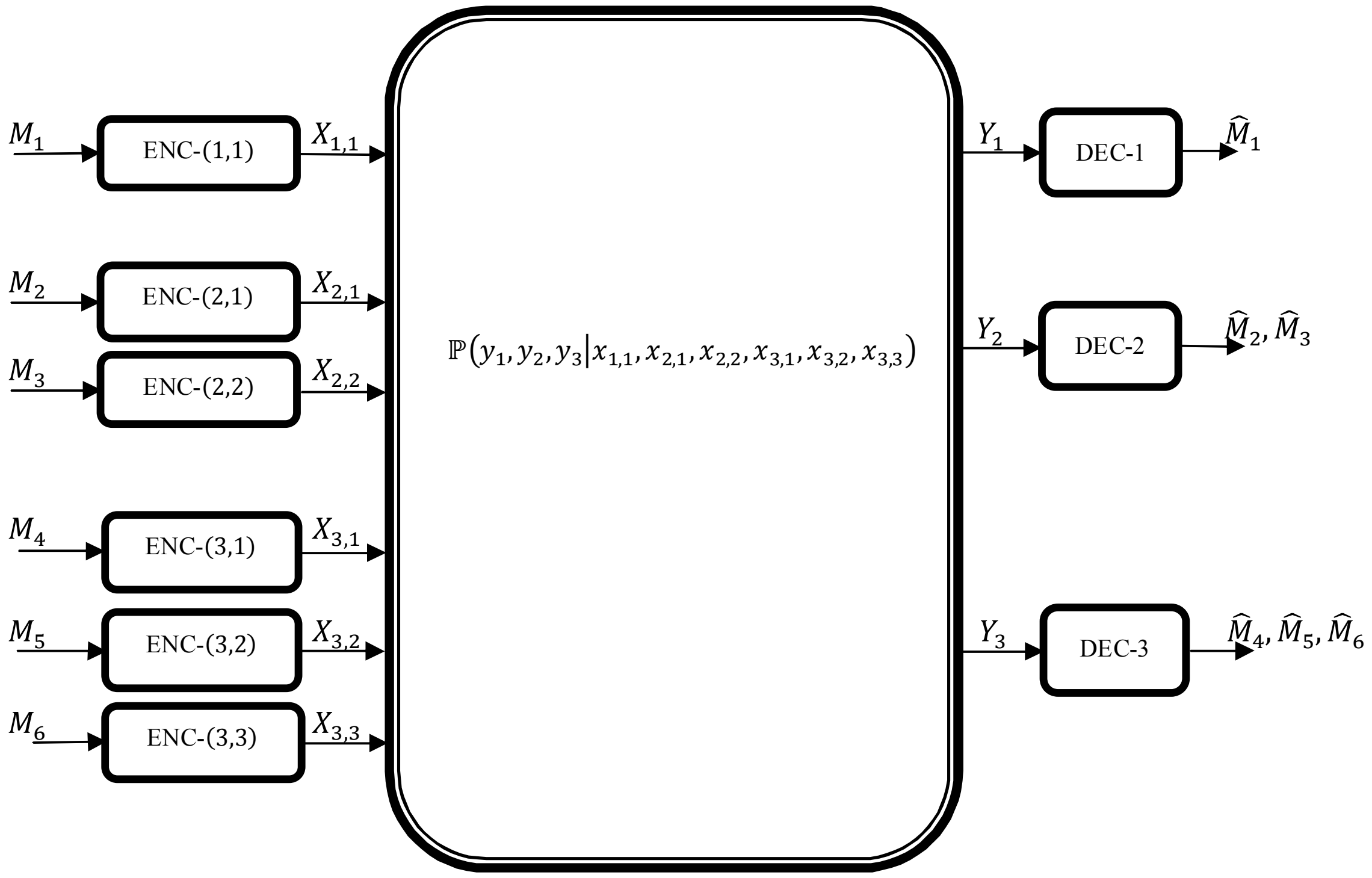

Figure 6. A three-receiver MAIN.

This is a three-receiver MAIN where there exist $i$ transmitters corresponding to the receiver $Y_i, i = 1,2,3$. Each transmitter sends a private message to its respective receiver. Consider the case where the transition probability function of the network is factorized as follows:

$$\mathbb{P}(y_1, y_2, y_3 | x_{1,1}, x_{2,1}, x_{2,2}, x_{3,1}, x_{3,2}, x_{3,3}) = \mathbb{P}(y_1 | x_{1,1}, x_{2,1}, x_{2,2}, x_{3,1}) \mathbb{P}(y_2 | x_{1,1}, x_{2,1}, x_{2,2}, x_{3,1}, x_{3,2}) \mathbb{P}(y_3 | x_{1,1}, x_{2,1}, x_{2,2}, x_{3,1}, x_{3,2}, x_{3,3}) \tag{120}$$

In this scenario, the receiver $Y_1$ is unconnected to the transmitters $X_{3,2}$ and $X_{3,3}$ and the receiver $Y_2$ is unconnected to $X_{3,3}$.

Consider the conditions given in Theorem 7 for the network shown in Fig. 6 with the factorization (120). Let $\lambda_2(1) = 1$ and $\lambda_3(1) = 1$ and $\lambda_3(2) = 2$. Thus, we have:

$$\mathbb{M}_{\lambda_2}^1 = \emptyset, \qquad \mathbb{M}_{\lambda_3}^1 = \{M_5, M_6\}, \qquad \mathbb{M}_{\lambda_3}^2 = \{M_6\} \tag{121}$$

Therefore, if the following conditions hold:

$$I(U;Y_2|X_{3,1},X_{3,2},X_{3,3}) \leq I(U;Y_1|X_{3,1},X_{3,2},X_{3,3}), \quad \text{for all joint PDFs} \quad P_{UX_{1,1}X_{2,1}X_{2,2}}P_{X_{3,1}}P_{X_{3,2}}P_{X_{3,3}}$$
$$I(U;Y_3|X_{3,2},X_{3,3}) \leq I(U;Y_1|X_{3,2},X_{3,3}), \quad \text{for all joint PDFs} \quad P_{UX_{1,1}X_{2,1}X_{2,2}X_{3,1}}P_{X_{3,2}}P_{X_{3,3}}$$
$$I(U;Y_3|X_{3,3}) \leq I(U;Y_2|X_{3,3}), \quad \text{for all joint PDFs} \quad P_{UX_{1,1}X_{2,1}X_{2,2}X_{3,1}X_{3,2}}P_{X_{3,3}},$$
(122)

then the sum-rate capacity is given by:

$$\max_{\substack{P_QP_{X_{1,1}|Q}P_{X_{2,1}|Q}P_{X_{2,2}|Q}\\ P_{X_{3,1}|Q}P_{X_{3,2}|Q}P_{X_{3,3}|Q}}} \left(I(X_{1,1};Y_1|X_{2,1},X_{2,2},X_{3,1},X_{3,2},X_{3,3},Q)+I(X_{2,1},X_{2,2};Y_2|X_{3,1},X_{3,2},X_{3,3},Q)+I(X_{3,1},X_{3,2},X_{3,3};Y_3|Q)\right)$$
(123)

Note that for the MAIN networks, in the expression (108), the messages with respect to each receiver can be replaced by the corresponding input signals. To achieve the sum-rate (123), the receiver $Y_3$ decodes the signal $X_{3,3}$ first, then the signal $X_{3,2}$, and finally the signal $X_{3,1}$; the receiver $Y_2$ decodes $X_{3,2}$ first, then $X_{3,1}$, then $X_{2,2}$, and finally $X_{2,1}$; the receiver $Y_1$ decodes $X_{3,1}$ first, then $X_{2,2}$, then $X_{2,1}$, and finally $X_{1,1}$.

The outer bound (75) derived in Theorem 5, which holds under the conditions (74), can be used to establish exact sum-rate capacity results for many scenarios other than those identified in Theorem 7 by the conditions (106)-(107). In fact, for a given network topology, by comparing the resultant sum-rate of the successive decoding scheme (or the successive-joint decoding scheme) with the outer bound (75), one may impose other appropriate conditions on the network transition probability function so that they coincide. The procedure for this development is similar to the one presented at the end of Section III for two-receiver networks.

## IV.B) Generalizations

In this subsection, we present some interesting generalizations of our unified outer bound derived in Theorem 5 for the sum-rate capacity of the general interference network under the conditions (74). These new outer bounds indeed can be used to derive explicit sum-rate capacity for even broader classes of network scenarios.

Our first result actually is a generalization of the outer bound given by (56) for the two-receiver networks (under the conditions (55)) to the networks with arbitrary number of receivers. To prove this result, we need to a new lemma as given below.

***Lemma 5)*** *Consider the general interference network shown in Fig. 1. Let $Y_A$ and $Y_B$ be two arbitrary receivers with the corresponding message sets $\mathbb{M}_A$ and $\mathbb{M}_B$. Also, let $\mathbb{M}_C$ be an arbitrary subset of messages. Assume that the following condition holds:*

$$I(U,\mathbb{X}_{\mathbb{M}_A\cup\mathbb{M}_B\cup\mathbb{M}_C};Y_A|\mathbb{X}_{\mathbb{M}_B\cup\mathbb{M}_C}) \leq I(U,\mathbb{X}_{\mathbb{M}_A\cup\mathbb{M}_B\cup\mathbb{M}_C};Y_B|\mathbb{X}_{\mathbb{M}_B\cup\mathbb{M}_C})$$
$$\textit{for all joint PDFs } P_{U,\mathbb{X}-\mathbb{X}_{\mathbb{M}_B\cup\mathbb{M}_C}}\prod_{X_i\in\mathbb{X}_{\mathbb{M}_B\cup\mathbb{M}_C}}P_{X_i}$$
(124)

*We recall that $\mathbb{X}$ is the set of all transmitters. Then, given a code of length $n$ for the network, we have:*

$$\boldsymbol{R}_{\Sigma\mathbb{M}_A\cup\mathbb{M}_B-\mathbb{M}_C} \leq \frac{1}{n}I(\mathbb{M}_A;Y_A^n|\mathbb{M}_B,\mathbb{M}_C)+\frac{1}{n}I(\mathbb{M}_B;Y_B^n|\mathbb{M}_C)+\epsilon_n \leq \frac{1}{n}I(\mathbb{M}_A\cup\mathbb{M}_B;Y_B^n|\mathbb{M}_C)+\epsilon_n$$
(125)

*where $\epsilon_n \to 0$ as $n \to \infty$.*

*Proof of Lemma 5)* First note that the condition (124) can be extended to:

$$I(U,\mathbb{X}_{\mathbb{M}_A\cup\mathbb{M}_B\cup\mathbb{M}_C};Y_A|\mathbb{X}_{\mathbb{M}_B\cup\mathbb{M}_C},D) \leq I(U,\mathbb{X}_{\mathbb{M}_A\cup\mathbb{M}_B\cup\mathbb{M}_C};Y_B|\mathbb{X}_{\mathbb{M}_B\cup\mathbb{M}_C},D), \quad \text{for all joint PDFs } P_{DU\mathbb{X}}$$
(126)

This can be proved by following the same arguments as the proof of Lemma 1 in Part III [3, Sec. II]. Now, consider a code of length $n$ for the network. The proof indeed involves very interesting computations including subtle applications of Csiszar-Korner identity. We have:

$$\begin{aligned}\boldsymbol{R}_{\Sigma\mathbb{M}_A\cup\mathbb{M}_B-\mathbb{M}_C} = \boldsymbol{R}_{\Sigma\mathbb{M}_A-\mathbb{M}_B\cup\mathbb{M}_C} + \boldsymbol{R}_{\Sigma\mathbb{M}_B-\mathbb{M}_C} &\overset{(a)}{\leq} \frac{1}{n}I(\mathbb{M}_A - \mathbb{M}_B\cup\mathbb{M}_C; Y_A^n) + \frac{1}{n}I(\mathbb{M}_B - \mathbb{M}_C; Y_B^n) + \epsilon_n \\ &\overset{(b)}{\leq} \frac{1}{n}I(\mathbb{M}_A - \mathbb{M}_B\cup\mathbb{M}_C; Y_A^n|\mathbb{M}_B,\mathbb{M}_C) + \frac{1}{n}I(\mathbb{M}_B - \mathbb{M}_C; Y_B^n|\mathbb{M}_C) + \epsilon_n \\ &= \frac{1}{n}\big(I(\mathbb{M}_A; Y_A^n|\mathbb{M}_B,\mathbb{M}_C) + I(\mathbb{M}_B; Y_B^n|\mathbb{M}_C)\big) + \epsilon_n\end{aligned} \tag{127}$$

where (a) is due to Fano's inequality and the inequality (b) holds because the messages in $\mathbb{M}_A - \mathbb{M}_B\cup\mathbb{M}_C$ are independent of those in $\mathbb{M}_B\cup\mathbb{M}_C$, and the ones in $\mathbb{M}_B - \mathbb{M}_C$ are independent of those in $\mathbb{M}_C$. Next, consider the last expression of (127). We have:

$$\begin{aligned}&I(\mathbb{M}_A; Y_A^n|\mathbb{M}_B,\mathbb{M}_C) + I(\mathbb{M}_B; Y_B^n|\mathbb{M}_C) \\ &\quad= \textstyle\sum_{t=1}^n I(\mathbb{M}_A; Y_{A,t}|\mathbb{M}_B,\mathbb{M}_C,Y_A^{t-1}) + \sum_{t=1}^n I(\mathbb{M}_B; Y_{B,t}|\mathbb{M}_C,Y_{B,t+1}^n) \\ &\quad= \textstyle\sum_{t=1}^n I(\mathbb{M}_A,Y_{B,t+1}^n; Y_{A,t}|\mathbb{M}_B,\mathbb{M}_C,Y_A^{t-1}) + \sum_{t=1}^n I(\mathbb{M}_B,Y_A^{t-1}; Y_{B,t}|\mathbb{M}_C,Y_{B,t+1}^n) \\ &\qquad\qquad \textstyle- \sum_{t=1}^n I(Y_{B,t+1}^n; Y_{A,t}|\mathbb{M}_A,\mathbb{M}_B,\mathbb{M}_C,Y_A^{t-1}) - \sum_{t=1}^n I(Y_A^{t-1}; Y_{B,t}|\mathbb{M}_B,\mathbb{M}_C,Y_{B,t+1}^n) \\ &\quad= \textstyle\sum_{t=1}^n I(\mathbb{M}_A; Y_{A,t}|\mathbb{M}_B,\mathbb{M}_C,Y_{B,t+1}^n,Y_A^{t-1}) + \sum_{t=1}^n I(Y_{B,t+1}^n; Y_{A,t}|\mathbb{M}_B,\mathbb{M}_C,Y_A^{t-1}) + \sum_{t=1}^n I(\mathbb{M}_B,Y_A^{t-1}; Y_{B,t}|\mathbb{M}_C,Y_{B,t+1}^n) \\ &\qquad\qquad \textstyle- \sum_{t=1}^n I(Y_{B,t+1}^n; Y_{A,t}|\mathbb{M}_A,\mathbb{M}_B,\mathbb{M}_C,Y_A^{t-1}) - \sum_{t=1}^n I(Y_A^{t-1}; Y_{B,t}|\mathbb{M}_B,\mathbb{M}_C,Y_{B,t+1}^n) \\ &\quad\overset{(a)}{=} \textstyle\sum_{t=1}^n I(\mathbb{M}_A; Y_{A,t}|\mathbb{M}_B,\mathbb{M}_C,Y_{B,t+1}^n,Y_A^{t-1}) + \sum_{t=1}^n I(\mathbb{M}_B,Y_A^{t-1}; Y_{B,t}|\mathbb{M}_C,Y_{B,t+1}^n) - \sum_{t=1}^n I(Y_{B,t+1}^n; Y_{A,t}|\mathbb{M}_A,\mathbb{M}_B,\mathbb{M}_C,Y_A^{t-1})\end{aligned} \tag{128}$$

where equality (a) holds because, according to the Csiszar-Korner identity, the second and the fifth ensembles on the left side of (a) are equal. Now consider the first ensemble in the last expression of (128). We intend to show that the condition (126) implies the following:

$$\textstyle\sum_{t=1}^n I(\mathbb{M}_A; Y_{A,t}|\mathbb{M}_B,\mathbb{M}_C,Y_{B,t+1}^n,Y_A^{t-1}) \leq \sum_{t=1}^n I(\mathbb{M}_A; Y_{B,t}|\mathbb{M}_B,\mathbb{M}_C,Y_{B,t+1}^n,Y_A^{t-1}) \tag{129}$$

To derive this inequality, first note that we have:

$$\begin{cases}\sum_{t=1}^n I(\mathbb{M}_A; Y_{A,t}|\mathbb{M}_B,\mathbb{M}_C,Y_{B,t+1}^n,Y_A^{t-1}) = \sum_{t=1}^n I(\mathbb{M}_A,\mathbb{X}_{\mathbb{M}_A\cup\mathbb{M}_B\cup\mathbb{M}_C,t}; Y_{A,t}|\mathbb{X}_{\mathbb{M}_B\cup\mathbb{M}_C,t},\mathbb{M}_B,\mathbb{M}_C Y_{B,t+1}^n,Y_A^{t-1}) \\ \sum_{t=1}^n I(\mathbb{M}_A; Y_{B,t}|\mathbb{M}_B,\mathbb{M}_C,Y_{B,t+1}^n,Y_A^{t-1}) = \sum_{t=1}^n I(\mathbb{M}_A,\mathbb{X}_{\mathbb{M}_A\cup\mathbb{M}_B\cup\mathbb{M}_C,t}; Y_{B,t}|\mathbb{X}_{\mathbb{M}_B\cup\mathbb{M}_C,t},\mathbb{M}_B,\mathbb{M}_C,Y_{B,t+1}^n,Y_A^{t-1})\end{cases} \tag{130}$$

Considering (130), the inequality (129) is derived from (126) by setting $U \equiv \mathbb{M}_A$ and $D \equiv (\mathbb{M}_B,\mathbb{M}_C,Y_{B,t+1}^n,Y_A^{t-1})$. Finally, by substituting (129) in (128), we obtain:

$$\begin{aligned}&I(\mathbb{M}_A; Y_A^n|\mathbb{M}_B,\mathbb{M}_C) + I(\mathbb{M}_B; Y_B^n|\mathbb{M}_C) \\ &\quad\leq \textstyle\sum_{t=1}^n I(\mathbb{M}_A; Y_{B,t}|\mathbb{M}_B,\mathbb{M}_C,Y_{B,t+1}^n,Y_A^{t-1}) + \sum_{t=1}^n I(\mathbb{M}_B,Y_A^{t-1}; Y_{B,t}|\mathbb{M}_C,Y_{B,t+1}^n) - \sum_{t=1}^n I(Y_{B,t+1}^n; Y_{A,t}|\mathbb{M}_A,\mathbb{M}_B,\mathbb{M}_C,Y_A^{t-1}) \\ &\quad= \textstyle\sum_{t=1}^n I(\mathbb{M}_A,\mathbb{M}_B,Y_A^{t-1}; Y_{B,t}|\mathbb{M}_C,Y_{B,t+1}^n) - \sum_{t=1}^n I(Y_{B,t+1}^n; Y_{A,t}|\mathbb{M}_A,\mathbb{M}_B,\mathbb{M}_C,Y_A^{t-1}) \\ &\quad= \textstyle\sum_{t=1}^n I(\mathbb{M}_A,\mathbb{M}_B; Y_{B,t}|\mathbb{M}_C,Y_{B,t+1}^n) + \sum_{t=1}^n I(Y_A^{t-1}; Y_{B,t}|\mathbb{M}_A,\mathbb{M}_B,\mathbb{M}_C,Y_{B,t+1}^n) - \sum_{t=1}^n I(Y_{B,t+1}^n; Y_{A,t}|\mathbb{M}_A,\mathbb{M}_B,\mathbb{M}_C,Y_A^{t-1}) \\ &\quad\overset{(a)}{=} \textstyle\sum_{t=1}^n I(\mathbb{M}_A,\mathbb{M}_B; Y_{B,t}|\mathbb{M}_C,Y_{B,t+1}^n) \\ &\quad= I(\mathbb{M}_A\cup\mathbb{M}_B; Y_B^n|\mathbb{M}_C)\end{aligned} \tag{131}$$

where equality (a) is due to the Csiszar-Korner identity. Now (125) is obtained by embedding (131) into (127). The proof is thus complete. ■

***Remark 5:*** Consider the condition (124). If $\mathbb{X}_{\mathbb{M}_A\cup\mathbb{M}_B\cup\mathbb{M}_C}=\mathbb{X}$, where $\mathbb{X}$ is the set of all inputs, then the auxiliary random variable $U$ in (124) is dropped. In other words, it is reduced as follows:

$$I\big(\mathbb{X};Y_A\big|\mathbb{X}_{\mathbb{M}_B\cup\mathbb{M}_C}\big)=I\big(\mathbb{X}-\mathbb{X}_{\mathbb{M}_B\cup\mathbb{M}_C};Y_A\big|\mathbb{X}_{\mathbb{M}_B\cup\mathbb{M}_C}\big)\leq I\big(\mathbb{X}-\mathbb{X}_{\mathbb{M}_B\cup\mathbb{M}_C};Y_B\big|\mathbb{X}_{\mathbb{M}_B\cup\mathbb{M}_C}\big)=I\big(\mathbb{X};Y_B\big|\mathbb{X}_{\mathbb{M}_B\cup\mathbb{M}_C}\big)$$

$$\text{for all joint PDFs } P_{\mathbb{X}-\mathbb{X}_{\mathbb{M}_B\cup\mathbb{M}_C}}\prod\nolimits_{X_i\in\mathbb{X}_{\mathbb{M}_B\cup\mathbb{M}_C}}P_{X_i}\tag{132}$$

This is because $U\to\mathbb{X}\to Y_A,Y_B$ form a Markov chain.

Now, using Lemma 5, we derive the following generalization of Theorem 5.

***Theorem 8)*** *Consider the general interference network with $K_1$ transmitters and $K_2$ receivers with the associated message set $\mathbb{M}$ as shown in Fig. 1. Let $\lambda(.)$ be a permutation of the elements of the set $\{1,\dots,K_2\}$. Let also $j_1,j_2,\dots,j_\mu$ be elements of the set $\{1,\dots,K_2\}$ with:*

$$1\leq j_1<j_2<\cdots<j_{\mu-1}<j_\mu=K_2,\tag{133}$$

*where $\mu$ is an arbitrary natural number less than or equal to $K_2$. Define:*

$$\mathbb{M}_{j_\theta}\triangleq\bigcup_{j_{\theta-1}+1\leq l\leq j_\theta}\mathbb{M}_{Y_{\lambda(l)}},\qquad\theta=1,\dots,\mu,\qquad(j_0\triangleq 0)\tag{134}$$

*Assume that the network transition probability function satisfies the following conditions:*

- ✓ *For* $\omega=1,\dots,j_1-1\Longrightarrow$

$$I\left(\mathbb{X};Y_{\lambda(\omega)}\middle|\mathbb{X}_{\cup_{\omega+1\leq l}\mathbb{M}_{Y_{\lambda(l)}}}\right)\leq I\left(\mathbb{X};Y_{\lambda(\omega+1)}\middle|\mathbb{X}_{\cup_{\omega+1\leq l}\mathbb{M}_{Y_{\lambda(l)}}}\right)$$

$$\textit{for all joint PDFs } P_{\mathbb{X}-\mathbb{X}_{\cup_{\omega+1\leq l}\mathbb{M}_{Y_{\lambda(l)}}}}\prod\nolimits_{X_i\in\mathbb{X}_{\cup_{\omega+1\leq l}\mathbb{M}_{Y_{\lambda(l)}}}}P_{X_i},\tag{135}$$

- ✓ *For* $\theta=1,\dots,\mu-1,\ \omega=1,\dots,j_{\theta+1}-j_\theta-1\Longrightarrow$

$$I\left(U,\mathbb{X}_{\cup_{j_\theta+1\leq l}\mathbb{M}_{Y_{\lambda(l)}}};Y_{\lambda(j_\theta+\omega)}\middle|\mathbb{X}_{\cup_{j_\theta+\omega+1\leq l}\mathbb{M}_{Y_{\lambda(l)}}}\right)\leq I\left(U,\mathbb{X}_{\cup_{j_\theta+1\leq l}\mathbb{M}_{Y_{\lambda(l)}}};Y_{\lambda(j_\theta+\omega+1)}\middle|\mathbb{X}_{\cup_{j_\theta+\omega+1\leq l}\mathbb{M}_{Y_{\lambda(l)}}}\right)$$

$$\textit{for all joint PDFs } P_{U,\mathbb{X}-\mathbb{X}_{\cup_{j_\theta+\omega+1\leq l}\mathbb{M}_{Y_{\lambda(l)}}}}\prod\nolimits_{X_i\in\mathbb{X}_{\cup_{j_\theta+\omega+1\leq l}\mathbb{M}_{Y_{\lambda(l)}}}}P_{X_i},\tag{136}$$

- ✓ *For* $\theta=2,\dots,\mu\Longrightarrow$

$$I\left(U;Y_{\lambda(j_\theta)}\middle|\mathbb{X}_{\cup_{l=\theta}^{\mu}\mathbb{M}_{j_l}}\right)\leq I\left(U;Y_{\lambda(j_{\theta-1})}\middle|\mathbb{X}_{\cup_{l=\theta}^{\mu}\mathbb{M}_{j_l}}\right)$$

$$\textit{for all joint PDFs } P_{U,\mathbb{X}-\mathbb{X}_{\cup_{l=\theta}^{\mu}\mathbb{M}_{j_l}}}\prod\nolimits_{X_i\in\mathbb{X}_{\cup_{l=\theta}^{\mu}\mathbb{M}_{j_l}}}P_{X_i},\tag{137}$$

*Then the sum-rate capacity is bounded-above by:*

$$\mathcal{C}_{sum}^{GIN}\leq\max_{\mathcal{P}^{GIN}}\left(I\left(\mathbb{M}_{j_1};Y_{\lambda(j_1)}\middle|\mathbb{M}_{j_2},\dots,\mathbb{M}_{j_{\mu-1}},\mathbb{M}_{j_\mu},Q\right)+I\left(\mathbb{M}_{j_2};Y_{\lambda(j_2)}\middle|\mathbb{M}_{j_3},\dots,\mathbb{M}_{j_{\mu-1}},\mathbb{M}_{j_\mu},Q\right)+\cdots+I\left(\mathbb{M}_{j_\mu};Y_{\lambda(j_\mu)}\middle|Q\right)\right)\tag{138}$$

where $\mathcal{P}^{GIN}$ *denotes the set of all joint PDFs given below:*

$$P_Q P_{M_1} \dots P_{M_K} \prod_{i=1}^{K_1} P_{X_i|\mathbb{M}_{X_i},Q}, \quad P_{X_i|\mathbb{M}_{X_i},Q} \in \{0,1\},\ i = 1, \dots, K_1 \tag{139}$$

*Proof of Theorem 8)* Consider a code of length $n$ for the network. Similar to (80), one can readily derive:

$$\begin{aligned} n\mathcal{C}_{sum}^{GIN} \leq\ & I\left(\mathbb{M}_{Y_{\lambda(1)}}; Y_{\lambda(1)}^n \middle| \cup_{2\leq l} \mathbb{M}_{Y_{\lambda(l)}}\right) + I\left(\mathbb{M}_{Y_{\lambda(2)}}; Y_{\lambda(2)}^n \middle| \cup_{3\leq l} \mathbb{M}_{Y_{\lambda(l)}}\right) + \dots + I\left(\mathbb{M}_{Y_{\lambda(j_1)}}; Y_{\lambda(j_1)}^n \middle| \cup_{j_1+1\leq l} \mathbb{M}_{Y_{\lambda(l)}}\right) \\ & + I\left(\mathbb{M}_{Y_{\lambda(j_1+1)}}; Y_{\lambda(j_1+1)}^n \middle| \cup_{j_1+2\leq l} \mathbb{M}_{Y_{\lambda(l)}}\right) + \dots + I\left(\mathbb{M}_{Y_{\lambda(j_2)}}; Y_{\lambda(j_2)}^n \middle| \cup_{j_2+1\leq l} \mathbb{M}_{Y_{\lambda(l)}}\right) \\ & + I\left(\mathbb{M}_{Y_{\lambda(j_2+1)}}; Y_{\lambda(j_2+1)}^n \middle| \cup_{j_2+2\leq l} \mathbb{M}_{Y_{\lambda(l)}}\right) + \dots + I\left(\mathbb{M}_{Y_{\lambda(j_3)}}; Y_{\lambda(j_3)}^n \middle| \cup_{j_3+1\leq l} \mathbb{M}_{Y_{\lambda(l)}}\right) \\ & \vdots \\ & + I\left(\mathbb{M}_{Y_{\lambda(j_{\mu-1}+1)}}; Y_{\lambda(j_{\mu-1}+1)}^n \middle| \cup_{j_{\mu-1}+2\leq l} \mathbb{M}_{Y_{\lambda(l)}}\right) + \dots + I\left(\mathbb{M}_{Y_{\lambda(j_\mu)}}; Y_{\lambda(j_\mu)}^n\right) \\ & + n\epsilon_n \end{aligned} \tag{140}$$

where $\epsilon_n \to 0$ as $n \to 0$. Now consider the first row of (140). The conditions (135), according to Lemma 5 (see also Remark 5), imply that:

$$\begin{aligned} & I\left(\mathbb{M}_{Y_{\lambda(1)}}; Y_{\lambda(1)}^n \middle| \cup_{2\leq l} \mathbb{M}_{Y_{\lambda(l)}}\right) + I\left(\mathbb{M}_{Y_{\lambda(2)}}; Y_{\lambda(2)}^n \middle| \cup_{3\leq l} \mathbb{M}_{Y_{\lambda(l)}}\right) + I\left(\mathbb{M}_{Y_{\lambda(3)}}; Y_{\lambda(3)}^n \middle| \cup_{4\leq l} \mathbb{M}_{Y_{\lambda(l)}}\right) + \dots + I\left(\mathbb{M}_{Y_{\lambda(j_1)}}; Y_{\lambda(j_1)}^n \middle| \cup_{j_1+1\leq l} \mathbb{M}_{Y_{\lambda(l)}}\right) \\ & \leq I\left(\mathbb{M}_{Y_{\lambda(1)}}; Y_{\lambda(2)}^n \middle| \cup_{2\leq l} \mathbb{M}_{Y_{\lambda(l)}}\right) + I\left(\mathbb{M}_{Y_{\lambda(2)}}; Y_{\lambda(2)}^n \middle| \cup_{3\leq l} \mathbb{M}_{Y_{\lambda(l)}}\right) \\ & \qquad + I\left(\mathbb{M}_{Y_{\lambda(3)}}; Y_{\lambda(3)}^n \middle| \cup_{4\leq l} \mathbb{M}_{Y_{\lambda(l)}}\right) + \dots + I\left(\mathbb{M}_{Y_{\lambda(j_1)}}; Y_{\lambda(j_1)}^n \middle| \cup_{j_1+1\leq l} \mathbb{M}_{Y_{\lambda(l)}}\right) \\ & = I\left(\mathbb{M}_{Y_{\lambda(1)}}, \mathbb{M}_{Y_{\lambda(2)}}; Y_{\lambda(2)}^n \middle| \cup_{3\leq l} \mathbb{M}_{Y_{\lambda(l)}}\right) + I\left(\mathbb{M}_{Y_{\lambda(3)}}; Y_{\lambda(3)}^n \middle| \cup_{4\leq l} \mathbb{M}_{Y_{\lambda(l)}}\right) + \dots + I\left(\mathbb{M}_{Y_{\lambda(j_1)}}; Y_{\lambda(j_1)}^n \middle| \cup_{j_1+1\leq l} \mathbb{M}_{Y_{\lambda(l)}}\right) \\ & \leq I\left(\mathbb{M}_{Y_{\lambda(1)}}, \mathbb{M}_{Y_{\lambda(2)}}; Y_{\lambda(3)}^n \middle| \cup_{3\leq l} \mathbb{M}_{Y_{\lambda(l)}}\right) + I\left(\mathbb{M}_{Y_{\lambda(3)}}; Y_{\lambda(3)}^n \middle| \cup_{4\leq l} \mathbb{M}_{Y_{\lambda(l)}}\right) + \dots + I\left(\mathbb{M}_{Y_{\lambda(j_1)}}; Y_{\lambda(j_1)}^n \middle| \cup_{j_1+1\leq l} \mathbb{M}_{Y_{\lambda(l)}}\right) \\ & = I\left(\mathbb{M}_{Y_{\lambda(1)}}, \mathbb{M}_{Y_{\lambda(2)}}, \mathbb{M}_{Y_{\lambda(3)}}; Y_{\lambda(3)}^n \middle| \cup_{4\leq l} \mathbb{M}_{Y_{\lambda(l)}}\right) + \dots + I\left(\mathbb{M}_{Y_{\lambda(j_1)}}; Y_{\lambda(j_1)}^n \middle| \cup_{j_1+1\leq l} \mathbb{M}_{Y_{\lambda(l)}}\right) \\ & \vdots \\ & \leq I\left(\mathbb{M}_{Y_{\lambda(1)}}, \mathbb{M}_{Y_{\lambda(2)}}, \dots, \mathbb{M}_{Y_{\lambda(j_1)}}; Y_{\lambda(j_1)}^n \middle| \cup_{j_1+1\leq l} \mathbb{M}_{Y_{\lambda(l)}}\right) = I\left(\mathbb{M}_{j_1}; Y_{\lambda(j_1)}^n \middle| \mathbb{M}_{j_2}, \dots, \mathbb{M}_{j_{\mu-1}}, \mathbb{M}_{j_\mu}\right) \end{aligned} \tag{141}$$

Similarly, for the other rows in (140), according to Lemma 5, the conditions (136) imply that:

$$\begin{aligned} & I\left(\mathbb{M}_{Y_{\lambda(j_1+1)}}; Y_{\lambda(j_1+1)}^n \middle| \cup_{j_1+2\leq l} \mathbb{M}_{Y_{\lambda(l)}}\right) + \dots + I\left(\mathbb{M}_{Y_{\lambda(j_2)}}; Y_{\lambda(j_2)}^n \middle| \cup_{j_2+1\leq l} \mathbb{M}_{Y_{\lambda(l)}}\right) \leq I\left(\mathbb{M}_{j_2}; Y_{\lambda(j_2)}^n \middle| \mathbb{M}_{j_3}, \dots, \mathbb{M}_{j_{\mu-1}}, \mathbb{M}_{j_\mu}\right) \\ & \vdots \\ & I\left(\mathbb{M}_{Y_{\lambda(j_{\mu-1}+1)}}; Y_{\lambda(j_{\mu-1}+1)}^n \middle| \cup_{j_{\mu-1}+2\leq l} \mathbb{M}_{Y_{\lambda(l)}}\right) + \dots + I\left(\mathbb{M}_{Y_{\lambda(j_\mu)}}; Y_{\lambda(j_\mu)}^n\right) \leq I\left(\mathbb{M}_{j_\mu}; Y_{\lambda(j_\mu)}^n\right) \end{aligned} \tag{142}$$

Thus, by substituting (141)-(142) in (140), we obtain:

$$n\mathcal{C}_{sum}^{GIN} \leq I\left(\mathbb{M}_{j_1}; Y_{\lambda(j_1)}^n \middle| \mathbb{M}_{j_2}, \dots, \mathbb{M}_{j_{\mu-1}}, \mathbb{M}_{j_\mu}\right) + \dots + I\left(\mathbb{M}_{j_2}; Y_{\lambda(j_2)}^n \middle| \mathbb{M}_{j_3}, \dots, \mathbb{M}_{j_{\mu-1}}, \mathbb{M}_{j_\mu}\right) + I\left(\mathbb{M}_{j_\mu}; Y_{\lambda(j_\mu)}^n\right) + n\epsilon_n \tag{143}$$

Finally, if the conditions (137) hold, by following the same lines as (80)-(96), one can derive the single-letter outer bound given in (138). The proof is thus complete. ■

Next, we demonstrate how one can derive the sum-rate capacity results for different interference networks using the outer bound established in Theorem 8. Consider the three-user CIC shown in Fig. 7.

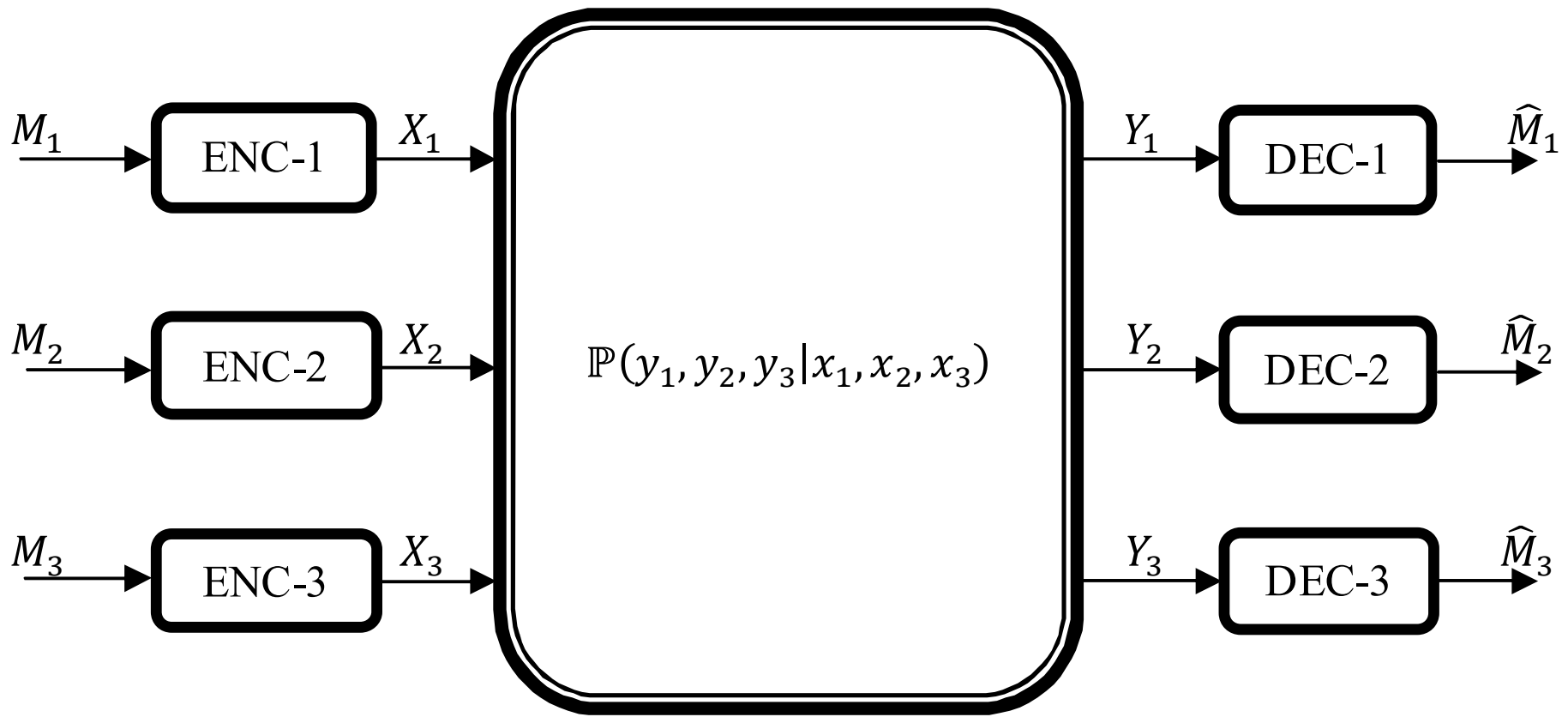

Figure 7. The three-user CIC.

The Gaussian channel is formulated as:

$$\begin{cases} Y_1 = X_1 + a_{12}X_2 + a_{13}X_3 + Z_1 \\ Y_2 = a_{21}X_1 + X_2 + a_{23}X_3 + Z_2 \\ Y_3 = a_{31}X_1 + a_{32}X_2 + X_3 + Z_3 \end{cases} \tag{144}$$

where $Z_1$, $Z_2$, and $Z_3$ are zero-mean unit-variance Gaussian RVs and the inputs are subject to power constraints $\mathbb{E}[X_i^2] \leq P_i, i = 1,2,3$.

First let us present the sum-rate achievable by the successive decoding scheme. It is given below:

$$\max_{P_Q P_{X_1|Q} P_{X_2|Q} P_{X_3|Q}} \min \left\{ \begin{array}{l} I(X_1; Y_1|X_2, X_3, Q) + I(X_2 : Y_2|X_3, Q) + I(X_3; Y_3|Q), \\ I(X_1; Y_1|X_2, X_3, Q) + I(X_2, X_3 : Y_2|Q), \\ I(X_1; Y_1|X_2, X_3, Q) + I(X_2 : Y_2|X_3, Q) + I(X_3; Y_1|Q), \\ I(X_1, X_2; Y_1|X_3, Q) + I(X_3; Y_3|Q), \\ I(X_1, X_2; Y_1|X_3, Q) + I(X_3; Y_2|Q), \\ I(X_1, X_2, X_3; Y_1|Q) \end{array} \right\} \tag{145}$$

To achieve this sum-rate:

- ✓ The receiver $Y_3$ decodes only its corresponding message $M_3$.
- ✓ The receiver $Y_2$ decodes the message $M_3$ first and then decodes its corresponding message $M_2$.
- ✓ The receiver $Y_1$ decodes the message $M_3$ first, then the message $M_2$ and lastly its corresponding message $M_1$.

As we see from (145), the sum-rate expression due to this achievability scheme is described by six constraints. We intend to explore less-noisy conditions so that this achievable sum-rate is optimal for the network. Note that some of the constraints in (145) do not have a structure similar to the expression of the outer bound (138); for example, the third and the fifth ones. Therefore, we need to impose appropriate conditions on the network probability function so that such constraints can be relaxed. Let the network satisfies the following conditions:

$$\begin{cases} I(X_2:Y_2|X_3,Q) \geq I(X_2:Y_1|X_3,Q) \\ I(X_3;Y_2|Q) \geq I(X_3;Y_1|Q) \end{cases}$$

(146)

for all joint PDFs which are a solution to the following maximization:

$$\max_{P_Q P_{X_1|Q} P_{X_2|Q} P_{X_3|Q}} \min \begin{pmatrix} I(X_1,X_2;Y_1|X_3,Q) + I(X_3;Y_3|Q), \\ I(X_1,X_2,X_3;Y_1|Q) \end{pmatrix}$$

(147)

In this case, the achievable sum-rate (145) is reduced to (147). In other words, the conditions (146) enable us to relax those constraints of (145) which are not given in (147). Now consider the outer bound (138) specialized for the three-user CIC. By setting $\lambda(1) = 2$, $\lambda(2) = 1$, $\lambda(3) = 3$, $\mu = 2$, $j_1 = 2$, and $j_2 = 3$, we obtain that the sum-rate capacity is bounded above by:

$$\mathcal{C}_{sum}^{CIC:3-user} \leq \max_{P_Q P_{X_1|Q} P_{X_2|Q} P_{X_3|Q}} \big(I(X_1,X_2;Y_1|X_3,Q) + I(X_3;Y_3|Q)\big)$$

(148)

If the following conditions hold:

$$\begin{cases} I(X_2;Y_2|X_1,X_3) \leq I(X_2;Y_1|X_1,X_3) & \text{for all joint PDFs } P_{X_1}P_{X_2}P_{X_3} \\ I(U;Y_3|X_3) \leq I(U;Y_1|X_3) & \text{for all joint PDFs } P_{UX_1X_2}P_{X_3} \end{cases}$$

(149)

Also, by setting $\lambda(1) = 3, \lambda(2) = 2, \lambda(3) = 1$, $\mu = 1$, and $j_1 = 3$, we obtain:

$$\mathcal{C}_{sum}^{CIC:3-user} \leq \max_{P_Q P_{X_1|Q} P_{X_2|Q} P_{X_3|Q}} \big(I(X_1,X_2,X_3;Y_1|Q)\big)$$

(150)

If the following conditions hold:

$$\begin{cases} I(X_3;Y_3|X_1,X_2) \leq I(X_3;Y_2|X_1,X_2) & \text{for all joint PDFs } P_{X_1}P_{X_2}P_{X_3} \\ I(X_2,X_3;Y_2|X_1) \leq I(X_2,X_3;Y_1|X_1) & \text{for all joint PDFs } P_{X_1}P_{X_2X_3} \end{cases}$$

(151)

Thus, if the collection of the conditions (146), (149) and (151) are satisfied, the sum-rate capacity of the network is given by (147). This capacity result could not be obtained using the outer bound (75) derived in Theorem 5. Let us consider the Gaussian channel given in (144). Lemmas 3 and 4 imply that if the channel gains satisfy:

$$\begin{cases} |a_{12}| \geq 1, & |a_{31}| \leq 1, & |a_{23}| \geq 1 \\ a_{31} = \dfrac{a_{32}}{a_{12}}, & a_{12} = \dfrac{a_{13}}{a_{23}} & \end{cases},$$

(152)

then (149) and (151) are also satisfied. Note that, according to Corollary 1, the second inequality of (151) implies the first inequality of (149). For a Gaussian channel satisfying (152), using the entropy power inequality [1, Lemma II.2], one can prove that Gaussian input distributions without time-sharing ($Q \equiv \emptyset$) is the solution to the maximization (147). Hence, if the inequalities (146) hold for Gaussian distributions, then the sum-rate capacity is given by (147). Considering (152), it is readily derived that both inequalities (146) hold for Gaussian distributions provided that:

$$P_1 + 1 \geq a_{12}^2(a_{21}^2 P_1 + 1)$$

(153)

Thus, for a three-user Gaussian CIC (144) with the conditions (152)-(153), the sum-rate capacity is given as follows:

$$\min\begin{pmatrix}\psi(P_1+a_{12}^2P_2)+\psi\left(\frac{P_3}{a_{31}^2P_1+a_{32}^2P_2+1}\right)\\ \psi(P_1+a_{12}^2P_2+a_{13}^2P_3)\end{pmatrix} \tag{154}$$

which is achieved by the successive decoding scheme.

It should be remarked that by applying more efficient coding strategies such as the successive-joint decoding scheme described in Section III, one can achieve further capacity results for the CICs. Indeed, our approach could be followed to obtain the sum-rate capacity for many other network topologies.

A second generalization of the result of Theorem 5 is given below.

***Theorem 9)*** *Consider the general interference network with $K_1$ transmitters and $K_2$ receivers with the associated message set $\mathbb{M}$ as shown in Fig. 1. Let $\boldsymbol{Y}_{G(1)},\boldsymbol{Y}_{G(2)},\dots,\boldsymbol{Y}_{G(\mu)}$ be nonempty subsets of the set of outputs $\{Y_1,\dots,Y_{K_2}\}$ so that:*

$$\begin{cases}\boldsymbol{Y}_{G(j_1)}\cap\boldsymbol{Y}_{G(j_2)}\neq\emptyset\\ \bigcup_{j=1}^{\mu}\boldsymbol{Y}_{G(j)}=\{Y_1,\dots,Y_{K_2}\}\end{cases}$$

*i.e., the collection $\boldsymbol{Y}_{G(1)},\boldsymbol{Y}_{G(2)},\dots,\boldsymbol{Y}_{G(\mu)}$ constitutes a nonempty partitioning for $\{Y_1,\dots,Y_{K_2}\}$. Define:*

$$\mathbb{M}_{\boldsymbol{Y}_{G(j)}}\triangleq\bigcup_{Y_l\in\boldsymbol{Y}_{G(l)}}\mathbb{M}_{Y_l},\qquad j=1,\dots,\mu \tag{155}$$

*Assume that the network transition probability function satisfies the following less-noisy conditions:*

$$I\left(U;\boldsymbol{Y}_{G(j)}\middle|\mathbb{X}_{\bigcup_{l=j}^{\mu}\mathbb{M}_{\boldsymbol{Y}_{G(l)}}}\right)\leq I\left(U;\boldsymbol{Y}_{G(j-1)}\middle|\mathbb{X}_{\bigcup_{l=j}^{\mu}\mathbb{M}_{\boldsymbol{Y}_{G(l)}}}\right),$$

$$\text{for all joint PDFs}\quad P_{U,\mathbb{X}-\mathbb{X}_{\bigcup_{l=j}^{\mu}\mathbb{M}_{\boldsymbol{Y}_{G(l)}}}}\prod_{X_i\in\mathbb{X}_{\bigcup_{l=j}^{\mu}\mathbb{M}_{\boldsymbol{Y}_{G(l)}}}}P_{X_i},\qquad j=2,\dots,\mu \tag{156}$$

*Then the sum-rate capacity of the network is bounded above as:*

$$\mathcal{C}_{sum}^{GIN}\leq\max_{\mathcal{P}^{GIN}}\left(I\left(\mathbb{M}_{\boldsymbol{Y}_{G(1)}};\boldsymbol{Y}_{G(1)}\middle|\mathbb{M}_{\boldsymbol{Y}_{G(2)}},\dots,\mathbb{M}_{\boldsymbol{Y}_{G(\mu-1)}},\mathbb{M}_{\boldsymbol{Y}_{G(\mu)}},Q\right)+\dots+I\left(\mathbb{M}_{\boldsymbol{Y}_{G(\mu-1)}};\boldsymbol{Y}_{G(\mu-1)}\middle|\mathbb{M}_{\boldsymbol{Y}_{G(\mu)}},Q\right)+I\left(\mathbb{M}_{\boldsymbol{Y}_{G(\mu)}};\boldsymbol{Y}_{G(\mu)}\middle|Q\right)\right) \tag{157}$$

where *$\mathcal{P}^{GIN}$ denotes the set of all joint PDFs given below:*

$$P_QP_{M_1}\dots P_{M_K}\prod_{i=1}^{K_1}P_{X_i|\mathbb{M}_{X_i},Q},\quad P_{X_i|\mathbb{M}_{X_i},Q}\in\{0,1\},\ i=1,\dots,K_1 \tag{158}$$

*Proof of Theorem 9)* First note that according to the definition (155), each of the messages in $\mathbb{M}_{\boldsymbol{Y}_{G(j)}}$ is decoded at least at one of the receivers belonging to $\boldsymbol{Y}_{G(j)}$. Thus, we can apply Fano's inequality to derive:

$$\boldsymbol{R}_{\Sigma\mathbb{M}_{\boldsymbol{Y}_{G(j)}}}\leq I\left(\mathbb{M}_{\boldsymbol{Y}_{G(j)}};\boldsymbol{Y}_{G(j)}^n\right)+n\epsilon_{G(j),n} \tag{159}$$

where $\epsilon_{G(j),n}\to 0$ as $n\to\infty$. Define the sets $\mathbb{M}_{\overset{\mapsto}{\boldsymbol{Y}_{G(j)}}}$, $j=1,\dots,\mu$, as follows:

$$\mathbb{M}_{\overset{\mapsto}{\boldsymbol{Y}_{G(j)}}}\triangleq\mathbb{M}_{\boldsymbol{Y}_{G(j)}}-\left(\mathbb{M}_{\boldsymbol{Y}_{G(j+1)}}\cup\dots\cup\mathbb{M}_{\boldsymbol{Y}_{G(\mu)}}\right),\qquad j=1,\dots,\mu \tag{160}$$

Therefore, we can write:

$$n\left(\sum_{l\in \underline{id}_{\mathbb{M}_{\overset{\mapsto}{Y_{G(j)}}}}} R_l\right) \leq I\left(\mathbb{M}_{\overset{\mapsto}{Y_{G(j)}}};\boldsymbol{Y}_{G(j)}^n\right)+n\epsilon_{G(j),n}$$

$$\overset{(a)}{\leq} I\left(\mathbb{M}_{\overset{\mapsto}{Y_{G(j)}}};\boldsymbol{Y}_{G(j)}^n\middle|\mathbb{M}_{\boldsymbol{Y}_{G(j+1)}}\cup\ldots\cup\mathbb{M}_{\boldsymbol{Y}_{G(\mu)}}\right)+n\epsilon_{G(j),n}$$

$$= I\left(\mathbb{M}_{\boldsymbol{Y}_{G(j)}};\boldsymbol{Y}_{G(j)}^n\middle|\mathbb{M}_{\boldsymbol{Y}_{G(j+1)}}\cup\ldots\cup\mathbb{M}_{\boldsymbol{Y}_{G(\mu)}}\right)+n\epsilon_{G(j),n}$$

(161)

Note that the inequality (a) in (161) holds because the messages that belong to $\mathbb{M}_{\overset{\mapsto}{Y_{G(j)}}}$ are independent of those in $\mathbb{M}_{\boldsymbol{Y}_{G(j+1)}}\cup\ldots\cup\mathbb{M}_{\boldsymbol{Y}_{G(\mu)}}$. Considering (155) and by adding the two sides of (161) for $j=1,\ldots,\mu$, we obtain:

$$n\mathcal{C}_{sum}^{GIN}=n\left(\sum_{j=1}^{\mu}\sum_{l\in \underline{id}_{\mathbb{M}_{\overset{\mapsto}{Y_{G(j)}}}}} R_l\right)$$

$$\leq I\left(\mathbb{M}_{\boldsymbol{Y}_{G(1)}};\boldsymbol{Y}_{G(1)}^n\middle|\mathbb{M}_{\boldsymbol{Y}_{G(2)}},\ldots,\mathbb{M}_{\boldsymbol{Y}_{G(\mu-1)}},\mathbb{M}_{\boldsymbol{Y}_{G(\mu)}}\right)+\cdots+I\left(\mathbb{M}_{\boldsymbol{Y}_{G(\mu-1)}};\boldsymbol{Y}_{G(\mu-1)}^n\middle|\mathbb{M}_{\boldsymbol{Y}_{G(\mu)}}\right)+I\left(\mathbb{M}_{\boldsymbol{Y}_{G(\mu)}};\boldsymbol{Y}_{G(\mu)}^n\right)+n\epsilon_n$$

(162)

where $\epsilon_n\to 0$ as $n\to\infty$. Now, if the conditions (156) hold, by following the same lines as (80)-(96), one can derive the single-letter outer bound given in (157). The proof is thus complete. ■

***Remark 6)*** It is clear that by setting $\mu=K_2$ and $\boldsymbol{Y}_{G(j)}=\{Y_j\}$, $j=1,\ldots,K_2$, the outer bound of Theorem 9 is reduced to the one given in Theorem 5, i.e., (75).

The outer bound given in Theorem 9 may be used to prove explicit sum-rate capacity results which are not necessarily derived from the bounds in Theorem 8. We conclude this subsection by providing an example in this regard. Consider the K-user *many-to-one interference channel* shown in Fig. 8.

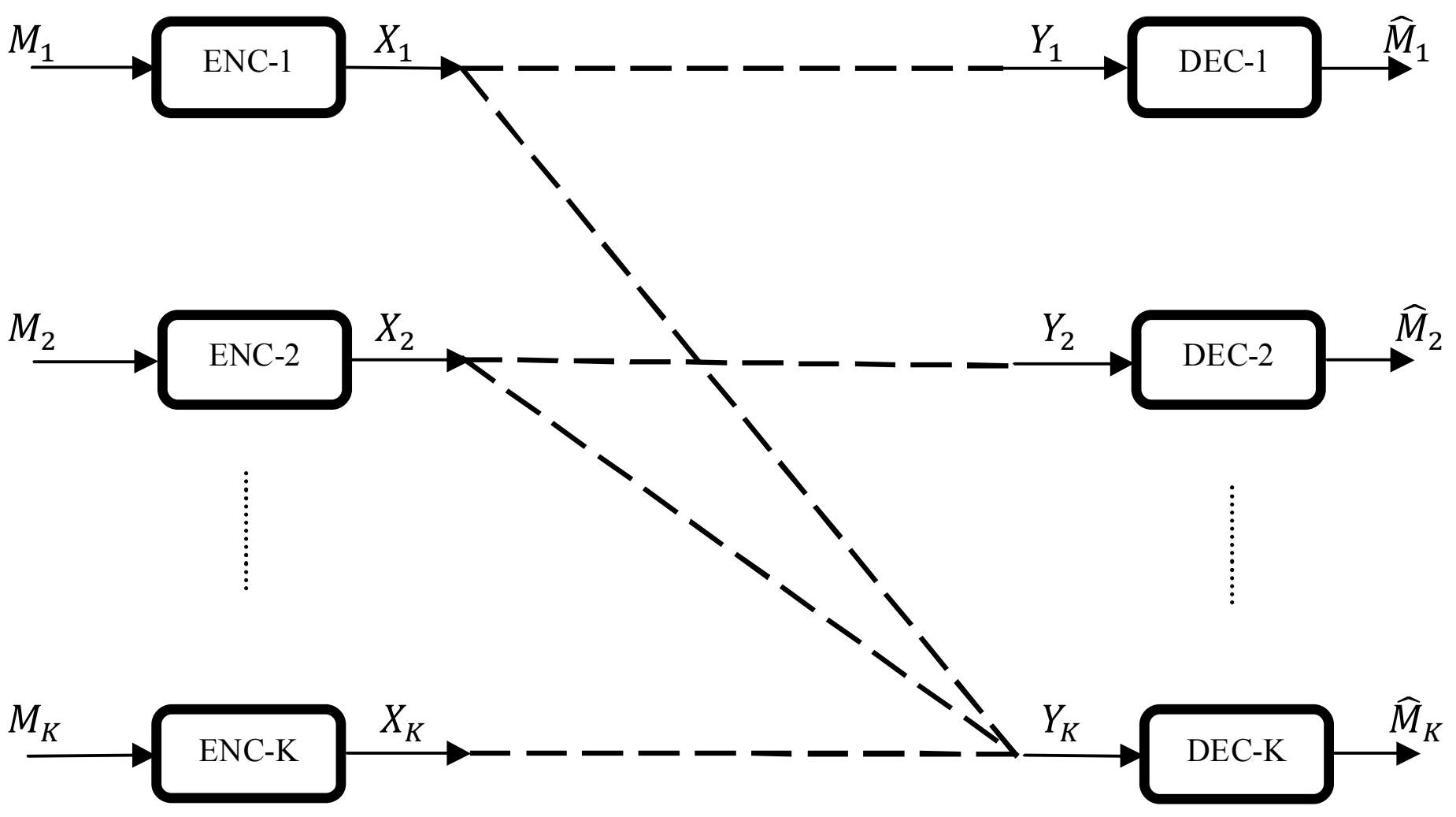

Figure 8. The K-user many-to-one interference channel. Each receiver has been linked to its connected transmitters by a dashed line.

This is a special class of the K-user interference channel where only one receiver experiences interference. In this case, the channel transition probability function is factorized as follows:

$$\mathbb{P}_{Y_1\ldots Y_K|X_1\ldots X_K}=\mathbb{P}_{Y_1|X_1}\mathbb{P}_{Y_2|X_2}\mathbb{P}_{Y_3|X_3}\ldots\mathbb{P}_{Y_K|X_K}\mathbb{P}_{Y_K|X_1X_2X_3\ldots X_K} \tag{163}$$

Define $\boldsymbol{Y}_{G(1)},\boldsymbol{Y}_{G(2)}$ as follows:

$$\boldsymbol{Y}_{G(1)}\triangleq\{Y_1,Y_2,\ldots,Y_{K-1}\},\qquad \boldsymbol{Y}_{G(2)}\triangleq\{Y_K\} \tag{164}$$

Now by letting $\mu=2$ and substituting (164) in Theorem 9, we obtain that if the following condition holds:

$$I(U;Y_K|X_K)\leq I(U;Y_1,Y_2,\ldots,Y_{K-1}|X_K),\quad \text{for all joint PDFs}\quad P_{UX_1X_2\ldots X_{K-1}}P_{X_K}, \tag{165}$$

Then the sum-rate capacity is bounded above as:

$$\mathcal{C}_{sum}^{m-t-o}\leq \max_{\substack{P_QP_{M_1}P_{M_2}\ldots P_{M_K}\\ \times\prod_{i=1}^{K}P_{X_i|M_iQ}}}\big(I(M_1,M_2,\ldots,M_{K-1};Y_1,Y_2,\ldots,Y_{K-1}|M_K,Q)+I(M_K;Y_K|Q)\big) \tag{166}$$

One can easily show that the expression (166) can be simplified as follows:

$$\begin{aligned}\mathcal{C}_{sum}^{m-t-o}&\leq \max_{P_QP_{X_1|Q}P_{X_2|Q}\ldots P_{X_K|Q}}\big(I(X_1,X_2,\ldots,X_{K-1};Y_1,Y_2,\ldots,Y_{K-1}|X_K,Q)+I(X_K;Y_K|Q)\big)\\ &=\max\nolimits_{P_QP_{X_1|Q}P_{X_2|Q}\ldots P_{X_K|Q}}\big(I(X_1,X_2,\ldots,X_{K-1};Y_1,Y_2,\ldots,Y_{K-1}|X_K,Q)+I(X_K;Y_K|Q)\big)\\ &\overset{(a)}{=}\max\nolimits_{P_QP_{X_1|Q}P_{X_2|Q}\ldots P_{X_K|Q}}\big(I(X_1;Y_1|Q)+I(X_2;Y_2|Q)+\cdots+I(X_K;Y_K|Q)\big)\end{aligned} \tag{167}$$

where equality (a) is due to the factorization (163). Now, one can achieve (167) by a simple treating interference as noise strategy, i.e., the outer bound is actually optimal. Thus, the sum-rate capacity of the many-to-one interference channel, if the less noisy condition (165) holds, is given by:

$$\mathcal{C}_{sum}^{m-t-o}=\max_{P_QP_{X_1|Q}P_{X_2|Q}\ldots P_{X_K|Q}}\big(I(X_1;Y_1|Q)+I(X_2;Y_2|Q)+\cdots+I(X_K;Y_K|Q)\big) \tag{168}$$

In fact, by following the same steps as Theorem 6, one can prove that the sum-rate capacity is still given by (168) if:

$$I(U;Y_K|X_K)\leq I(U;Y_1,Y_2,\ldots,Y_{K-1}|X_K),\qquad \text{for all joint PDFs}\quad P_{X_1}P_{X_2}\ldots P_{X_K}P_{U|X_1X_2\ldots X_K} \tag{169}$$

This is a weaker condition than (165).

Similarly, many other scenarios can be identified for which the outer bound derived in Theorem 9 yields the exact sum-rate capacity.

## IV.C) A Note on Strong Interference Regime

In Part III of our multi-part papers [3], we identified strong interference regimes for arbitrary interference networks. Using Lemma 5 derived in Subsection IV.B, one may obtain new strong interference regimes for networks with more than two-receivers. Let us provide an example. Again consider the three-user CIC shown in Fig. 7. In Part III [3, Sec. V.B.1], we proved that if either one of the following conditions hold:

$$\begin{cases} I(X_2;Y_2|X_1,X_3) \le I(X_2;Y_3|X_1,X_3) & \text{for all joint PDFs} \quad P_{X_1}P_{X_2}P_{X_3} \\ I(X_2,X_3;Y_3|X_1) \le I(X_2,X_3;Y_1|X_1) & \text{for all joint PDFs} \quad P_{X_1}P_{X_2X_3} \end{cases}, \tag{170}$$

$$\begin{cases} I(X_3;Y_3|X_1,X_2) \le I(X_3;Y_2|X_1,X_2) & \text{for all joint PDFs} \quad P_{X_1}P_{X_2}P_{X_3} \\ I(X_2,X_3;Y_2|X_1) \le I(X_2,X_3;Y_1|X_1) & \text{for all joint PDFs} \quad P_{X_1}P_{X_2X_3} \end{cases}, \tag{171}$$

then the receiver $Y_1$ experiences strong interference, i.e., it is optimal to decode all messages at this receiver. Using Lemma 5, we now provide a new set of conditions for this purpose. Specifically, consider the following conditions:

$$\begin{cases} I(U,X_2;Y_2|X_1) \le I(U,X_2;Y_1|X_1), & \text{for all joint PDFs} \quad P_{X_1}P_{UX_2X_3} \\ I(X_3;Y_3|X_1,X_2) \le I(X_3;Y_1|X_1,X_2), & \text{for all joint PDFs} \quad P_{X_1}P_{X_2}P_{X_3} \end{cases} \tag{172}$$

Note that the conditions (172) are different from both (170) and (171). We intend to prove that if these new conditions hold, then there exist a joint PDF $P_Q P_{X_1|Q} P_{X_2|Q} P_{X_3|Q}$ such that:

$$\begin{aligned} R_1 &\le I(X_1;Y_1|X_2,X_3,Q) \\ R_2 &\le I(X_2;Y_1|X_1,X_3,Q) \\ R_3 &\le I(X_3;Y_1|X_1,X_2,Q) \\ R_1+R_2 &\le I(X_1,X_2;Y_1|X_3,Q) \\ R_2+R_3 &\le I(X_2,X_3;Y_1|X_1,Q) \\ R_1+R_3 &\le I(X_1,X_3;Y_1|X_2,Q) \\ R_1+R_2+R_3 &\le I(X_1,X_2,X_3;Y_1|Q) \end{aligned} \tag{173}$$

In what follows, we only derive the last bound on the sum-rate; the derivation of the other bound is rather similar. Consider a code of length $n$ for the network with vanishing average error probability. Using the Fano's inequality, we can write:

$$\begin{aligned} n(R_1+R_2+R_3) &\le I(M_3;Y_3^n) + I(M_2;Y_2^n) + I(M_1;Y_1^n) + n\epsilon_n \\ &\overset{(a)}{\le} I(M_3;Y_3^n|M_1,M_2) + I(M_2;Y_2^n|M_1) + I(M_1;Y_1^n) + n\epsilon_n \\ &\overset{(b)}{\le} I(M_3;Y_3^n|M_1,M_2) + I(M_1,M_2;Y_1^n) + n\epsilon_n \\ &\overset{(c)}{\le} I(M_1,M_2,M_3;Y_1^n) + n\epsilon_n \le \textstyle\sum_{t=1}^n I\left(X_{1,t},X_{2,t},X_{3,t};Y_{1,t}\right) + n\epsilon_n \end{aligned} \tag{174}$$

where $\epsilon_n \to 0$ as $n \to \infty$; the inequality (a) holds because the messages are independent, the inequality (b) is derived by the first condition of (172) and Lemma 5, and the inequality (c) is derived by the second condition of (172) and Lemma 5 (see also Remark 5).

Thus, the conditions (172) also represent a strong interference regime for the receiver $Y_1$. By combining these conditions with appropriate strong interference conditions for the other receivers, one may obtain new strong interference regimes for the three-user CIC. However, one can verify that such regimes do not yield significant situations other than those derived in Part III [3, Sec. V.B.1]. Moreover, these new strong interference conditions always include some auxiliary random variables in their characterizations. Therefore, although it is possible to develop the above approach to derive new strong interference regimes for large multi-message networks, but this is not of sufficient interest.

## CONCLUSION

In this part of our multi-part papers, we identified classes of interference networks with a sequence of less-noisy receivers for which a successive decoding scheme achieves the sum-rate capacity. First, in Section III, we analyzed the two-receiver networks. We demonstrated that the unified outer bounds derived in Part III are sum-rate optimal for network scenarios which satisfy certain less-noisy conditions. Next, we considered the multi-receiver networks. One of the main difficulties in analysis of such scenarios is how to establish useful capacity outer bounds. In Subsection IV.A, we developed a novel technique requiring a sequential application of the Csiszar-Korner identity to establish powerful single-letter outer bounds on the sum-rate capacity of multi-receiver interference networks which satisfy certain less-noisy conditions. By using these outer bounds, we derived a full characterization of the sum-rate capacity for general interference networks of arbitrary large sizes with a sequence of less-noisy receivers. Finally, in Subsection IV.B, we presented some generalizations of our outer bounds and showed that they are useful to obtain exact sum-rate capacity for various scenarios.

Our systematic study of fundamental limits of communications in interference networks will be followed in Part V [5] (see also [7, 8]), where we will design a random coding scheme for the networks with arbitrary configurations.

## ACKNOWLEDGEMENT

The author would like to appreciate Dr. H. Estekey, head of School of Cognitive Sciences, IPM, Tehran, Iran, and Dr. R. Ebrahimpour, faculty member at School of Cognitive Sciences, for their kindly support of the first author during this research. He also greatly thanks his mother whose love made this research possible for him. Lastly, F. Marvasti is acknowledged whose editing comments improved the language of this work.

## REFERENCES

[1] R. K. Farsani, "Fundamental limits of communications in interference networks-Part I: Basic structures," IEEE Trans. *Information Theory, Submitted for Publication, 2012, available at* http://arxiv.org/abs/1207.3018.

[2] __________, "Fundamental limits of communications in interference networks-Part II: Information flow in degraded networks," IEEE Trans. *Information Theory, Submitted for Publication, 2012, available at* http://arxiv.org/abs/1207.3027.

[3] __________, "Fundamental limits of communications in interference networks-Part III: Information flow in strong interference regime," IEEE *Trans. Information Theory, Submitted for Publication, 2012, available at* http://arxiv.org/abs/1207.3035.

[4] __________, "Fundamental limits of communications in interference networks-Part IV: Networks with a sequence of less-noisy receivers," IEEE *Trans. Information Theory, Submitted for Publication, 2012, available at* http://arxiv.org/abs/1207.3040.

[5] __________, "Fundamental limits of communications in interference networks-Part V: A random coding scheme for transmission of general message sets," IEEE *Trans. Information Theory, To be submitted, preprint available at* http://arxiv.org/abs/1107.1839.

[6] __________, " Capacity theorems for the cognitive radio channel with confidential messages," 2012, *available at* http://arxiv.org/abs/1207.5040.

[7] R. K. Farsani and F. Marvasti, "Interference networks with general message sets: A random coding scheme," [Online] *available at*: http://arxiv.org/abs/1107.1839.

[8] ______________________, "Interference networks with general message sets: A random coding scheme", 49th *Annual Allerton Conference on Communication, Control, and Computing*., Monticello, IL, Sep. 2011.

[9] A. El Gamal and Y.-H. Kim, "*Lecture notes on network information theory*," *arXiv:1001.3404*, 2010.

[10] M. Maddah-Ali, A. Motahari, and A. Khandani, "Communication over MIMO X channels: Interference alignment, decomposition, and performance analysis," *IEEE Trans. Inf. Theory*, vol. 54, no. 8, pp. 3457–3470, Aug. 2008.

[11] V. R. Cadambe and S. A. Jafar, "Interference alignment and degrees of freedom of the K-user interference channel," *Information Theory, IEEE Transactions on*, vol. 54, no. 8, pp. 3425–3441, 2008.

[12] S. A. Jafar, "Interference alignment: A new look at signal dimensions in a communication network," *Foundations and Trends in Communications and Information Theory*, vol. 7, no. 1, pp. 1–136, 2011.

[13] J. Körner and K. Marton, "A source network problem involving the comparison of two channels II," in *Trans. Colloquium Inform. Theory*, Keszthely, Hungary, Aug. 1975.

[14] A. El Gamal, "The capacity of a class of broadcast channels," *IEEE Trans. Inf. Theory*, vol. IT-25, no. 2, pp. 166–169, Mar. 1979.

[15] I. Csiszár and J. Körner, "Broadcast channels with confidential messages," *IEEE Trans. Inf. Theory*, vol. IT-24, no. 3, pp. 339–348, May 1978.

[16] R. K. Farsani, "The K-user interference channel: strong interference regime," *2012, available at* http://arxiv.org/abs/1207.3045.

[17] __________, "How much rate splitting is required for a random coding scheme? A new achievable rate region for the broadcast channel with cognitive relays," 50th *Annual Allerton Conference on Communication, Control, and Computing*., Monticello, IL, Oct. 2012.